\documentclass[]{elsarticle}
\usepackage[scriptsize]{subfigure}
\usepackage{color}
\usepackage{tikz}
\usepackage{newlfont}
\usepackage{multirow}
\usepackage{array}
\usepackage[font=footnotesize]{caption} 
\DeclareCaptionStyle{period-newline}%
[labelsep=period,justification=centering]%
{labelsep=period-newline}
\usepackage{longtable} 
\usepackage{ifthen}
\usepackage{alltt}
\usepackage{float}
\usepackage{enumerate}
\usepackage[text={6.5in,8.5in},centering]{geometry}
\newcolumntype{C}[1]{>{\centering\let\newline\\\arraybackslash\hspace{0pt}}m{#1}}
\newcolumntype{L}[1]{>{\raggedright\let\newline\\\arraybackslash\hspace{0pt}}m{#1}}
\newcolumntype{R}[1]{>{\raggedleft\let\newline\\\arraybackslash\hspace{0pt}}m{#1}}
\newcommand{\ba}{\begin{array} }
\newcommand{\ea}{\end{array} }
\newcommand{\bae}{\begin{eqnarray}}
\newcommand{\eae}{\end{eqnarray}}
\newcommand{\bea}{\begin{eqnarray*}}
\newcommand{\eea}{\end{eqnarray*}}
\newcommand{\be}{\begin{equation}}
\newcommand{\ee}{\end{equation}}

\newcommand{\pr}{{\bf Proof}~~}

\def\to{{\rightarrow}}

\def\v{\vspace{1ex}}

\def\to{{\rightarrow}}

\usepackage{amsfonts}
\usepackage{amssymb}
\usepackage{amsthm}
\usepackage{ mathrsfs }
 \usepackage{ae} 
\usepackage[T1]{fontenc}
\usepackage[ansinew]{inputenc}
\usepackage{amsmath}
\usepackage{eucal}

\usepackage[english]{babel}

\usepackage{color}
\usepackage{hyperref}
\hypersetup{
     colorlinks   = true,
     citecolor    = blue
}
\usepackage{lscape}
\usepackage{graphicx}
\usepackage{epstopdf}	
\newtheorem{theorem}{\hskip\parindent\bf Theorem}[section]

\usepackage[title]{appendix}
\usepackage{etoolbox}

\patchcmd{\appendices}{\quad}{: }{}{}

\usepackage{booktabs}
\usepackage{bm} 
\usepackage[cal=cm]{mathalfa} 

\makeatletter
\ifcase \@ptsize \relax
  \newcommand{\miniscule}{\@setfontsize\miniscule{2}{3}}
\or
  \newcommand{\miniscule}{\@setfontsize\miniscule{2}{3}}
\or
  \newcommand{\miniscule}{\@setfontsize\miniscule{2}{3}}
\fi
\makeatother
\usepackage{lipsum}

\usepackage{float}

\usepackage{titlesec}
\titleformat*{\section}{\large\bfseries}
\titleformat*{\subsection}{\large\bfseries}

\usepackage{relsize}

\usepackage{pbox} 

\usepackage{pifont}
\begin{document}

\begin{frontmatter}

\title{Population dynamics of {\it Varroa} mite and honeybee: Effects of parasitism with age structure and seasonality}
\author[2]{Komi Messan} \ead{Komi.S.Messan@erdc.dren.mil}
\author[3]{Marisabel Rodriguez Messan}\ead{Marisabel@asu.edu}
\address[2]{Simon A. Levin Mathematical and Computational Modeling Sciences Center, Arizona State University, \\Tempe, AZ 85281, USA.}
\address[3]{Department of Ecology and Evolutionary Biology, Brown University, Providence RI 02912 USA}
\author[2]{Jun Chen} \ead{jchen152@asu.edu}
\author[4]{Gloria DeGrandi-Hoffman } \ead{gloria.hoffman@ars.usda.gov}
\address[4]{Carl Hayden Bee Research Center, United States Department of Agriculture-Agricultural Research Service,\\ Tucson, AZ  85719 , USA.}
\author[6]{Yun Kang} \ead{yun.kang@asu.edu}
\address[6]{Sciences and Mathematics Faculty, College of Integrative Sciences and Arts, Arizona State University, \\Mesa, AZ 85212, USA.}





\begin{abstract} \\
 Honeybees play an important role in the production of many agricultural crops and in sustaining plant diversity in undisturbed ecosystems. The rapid decline of honeybee populations have sparked great concern worldwide. Field and theoretical studies have shown that the parasitic {\it Varroa} mite (\emph{Varroa destructor} Anderson and Trueman) could be the main reason for colony losses. In order to understand how mites affect population dynamics of honeybees and the health of a colony, we propose a brood-adult bee-mite interaction model in which the time lag from brood to adult bee is taken into account. Noting that the temporal dynamics of a honeybee colony varies with respect to season, we validate the model and perform parameter estimations under both constant and fluctuating seasonality scenarios. Our analytical and numerical studies reveal the following: (a) In the presence of parasite mites, the large time lag from brood to adult bee could destabilize population dynamics and drive the colony to collapse; however the small natural mortality of the adult bee population can promote a mite-free colony when time lag is small or at an  intermediate level; (b) Small brood' infestation rates could stabilize all populations at the unique interior equilibrium under constant seasonality while may drive the mite population to die out when seasonality is taken into account; (c) High brood' infestation rates can destabilize the colony dynamics leading to population collapse depending on initial population size under constant and seasonal conditions; (d) Results from our sensitivity analysis indicate that the queen's egg-laying may have the greatest effect on colony population size. The  death rate of the brood and the colony size at which brood survivability is the half maximal were also shown to be highly sensitive with an inverse correlation to the colony population size. Our results provide insights on the effects of seasonality on the dynamics. For example, mites may die out leaving a healthy colony with brood and adult bees in the presence of seasonality while the colony collapses without seasonality.\\


 \end{abstract}


\end{frontmatter}

\section{Introduction}
Honeybees (\emph{Apis mellifera}) are exemplars of social evolution, known for their complex social organization \citep{linksvayer2009honeybee}, and are the most economically valuable pollinators of crops in the world  \citep{doublet2015bees,klein2007importance}. However, honeybees colonies are being lost at alarming rates particularly over winter \citep{doublet2015bees,vanengelsdorp2012national,neumann2010honey}. Honeybee colony health is challenged by different factors including diseases such as American and European foulbrood, Chalkbrood, Stonebrood, and Nosema, parasitism (e.g. mites), and nutritional stress \citep{oldroyd2007s}. Most notably, the {\it Varroa} mite has posed a huge threat on the honeybees well-being \citep{degrandi2017dispersal,koleoglu2017effect,genersch2010german, guzman2010varroa,van2012winter,  kang2016disease, degrandi2016population, sumpter2004dynamics}. \\

The dynamics within a honeybee colony are complicated and characterized by different behaviors performed by worker bees at different ages. Referred to as division of labor, the behaviors include caring for the queen, brood rearing, foraging, food storing, and nest defense. Specifically, brood rearing and colony growth depend on queen's egg-laying activity that relies upon successful foraging by the workers thus creating the dynamics of a feedback system of interdependent elements \citep{degrandi1989beepop}. Honeybees must go through an optimal collective-decision making process in order to sustain the colony. \\

The {\it Varroa} mite is the most adverse parasite of the honeybee associated with a high percentage of colony losses over the winter \citep{koleoglu2017effect}. Mites are known to parasitize both brood and adult bees. Mites are rarely found attached to queens \citep{degrandi2017dispersal}. Mites affect honeybees in different ways, either by direct physical damage or activation of viruses \citep{koleoglu2017effect}. For instance, parasitized bee brood develop into adults with shorter abdomens, deformed wings and shorter lifespans \citep{koleoglu2017effect,de1982weight,allen1996incidence} due to suppressed expression of genes related to longevity and development (e.g. protein storage, vitellogenin, etc.) \citep{navajas2008differential}. Also, parasitized foragers are more likely to get lost and wander between colonies \citep{kralj2007parasitic,koleoglu2017effect}. \\

The spread of mites through the bee population can be both vertically, which occurs when phoretic mites travel upon the swarming bees, and horizontally, which occurs when infested brood is moved between colonies by beekeepers, or when foragers are robbing honey from other colonies \citep{peck2016varroa}. Availability of brood in the colony is paramount for {\it Varroa} mites reproduction given that it occurs in capped worker and drone brood cells when a mature female mite (foundress) enters the cell preceding capping \citep{degrandi2017dispersal,rosenkranz2010biology}. The foundress starts feeding on the brood and continues to feed regularly thereafter \citep{donze1996effect}. The foundress lays its first egg which develops into a male and the second one into a female mite that mates with the male \citep{degrandi2017dispersal}. The mother mite keeps feeding on the developing larva, and in the process, transmits several viruses \citep{degrandi2017dispersal}. After the bee is fully developed and emerges from the capped cell, the mother mite and offspring emerge with it and attach to other adult bees as ``phoretic mites" \citep{degrandi2017dispersal}. In general, phoretic mites target nurse bees \citep{del2010selection,cervo2014high} because they remain in the brood area and serve as a medium to transport mites to brood cells where they can reproduce \citep{degrandi2017dispersal}. In recent years, the population growth rates of {\it Varroa} have exceeded those expected \citep{degrandi2014population,degrandi2016population,degrandi2017dispersal}, causing this pest to be difficult to control and a major factors in colony losses. \\

Mathematical models have been powerful tools to help us understand the effects of mites (e.g.,\cite{kang2016disease}) , disease (e.g. \cite{ratti2012mathematical,kang2016disease}), and pesticides (e.g.,\cite{MagalWebbWu2019,MagalWebbWu2020}) on honeybee population dynamics. There are some models that are introduced to explore the role of mite infestation in honeybee colonies  (see the work of \cite{ratti2012mathematical, kribs2014modeling, ratti2015mathematical, messan2017migration}). Kang et al. \cite{kang2016disease} proposed a honeybee-mite-virus model that incorporates parasitic interactions between honeybees and {\it Varroa} mites in addition to a virus transmission dynamics. In this study, it was found that low adult bees to brood ratios have destabilizing effects on the system, can generate fluctuating dynamics, and potentially lead to a catastrophic event where both honeybees and mites suddenly become extinct within a colony. However, \citep{kang2016disease} did not explicitly model the brood population thus omitting the role of brood population size on mite's proliferation. Becher et al. \cite{becher2014beehave} constructed an agent-based model to explore how various stressors (including {\it Varroa} mites, virus infections, impaired foraging behavior, changes in landscape structure, pesticides, etc.) affect the performance of single managed honeybee colonies. While the latter work provides valuable results on different mechanisms that may induce the decline of honeybee population in a colony, brood population was not taken into account explicitly. An another approach studying the effects of {\it Varroa} mites infestation on honeybees through dispersal mechanisms was done in \citep{messan2017migration}. This study contains a complete analysis of local and global dynamics of a two-patch model that incorporates mite migration through bees' foraging activities. The results of this study provide insight on different scenarios where different migration rates can affect the bee population negatively or drive the mite population extinct within the colony. However, this study lacks focus on the different population dynamics that can arise within a colony from mechanisms such as reproduction and parasitism of mites on brood and adult bees.\\

Motivated by work of \citep{degrandi1989beepop, messan2017migration}, we propose a single-patch stage-structure delay differential equation model that considers the time lag from brood to adult bees with parasite mites in seasonal environment. The modeling framework used here is inspired by the work of \citep{aiello1990time}. In the proposed model, we focus on specific mechanisms related to mite reproduction and parasitism effects on the bees' life cycle. Unlike other studies, such as some of the ones mentioned earlier, our proposed model incorporates individual mechanisms of brood and adult bees, and their interactions with mites that give rise to different population dynamics at the colony level. We  assess the effects of different parameters affecting the population dynamics inside the colony. More specifically, we aim to use this model to explore how the synergistic  effects of age structure and parasitism affect the colony dynamics in seasonal environment. The seasonality in our model is reflected through the seasoning varying egg-laying rate by the queen bee.\\


\section{Model Derivation}\label{ModelDerivation}

\begin{figure}[ht]
 \vspace{-10pt}
\begin{center}
\includegraphics[scale=0.55]{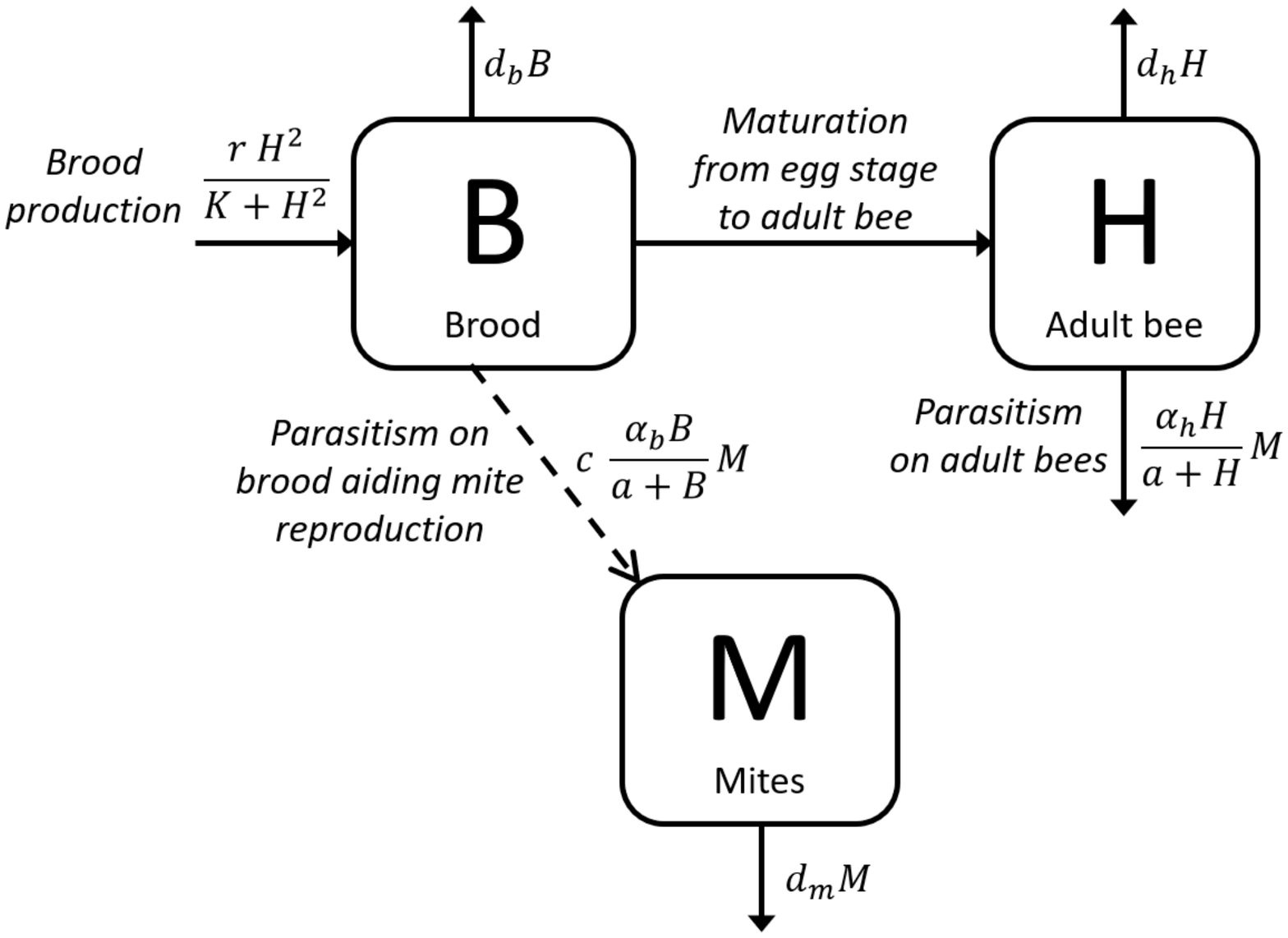}
  \end{center}
\caption{Schematic diagram for the honeybee-mite parasitic interaction. The solid lines represent direct interactions. The dashed line represents indirect interactions, where one population affects the \emph{rate} of another interaction.}
\label{fig:Model_Diagram1}
\end{figure}

Let $B(t)$, $H(t)$, and $M(t)$ denote the total population of brood, adult honeybees, and mites at time $t$, respectively. Following the schematic diagram in Figure \eqref{fig:Model_Diagram1}, our model has the following assumptions:\\


%

Let $\tau>0$ be the time interval in which the population entering an homogeneous environment equals to the length of time from egg to fully developed adult bee. We assume that  all populations are known during the interval  $-\tau\leq t\leq 0$. 
More specifically, let $B_0(t)$ be the observed or assumed egg-laying rate of the queen during the time interval $-\tau\leq t\leq0$.   
The population of adult honeybee $H(t)$ and mite $M(t)$ are being constant during the time interval $-\tau\leq t\leq0$, and we denote that $H(t)=H(0)$ and $M(t)=M(0)$ for all time $t$ in $-\tau\leq t\leq0$. Our model is derived as follows and takes one form on the interval $0<t\leq\tau$ and a second form on the interval $t>\tau$:\\
\begin{enumerate}
\item We assume that at any time $t>0$, the brood population, $B$, increases through the successful survivability of an egg into pupae stage represented by the term $\frac{H^2}{K+H^2}$, which incorporates the collaborative efforts of adult workers, via division of labor. This term assumes that successful colonies produce more brood and efficient workers, an assumption supported by  \citep{schmickl2007hopomo, kang2016disease, eischen1984some}. 
\item The brood, adult bee, and mite populations are assumed to have a natural average death rate proportional to the existing population denoted with constant parameters $d_{b}$, $d_h$, and $d_{m}$, respectively.

\item Both the brood and adult bee populations decrease through the parasitism effect of mites. The probability of mites attaching to brood and adult bees is modeled with the terms $\frac{B}{a+B}$ and $\frac{H}{a+H}$, respectively, where $a$ is the size of the brood or adult bee population, accordingly, at which the rate of attachment is half maximal (see a similar approach in \cite{sumpter2004dynamics, betti2014effects}). The parameters $\alpha_b$ and $\alpha_h$ measure the parasitism rate of mites on the brood and adult bees, respectively.

\item The mite population increases through the parasitism effects of brood that aids mite reproduction. The work of \citep{steiner1994first, garrido2003reproductive, boot1997reproductive} suggests that initiation of oocyte development in {\it Varroa} depends on whether the female enters the brood cells of {\it Apis mellifera} before operculation, thus, the term $\frac{c\alpha_bB}{a+B}$ accounts for the production of new mites, where $c$ is the conversion factor from brood to mite population. The mite equation could hence be described by:
\begin{align}
\frac{dM}{dt} = \left[\frac{c\alpha_bB}{a+B}-d_m\right]M
\label{MiteModel}
\end{align}

\item The life cycle of the female {\it Varroa} is normally subdivided into a phoretic phase in which it lives on adult bees and a reproductive phase occurring within worker or drone brood cells. Thus, the two life stages should be modeled explicitly. However, from the work of \citep{kang2016disease, messan2017migration}, we assume an implicit age structure for the mite population where the ratio of different stages are constant. For example, consider $\xi \in [0,1]$ to be the percentage of mites at the non-phoretic stage, then $(1-\xi)M$ is the phoretic mite population. Let $\hat{d}_{m} = d_{m}(1-\xi)$, then the phoretic mite equation becomes
$$ \frac{dM}{dt} =  \frac{c\alpha_bBM}{a+B}-d_{m}(1-\xi)M =   \frac{c\alpha_bBM}{a+B}-\hat{d}_{m}M.$$
Similar approach can be followed to find the reproductive mite population and by grouping the reproductive and phoretic mites together, we obtain the mite model defined as in in \eqref{MiteModel}.

\item We assume that eggs laid by the queen at time $t-\tau$, which survive to time $t$, i.e. $\frac{rH(t-\tau)^2}{K+H(t-\tau)^2}$, exit (or mature) from the brood population $B$ and enter the adult bee population $H$. The survival of the brood depends on their own natural death and if they survive the mite infestation. Therefore, the probability of survival is $e^{-\int_{t-\tau}^{t}\left( d_b+\frac{\alpha_bM(s)}{a+B(s)}\right)ds}$ when $0 < t \leq \tau$ and $t>\tau$. We follow a similar approach and derivation as in \citep{aiello1990time} and obtain the number of brood that survive into adult bees:
$$B_0(t-\tau)e^{-\int_{t-\tau}^{t}\left( d_b+\frac{\alpha_bM(s)}{a+B(s)}\right)ds} \quad \quad \mbox{when}  \quad \quad  0 < t \leq \tau$$
and
$$\frac{rH(t-\tau)^2}{K+H(t-\tau)^2}e^{-\int_{t-\tau}^{t}\left( d_b+\frac{\alpha_bM(s)}{a+B(s)}\right)ds}  \quad \quad \mbox{when}  \quad \quad t>\tau.$$

\item For continuity of initial conditions and following a similar approach as in \citep{aiello1990time}, the total surviving brood population from the observed eggs laid on $-\tau\leq t\leq0$ is
\[B(0) = \int_{-\tau}^{0}B_0(t)dt>0.\]\\

%


\end{enumerate}

The model formulated by the assumptions provided above is therefore composed by two time intervals, $0 < t \leq \tau$ and $t > \tau$, in the following form:\\

 For  $t\in(0  , \tau]$,
{\small
\begin{equation}
\begin{aligned}
\frac{dB}{dt} &= \underbrace{\frac{rH^2}{K+ H^2}}_{\text{egg production}} - \quad\underbrace{\alpha_b  \overbrace{\frac{B}{a+B}}^{\substack{\text{probability of $M$}\\\text{attaching to $B$}}} M}_{\text{parasitism on brood}}\quad - \underbrace{d_b B}_{\text{natural death}} 
- \quad \underbrace{e^{-\int_{t-\tau}^t \left[d_b+\frac{\alpha_b M(s)}{a+B(s)}\right] ds}B_0(t-\tau)}_{\text{maturation from egg to adult}} \\
\frac{dH}{dt} &= \underbrace{e^{-\int_{t-\tau}^t \left[d_b+\frac{\alpha_b M(s)}{a+B(s)}\right] ds}B_0(t-\tau)}_{\text{transition from brood}}\quad - \quad \underbrace{\alpha_h  \overbrace{\frac{H}{a+H}}^{\substack{\text{probability of $M$}\\\text{attaching to $H$}}} M}_{\text{parasitism on adult bee}}\quad - \underbrace{d_h H}_{\text{natural death}}\\[.2cm]
\frac{dM}{dt} &= \underbrace{c\alpha_b \frac{B}{a+B}M}_{\text{ newborns from brood parasitism}}- \underbrace{d_m M }_{\text{natural death}}
\end{aligned} \label{BHM1}
\end{equation}
}
For $t > \tau$,
{\small
\begin{equation}
\begin{aligned}
\frac{dB}{dt} &= \underbrace{\frac{rH^2}{K+ H^2}}_{\text{egg production}} - \quad \underbrace{\alpha_b  \overbrace{\frac{B}{a+B}}^{\substack{\text{probability of $M$}\\\text{attaching to $B$}}} M}_{\text{parasitism on brood}} \quad -\quad  \underbrace{d_b B}_{\text{natural death}} 
- \quad \underbrace{\frac{e^{-\int_{t-\tau}^t \left[d_b+\frac{\alpha_b M(s)}{a+B(s)}\right] ds}rH(t-\tau)^2}{K+ H(t-\tau)^2}}_{\text{maturation from egg to adult}} \\[.2cm]
\frac{dH}{dt} &= \underbrace{\frac{e^{-\int_{t-\tau}^t \left[d_b+\frac{\alpha_b M(s)}{a+B(s)}\right] ds}rH(t-\tau)^2}{K+ H(t-\tau)^2}}_{\text{transition from brood}}\quad -\quad\underbrace{\alpha_h  \overbrace{\frac{H}{a+H}}^{\substack{\text{probability of $M$}\\\text{attaching to $H$}}} M}_{\text{parasitism on adult bee}} \quad - \underbrace{d_h H}_{\text{natural death}}\\[.2cm]
\frac{dM}{dt} &= \underbrace{c\alpha_b \frac{B}{a+B}M}_{\text{newborns from brood parasitism}}- \underbrace{d_m M }_{\text{natural death}}
\end{aligned} \label{BHM2}
\end{equation}
}\\

with initial conditions
   \begin{equation}\label{initial1}
   B(t)=B_0(t)>0,~ t\in[-\tau,0], ~ H(0)>0, ~M(0)\geq0,
   \end{equation}
   
where $B_0(t)\in \mathcal{C}:=C([-\tau,0],[0,+\infty))$ is the brood population at $t\in[-\tau,0]$. With these initial conditions, our model describes a scenario where mite reproduction and parasitism initiate at time $t>0$. We note that the age structure of mite population was not taken into account.\\

 
  In the next section, we compare the dynamics of our proposed model with, and without, parasite $M(t)$ to gain insights on the effects of mites.\\


\section{Mathematical Analysis}\label{sec_Math_Analysis}
We first provide the basic dynamical properties of Model \eqref{BHM1}-\eqref{BHM2} as follows:

\begin{theorem}\label{th1:pb}
Solution $(B(t), H(t), M(t))$ of System \eqref{BHM1}-\eqref{BHM2} satisfying \eqref{initial1} is positive for all $t>0$. In addition,
\begin{equation*}
\limsup_{t\rightarrow\infty}(B(t)+H(t)+M(t))\leq \frac{cr}{\min\{d_b,d_h, d_m\}}.
\end{equation*}
\end{theorem}


 \noindent\textbf{Biological Implications:}  Theorem \ref{th1:pb} implies that Model \eqref{BHM1}-\eqref{BHM2} is well-defined biologically as it is positively invariant and bounded.
 
\subsection{Dynamics of Honeybees}

\noindent The honeybee-mite system \eqref{BHM1}-\eqref{BHM2} reduces to the following honeybee-only subsystem \eqref{subBH1}-\eqref{subBH2} when $M(0)=0$:\\

For $t \in(0,\tau]$,
\begin{equation}
\begin{aligned}
\frac{dB}{dt}&= \frac{rH^2(t)}{K+H^2(t)}-d_b B(t)- e^{- d_b \tau} B_0(t-\tau)  \\
\frac{dH}{dt}&= e^{- d_b \tau} B_0(t-\tau)-d_h H(t)
\label{subBH1}
\end{aligned}
\end{equation}

For $t > \tau$,
\begin{equation}
\begin{aligned}
\frac{dB}{dt}&= \frac{rH^2(t)}{K+H^2(t)}-d_b B(t)- e^{- d_b \tau} \frac{rH^2(t-\tau)}{K+H^2(t-\tau)}  \\
\frac{dH}{dt}&= e^{- d_b \tau} \frac{rH^2(t-\tau)}{K+H^2(t-\tau)}-d_h H(t)
\label{subBH2}
\end{aligned}
\end{equation}

The detailed dynamics of the honeybee-only subsystem \eqref{subBH1}-\eqref{subBH2} have been studied in \cite{chen2020model}. 
Note that its extinction equilibrium $E_{e}=(0,0)$ always exists. Let
\begin{eqnarray}\label{eqbt5}
H_{1,2}^*=\frac{r e^{-d_b \tau }}{2 d_h} \left(1\pm\sqrt{1-\left(\frac{2d_h e^{d_b \tau}}{r}\right)^2 K}\right),
\end{eqnarray} with $H_1^*\leq H_2^*$, hence the subsystem \eqref{subBH1}-\eqref{subBH2} has two interior equilibria $E_{i}=(B_i^*, H_i^*)$ with
\begin{eqnarray}\label{eqbt6}
B_i^*=\frac{1}{d_b}(1-e^{-d_b \tau})\frac{r(H_i^*)^2}{K+(H_i^*)^2}=\frac{d_h\left(e^{d_b\tau}-1\right)}{d_b} H^*_i, i=1,2.
\end{eqnarray}


Based on the work of Chen et al. \cite{chen2020model}, the summarized dynamical results of the subsystem \eqref{subBH1}-\eqref{subBH2} are as follows (also see Table \ref{tab:stability}):
\begin{enumerate}
\item The extinction equilibrium $E_e$ of the subsystem \eqref{subBH1}-\eqref{subBH2} always exists and is always locally asymptotically stable.
\item If $\frac{r}{d_h} < 2e^{d_b\tau}\sqrt{K}$, the subsystem \eqref{subBH1}-\eqref{subBH2} has global stability at the extinction equilibrium $E_e$.
\item If $\frac{r}{d_h} = 2e^{d_b\tau}\sqrt{K}$, the subsystem \eqref{subBH1}-\eqref{subBH2} has a unique interior equilibrium $E=(B^{*}, H^{*})=\left( \frac{r\left(1 - e^{-d_b\tau}\right)}{2d_b},\sqrt{K}\right)$, which is always locally asymptotically stable for any delay $\tau>0$.
\item If $\frac{r}{d_h} > 2e^{d_b\tau}\sqrt{K}$, the subsystem \eqref{subBH1}-\eqref{subBH2} has two attractors: the extinction equilibrium $E_e$ and the interior equilibrium $E_2=(B^*_2,H^*_2)$ which are locally asymptotically stable.
\end{enumerate}

\begin{table}[h]
\begin{center}
{\begin{tabular}{ | c | c | c | }
		\hline
		\textbf{Equilibrium} & \textbf{Existence} & \textbf{Stability}  \\
		\hline
		$E_{00}$ & Always & LAS and GAS if  $d_h > \frac{re^{-d_b\tau}}{2\sqrt{K}}$\\
		\hline
		$E_{B^* H^*}$ & $d_h =\frac{re^{-d_b\tau}}{2\sqrt{K}}$ &LAS\\
		\hline
		$E_{B_1^* H_1^*}$ and $E_{B_2^* H_2^*}$& $d_h < \frac{re^{-d_b\tau}}{2\sqrt{K}}$ & $E_{B_1^* H_1^*}$ is unstable and $E_{B_2^* H_2^*}$ is LAS \\
		\hline
	\end{tabular}}
	\caption{Summary dynamics of Model \eqref{subBH1}-\eqref{subBH2} where LAS: Locally Asymptotically Stable; GAS: Globally Asymptotically Stable.}\label{tab:stability}
\end{center}
	\end{table}

\subsection{Dynamics of the full system}

 First, we look at the equilibria of Model \eqref{BHM1}-\eqref{BHM2} by setting $\frac{dB}{dt}=\frac{dH}{dt}=\frac{dM}{dt}=0$. We obtain the subsequent equations:
 
\begin{subequations}
\begin{align}
& \frac{rH^2}{K+H^2} - \frac{\alpha_bBM}{a+B}-d_bB-\frac{rH^2}{K+H^2}e^{-\left(d_b+\frac{\alpha_bM}{a+B}\right)\tau}=0 \label{Equilibrium1a} \\[.2cm]
& \frac{rH^2}{K+H^2}e^{-\left(d_b+\frac{\alpha_bM}{a+B}\right)\tau} - \frac{\alpha_hHM}{a+H}-d_hH=0 \label{Equilibrium1b}\\[.2cm]
&\frac{c\alpha_bBM}{a+B}-d_mM=0 \label{Equilibrium1c}
\end{align}\label{Equilibrium1}
\end{subequations}

{From equations \eqref{Equilibrium1a}-\eqref{Equilibrium1c}, we know that if $\tau=0$, then system \eqref{BHM1}-\eqref{BHM2} has only the trivial boundary equilibrium $E_{000} = (0,0,0)$. Moreover, if $d_h < \frac{re^{-d_b\tau}}{2\sqrt{K}}$, then Model \eqref{BHM1}-\eqref{BHM2} has the following two boundary equilibria:

\[E_{B^{*}_1H^{*}_10} =(B^{*}_{1},H^{*}_1,0), \quad\mbox{  and  } \quad E_{B^{*}_2H^{*}_20}  =(B^{*}_{2},H^{*}_2,0)\]

where $B^*_i$ and $H^*_i$, $i=1,2$ are shown in \eqref{eqbt6} and \eqref{eqbt5} with $H_1^*\le H_2^*$. For the convenience of the reader, we show their expressions as follows:
\[B_i^*=\frac{d_h[e^{d_b\tau}-1]}{d_b} H^*_i, \quad\mbox{ and } \quad
H_i^*=\frac{e^{-d_b \tau } \left(d_b r\pm\sqrt{\left(d_b r\right)^2-4 d_b^2 d_h^2 K e^{2 d_b \tau }}\right)}{2 d_b d_h}.\]
The following theorem concerns the stability of these boundary equilibria.\\

\begin{theorem}\label{th2:bq}[Boundary equilibria dynamics]  Model \eqref{BHM1}-\eqref{BHM2} always has the extinction equilibrium $E_{000}$ which is always locally asymptotically stable.  If $d_h < \frac{re^{-d_b\tau}}{2\sqrt{K}}$ Model \eqref{BHM1}-\eqref{BHM2} has additionally two boundary equilibria $E_{B^{*}_1H^{*}_10}$ and $E_{B^{*}_2H^{*}_20}$ where $E_{B^{*}_1H^{*}_10}$ is always unstable. The equilibrium $E_{B^{*}_2H^{*}_20}$ is however locally asymptotically stable when $d_m>\frac{c\alpha_b B_2^*}{a+B_2^*}$ and unstable when $d_m<\frac{c\alpha_b B_2^*}{a+B_2^*}$.

\end{theorem}

\noindent\textbf{Notes:} By comparing the local stability condition of the equilibrium $E_{B^{*}_2H^{*}_20}$ of Model \eqref{BHM1}-\eqref{BHM2} to the local stability condition of the equilibrium $E_{B_2^* H_2^*}$ of Model  \eqref{subBH1}-\eqref{subBH2}  (see Table \ref{tab:stability}), it implies that parasitism with smaller mortality rates, e.g., $d_m<\frac{c\alpha_b B_2^*}{a+B_2^*}$, can destabilize the full system such that $E_{B^{*}_2H^{*}_20}$ becomes unstable. The destabilization due to the introduction of parasites $M$ into the honeybee colony has been observed in our model's simulations as well (see Figure \ref{fig:2D-3D_M} in the next section). The following theorem focuses on the global stability of the extinction equilibrium $E_{000}$ of Model \eqref{BHM1}-\eqref{BHM2}.\\

\begin{theorem}\label{globalstability-fullsystem}[Global stability of full system \eqref{BHM1}-\eqref{BHM2}] .
If $d_h > \frac{re^{-d_b\tau}}{2\sqrt{K}}$ and $d_m>c\alpha_b$, the extinction equilibrium $E_{000}=(0,0,0)$  is globally asymptotically stable.
\end{theorem}
\noindent\textbf{Notes:} Theorem \ref{globalstability-fullsystem} indicates that the large mortality rate of honeybees $d_h$ and mites $d_m$ can lead to the colony collapsing. 
Next, we focus on the existence of interior equilibria of Model \eqref{BHM1}-\eqref{BHM2} that could lead to colony survival. \\

Model \eqref{BHM1}-\eqref{BHM2} has no  interior equilibria when $\tau=0$. Therefore, the existence of  interior equilibria requires the delay $\tau>0$. Our aim is to find sufficient conditions such that Model \eqref{BHM1}-\eqref{BHM2} has interior equilibria that can lead to the survival of the honeybee colony. To begin our analysis, note that from equation \eqref{Equilibrium1c}, $B^{*} = \frac{a}{\frac{c\alpha_b}{d_m}-1}>0$, that is, the inequality $\frac{c \alpha_b}{d_m}>1$ is required. Then using equation \eqref{Equilibrium1a} and \eqref{Equilibrium1b}, we obtain follows:
\begin{align}
 \frac{rH^2}{K+H^2} - \frac{\alpha_bBM}{a+B}-d_bB =  \frac{\alpha_hHM}{a+H}+d_hH\label{EqM}.
\end{align}

Let $f_1(H)=\frac{rH{^2}}{K+H{^2}}$ and $f_2(H)=\frac{B\left(\frac{\alpha_b M}{a+B}+d_b\right)}{1-e^{-\left(\frac{\alpha_b M}{a+B}+d_b\right)\tau}}$, where
$
B=B^*,\ M=\frac{\frac{rH^2}{K+H^2}-d_bB^*-d_hH}{\frac{\alpha_hH}{a+H}+\frac{\alpha_bB^*}{a+B^*}},
$
and
\begin{eqnarray}\label{Q}
Q(H)=-d_hH{^3}+(r-B^*d_b)H^2-d_hKH-d_bKB^*,
\end{eqnarray}
with two positive critical points:
\begin{eqnarray*}
H_{1}^c=\frac{(r-d_b B^*)-\sqrt{(r-d_b B^*)^2-3Kd_h^2}}{3d_h},\ H_{2}^c=\frac{(r-d_b B^*)+\sqrt{(r-d_b B^*)^2-3Kd_h^2}}{3d_h},
\end{eqnarray*}
if $d_h<\frac{r-B^*d_b}{\sqrt{3K}}$. Sufficient conditions for the existence of an interior equilibrium of Model \eqref{BHM1}-\eqref{BHM2} is provided in the following theorem.

 \begin{theorem}[Existence of interior equilibria]\label{pr1:IntEq}  Let $a, \alpha_b,\alpha_h, c, K, r, d_b,d_h,d_m$ and $\tau$ be positive parameters. Assume $\frac{c \alpha_b}{d_m}>1$,  $d_h<\frac{r-B^*d_b}{\sqrt{3K}}$ and $Q(H_2^c)> 0$.
Then $Q(H)$ has two positive roots $H_1^r$ and $H_2^r$($H_1^r<H_2^r$), and  Model \eqref{BHM2} has at least one interior equilibria $E_{B^*H^*M^*}=(B^*, H^*, M^*)$ with $B^*=\frac{a}{\frac{c \alpha_b}{d_m}-1}$ when $\tau\in(\beta_1, \beta_2)$, where
\begin{eqnarray*}
\beta_1=\frac{1}{d_b}\ln\left(\frac{f_1(H_2^r)}{f_1(H_2^r)-B^*d_b}\right),\quad \beta_2=\frac{1}{d_b}\ln\left(\frac{f_1(H_1^r)}{f_1(H_1^r)-B^*d_b}\right).
\end{eqnarray*}
In addition, if $d_h$ is sufficiently small such that
\begin{eqnarray*}
d_h<\frac{(r-B^*d_b)\sqrt{2(r-B^*d_b)}}{3\sqrt{K(r-B^*d_b)+3d_bKB^*}}
\end{eqnarray*}
then $Q(H_2^c)> 0$.
\end{theorem}

 \noindent\textbf{Notes.} Theorem \ref{pr1:IntEq} implies that even if the Model \eqref{BHM1}-\eqref{BHM2} is biologically relevant such that  $\frac{c\alpha_b}{d_m}>\frac{a d_b+r}{r}>1$ and the queen's egg production is sufficiently large satisfying $r>d_bB^*$, Model \eqref{BHM1}-\eqref{BHM2} may have no interior equilibrium unless the conditions $Q(H^c_2)>0$, $f_2(H^*_1)>f_1(H^*_1)$ and $f_2(H^*_2)<f_1(H^*_2)$ are satisfied in which case, a unique interior equilibrium emerges. The expressions of the interior equilibria are too complicated to solve. Thus, we seek help from numerical simulations to explore the stability of the interior equilibria.\\

\section{Effects of parasitism and seasonality}

In this section, we focus on the dynamical effects of parasitism and seasonality on colony survival. To explore the effects of parasites $M$, we compare the typical long term dynamics of system \eqref{BHM1}-\eqref{BHM2} when $M(0)=0$ and $M(0)>0$ through simulations. Figure \ref{fig:2D-3D_M} shows that: 1) when $M(0)=0$, the honeybee colony has equilibrium dynamics and the colony can survive; while 2) when $M(0)=1$, we can see that both honeybees and mites coexist through oscillating dynamics. Thus, we could deduce that the introduction of mites (i.e. $M>0$) can have a destabilizing effect on the system and produce fluctuating dynamics of the honeybee population (brood and adult bees) and mites when the time delay $\tau$ is large enough. For these simulations, we are using $\tau=21$ which corresponds to the time it takes for an egg to become an adult bee. We would like to point out that  varying $\alpha_b$ (i.e. parasitism rate on brood) has a potential to destabilize the dynamics, thus drive the population through oscillating dynamics (see Figures \ref{fig:TimeSerieOscillation} and \ref{fig:TimeSerieNonperiodic} in Appendix \ref{AppendixB}). \\

\begin{center} \textbf{Long term dynamics of Model \eqref{BHM1}-\eqref{BHM2} with $M=0$ and $M>0$}\end{center}
\begin{figure}[H]
\centering
\includegraphics[scale=0.35]{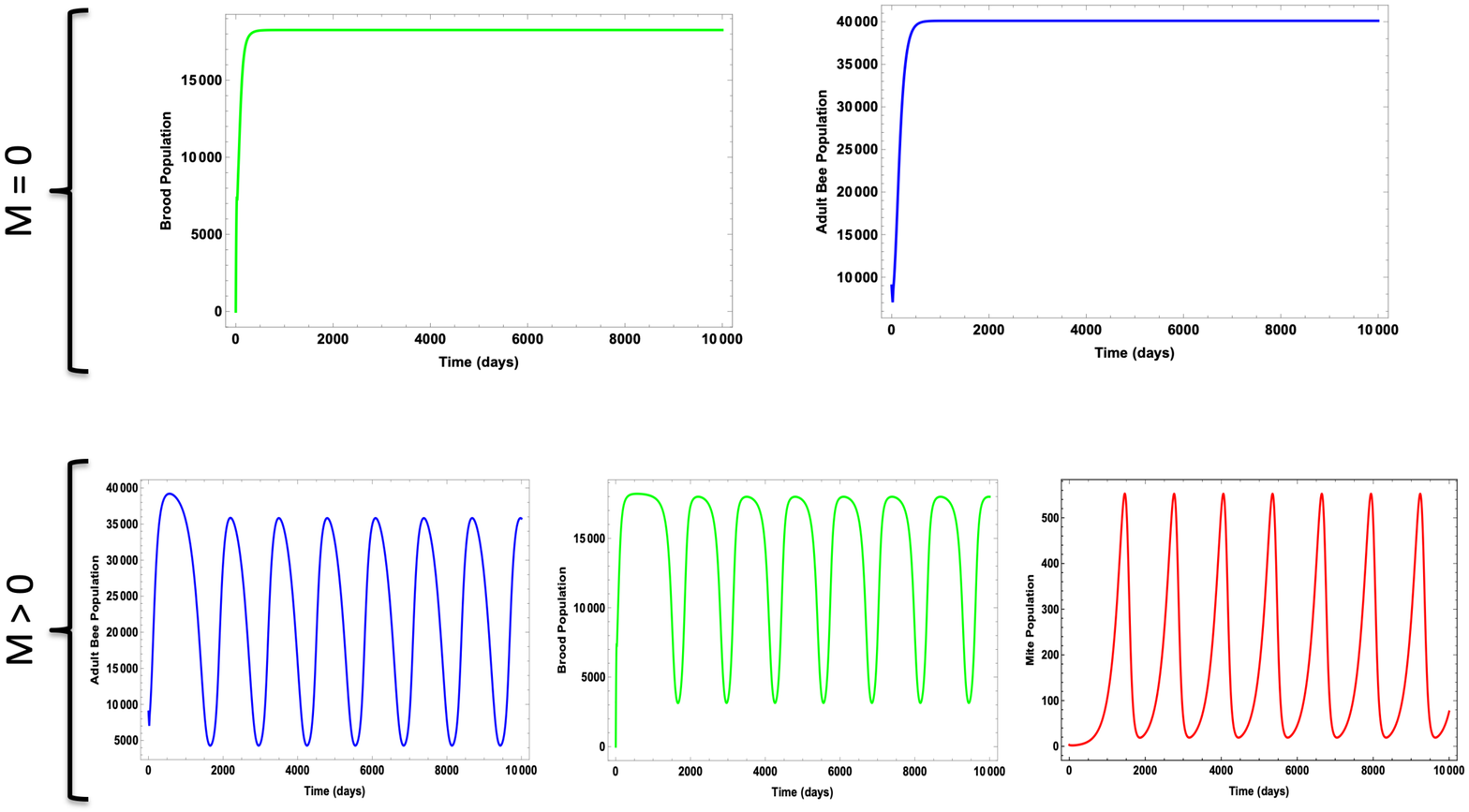}
\caption[Time Series: Brood, Adult Bee, and Mite with $\alpha_b=0.024$.]{{\small  Time series of the brood, adult bee, and mite population using $\tau=21$, $r=1500$, $K=95000000$, $d_b =0.051$, $d_h= 0.0121$, $d_m=0.027$, $\alpha_b =0.024$, $\alpha_h=0.8$, $c=1.9$, $a=8050$, $\tau=21$, $B_0(t)=B(0)=0$, $H(0)= 9000$, and $M(0) = 3$. }}\label{fig:2D-3D_M}
\end{figure}


\noindent\textbf{Effects of delay with parasitism:} from Theorem  \ref{pr1:IntEq}, we can confirm that there is a unique interior equilibrium when we choose parameter values $r=1500$,~$K=95000000$,~$d_b =0.051$,~ $d_h= 0.0121$,~$d_m=0.027$, ~$\alpha_b =0.024$, ~$\alpha_h=0.8$, ~$c=1.9$,~ $a=8050$.We take initial conditions $B_0(t)=B(0)=0$, ~$H(0)= 9000$, and $M(0) = 3$ and vary the maturation time $\tau\in (0,26]$ in Figure \ref{fig:3D_taudynamics}.
Then we have the following dynamics:
\[\mbox{with} \quad \mathbf{\tau=15}, \quad  E_{B^*H^*M^*}=\left(11~685.5,16~727.2,579.905\right); \mbox{\textbf{ stable} equilibrium  (see Fig. \ref{fig:3D_taudynamics}(a)-(c))}\]
\[ \mbox{with} \quad  \mathbf{\tau=21}, \quad E_{B^*H^*M^*}=\left(11~685.5,12~102.5,338.558\right); \mbox{\textbf{ periodic} solutions   (see Fig. \ref{fig:3D_taudynamics}(d)-(f))}\]
\[\mbox{with} \quad \mathbf{\tau=26},\quad  E_{B^*H^*M^*}=\left(11~685.5,10~607.8,189.793\right); \mbox{\textbf{ unstable} equilibrium (see Fig. \ref{fig:3D_taudynamics}(g)-(i))}\]
These simulations suggest that as $\tau$ increases, $(0<\tau<16)$, the interior equilibrium is asymptotically stable with our choice of parameter values above. Then, for $16<\tau<26$, the system has periodic solutions which could be due to a possible Hopf bifurcation. Lastly, for $\tau>26$ the interior equilibrium becomes unstable with a large oscillating cycle that hits the stable manifold of  the extinction equilibrium $E_{000}$ such that both honeybee and mite populations die out. \\



\begin{center}\textbf{Delay effects on Model \eqref{BHM1}-\eqref{BHM2} long term dynamics}\end{center}
\begin{figure}[H]
\centering
\subfigure[$\tau=15$]{\includegraphics[height = 40mm, width = 45mm]{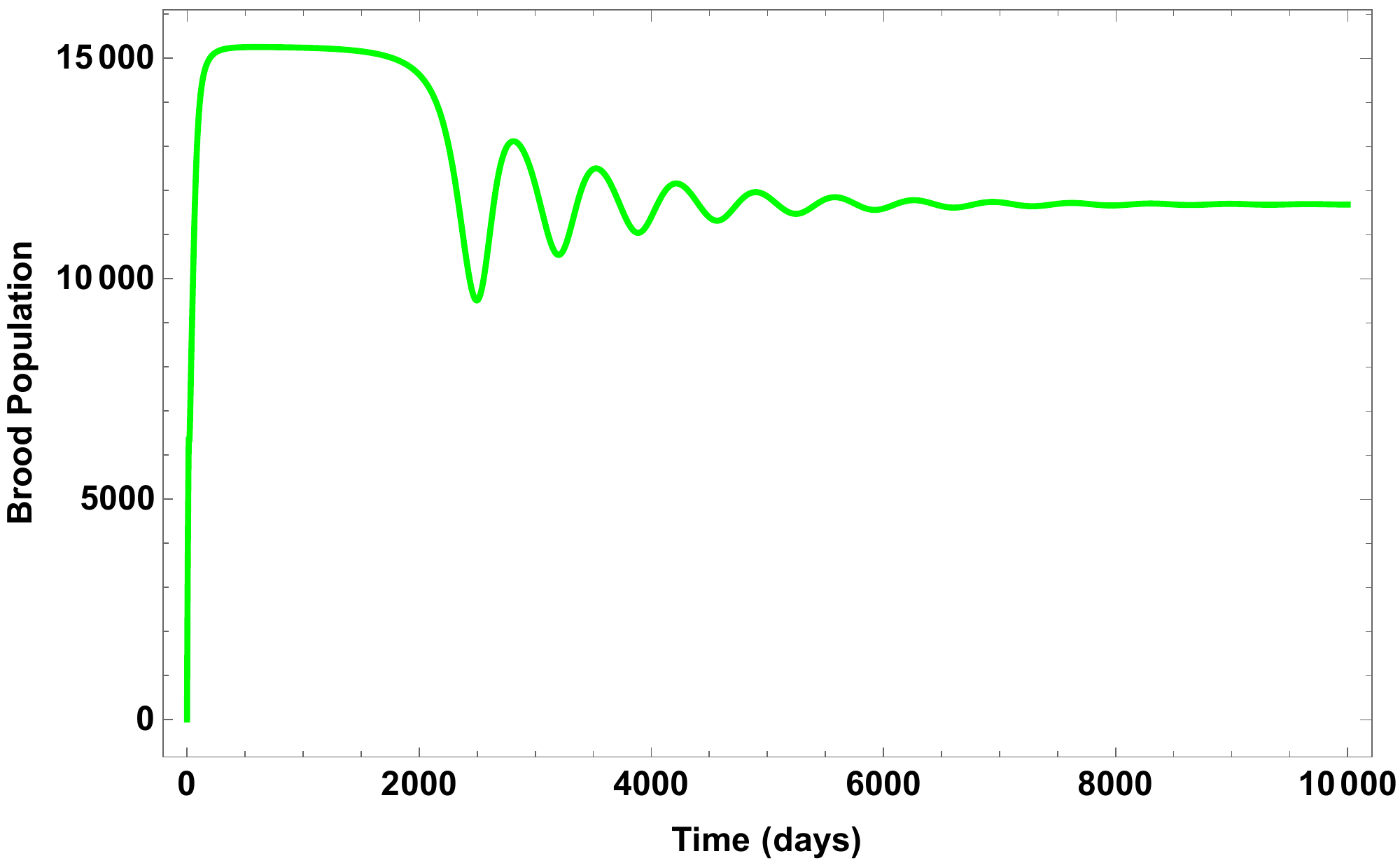}\label{fig:3D_taudynamics_a}}
\subfigure[$\tau=15$]{\includegraphics[height = 40mm, width = 45mm]{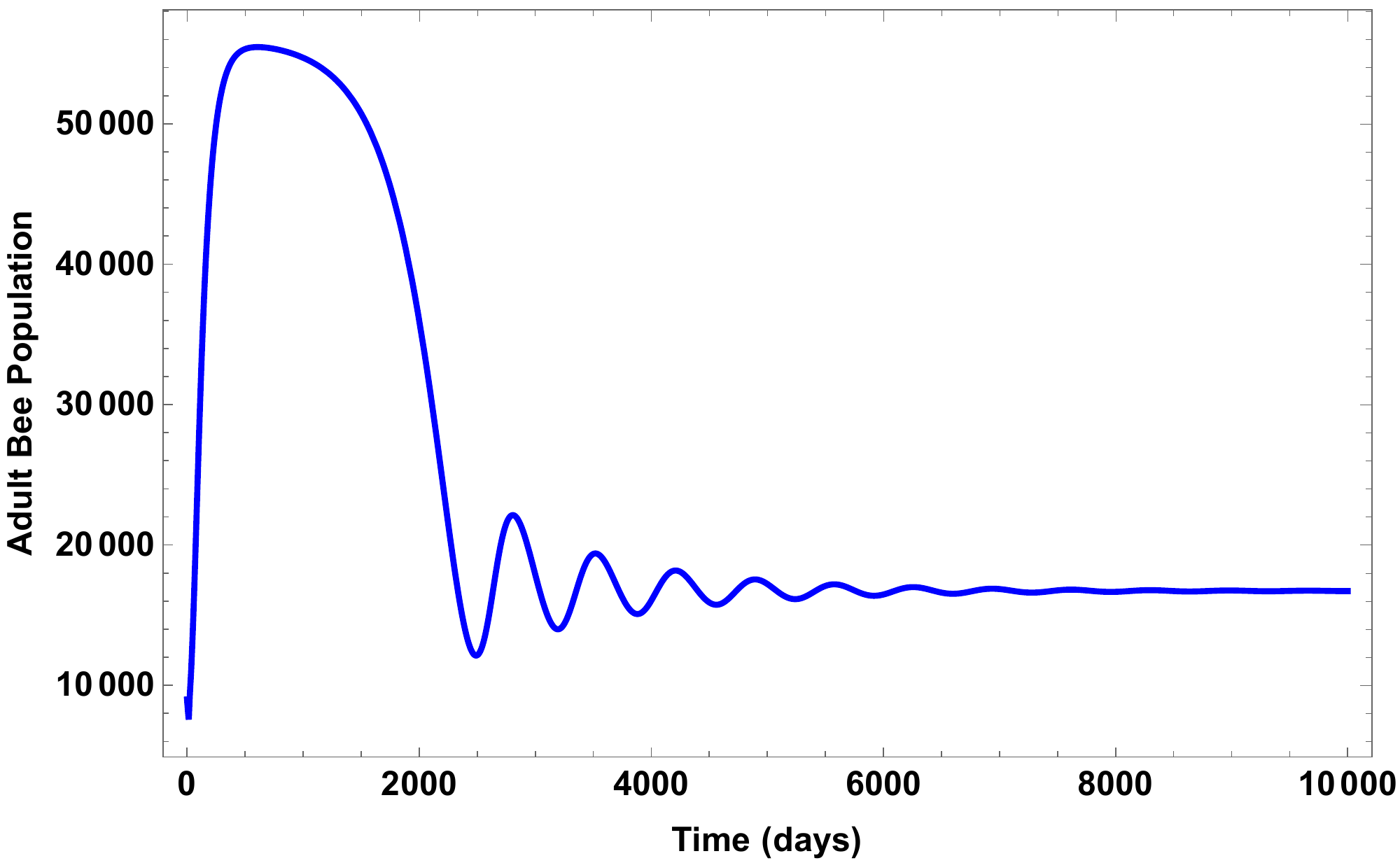}}
\subfigure[$\tau=15$]{\includegraphics[height = 40mm, width = 45mm]{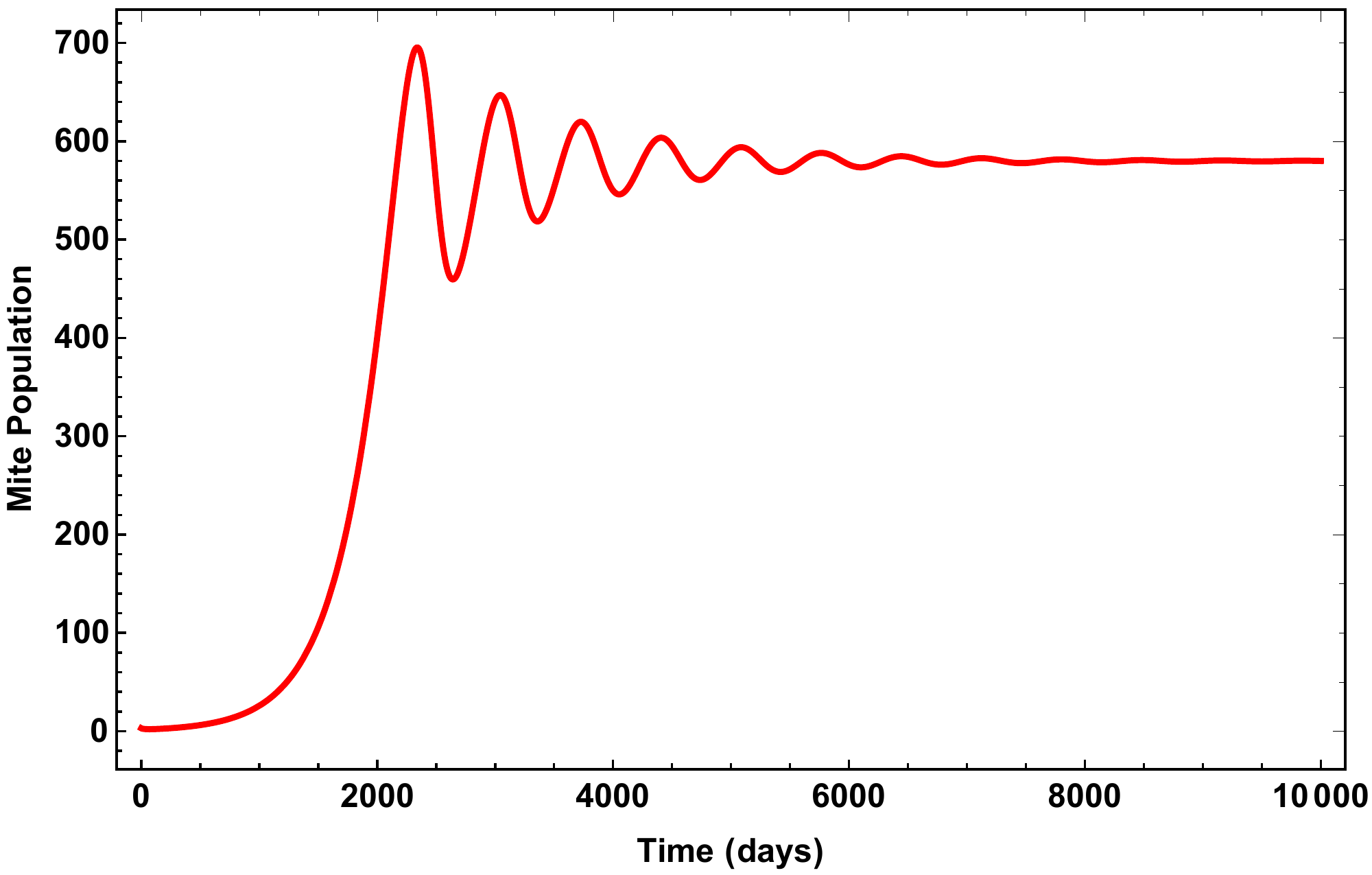}}
\subfigure[$\tau=21$]{\includegraphics[height = 40mm, width = 45mm]{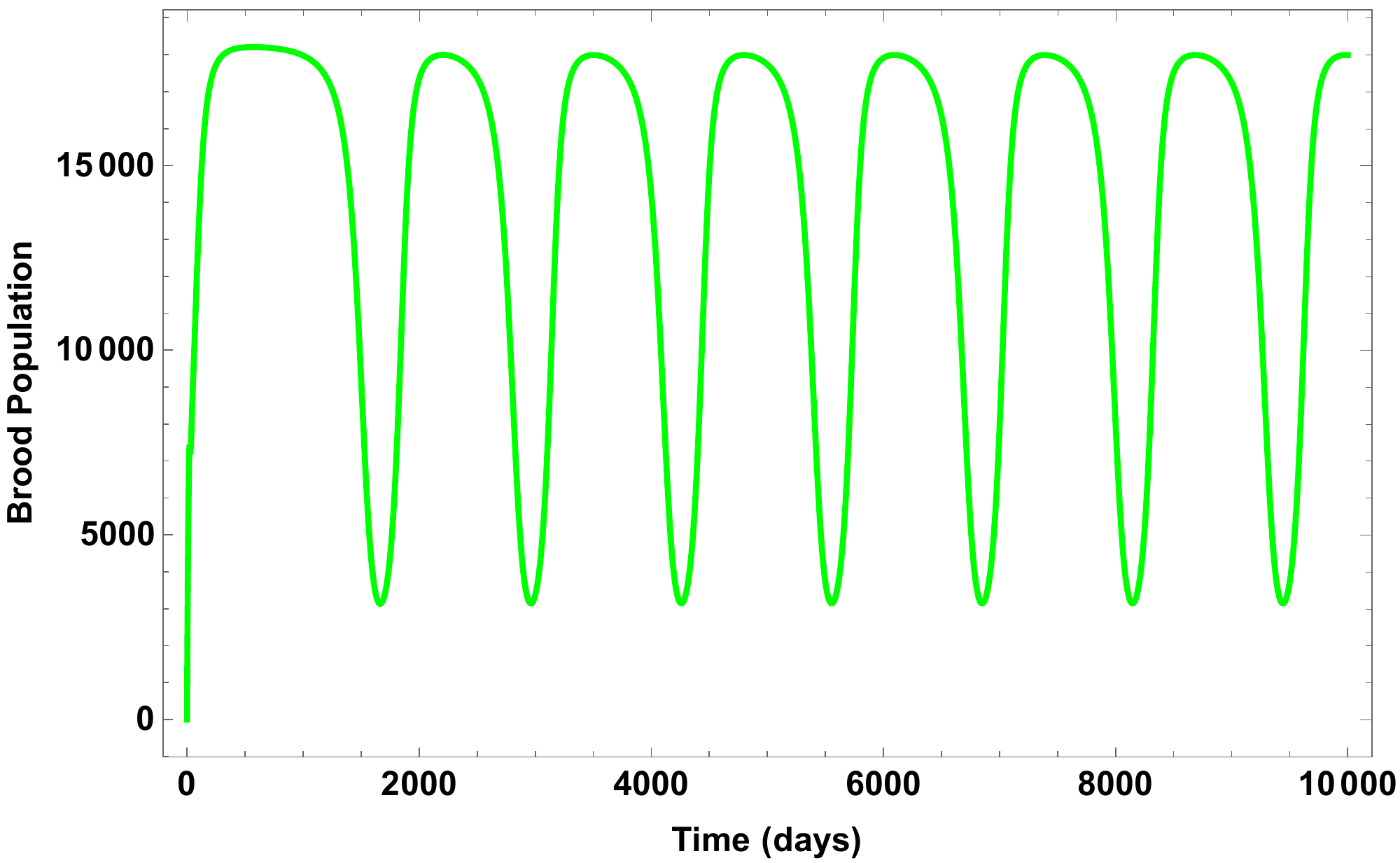}}
\subfigure[$\tau=21$]{\includegraphics[height = 40mm, width = 45mm]{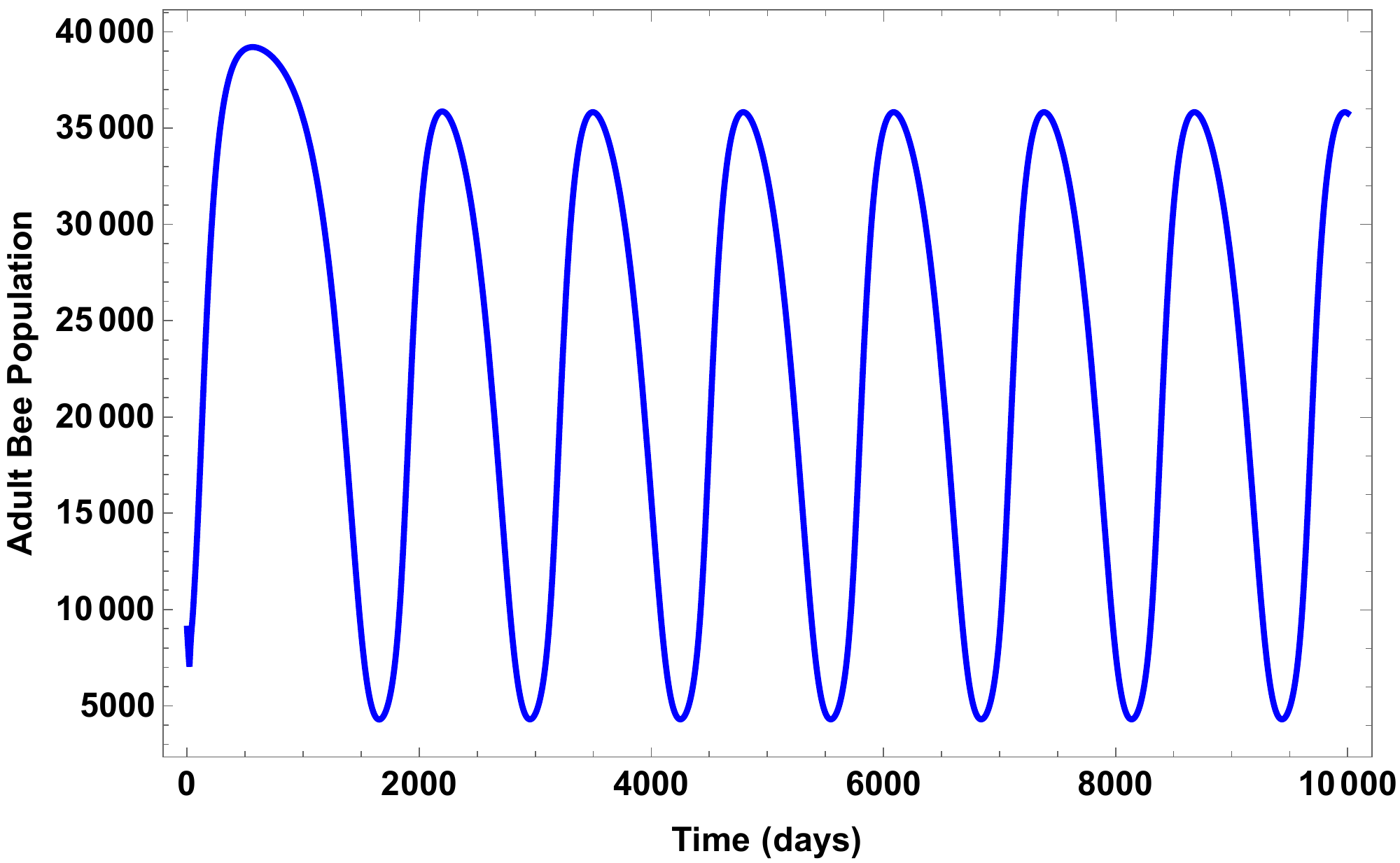}}
\subfigure[$\tau=21$]{\includegraphics[height = 40mm, width = 45mm]{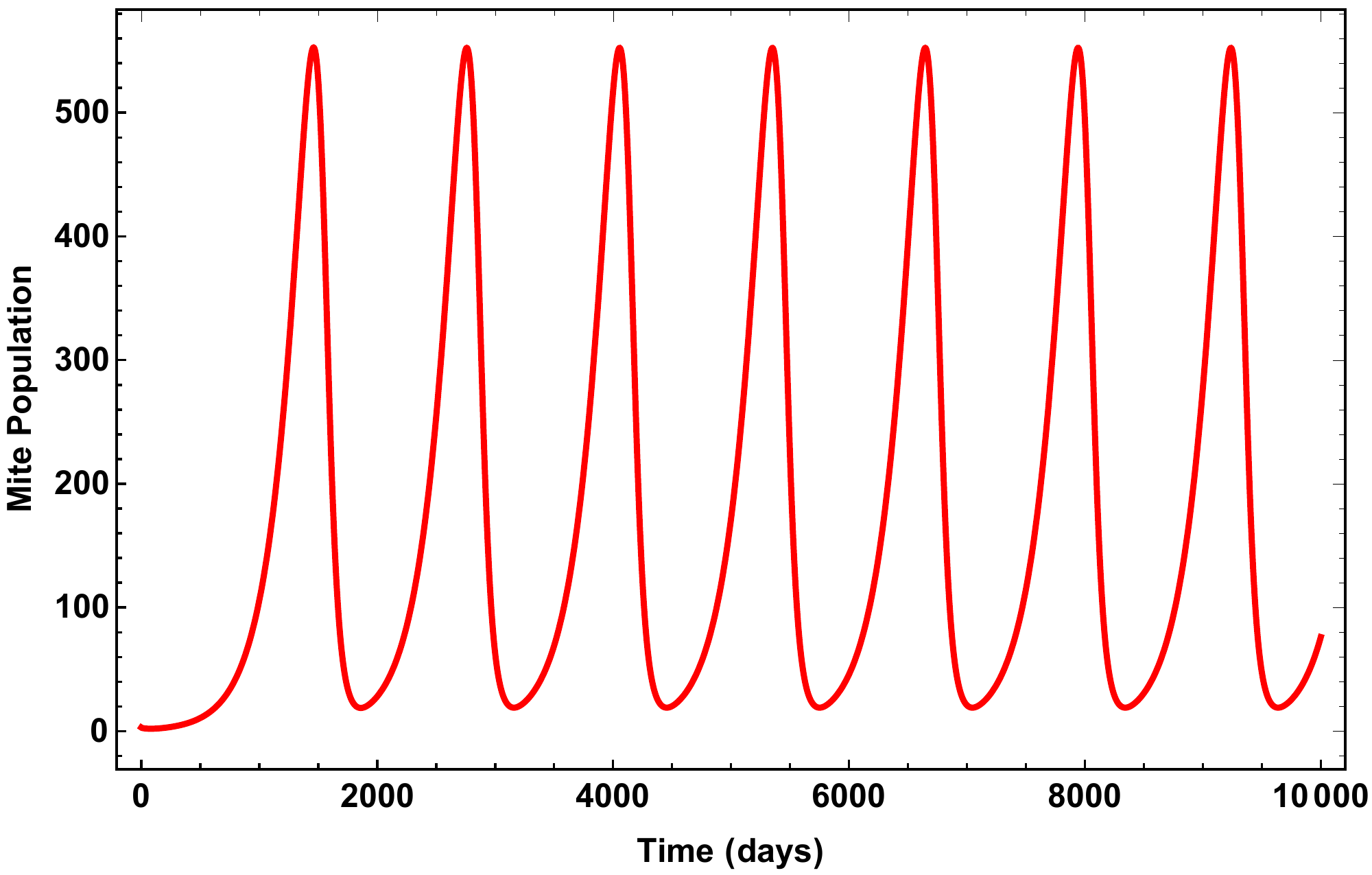}}
\subfigure[$\tau=26$]{\includegraphics[height = 40mm, width = 45mm]{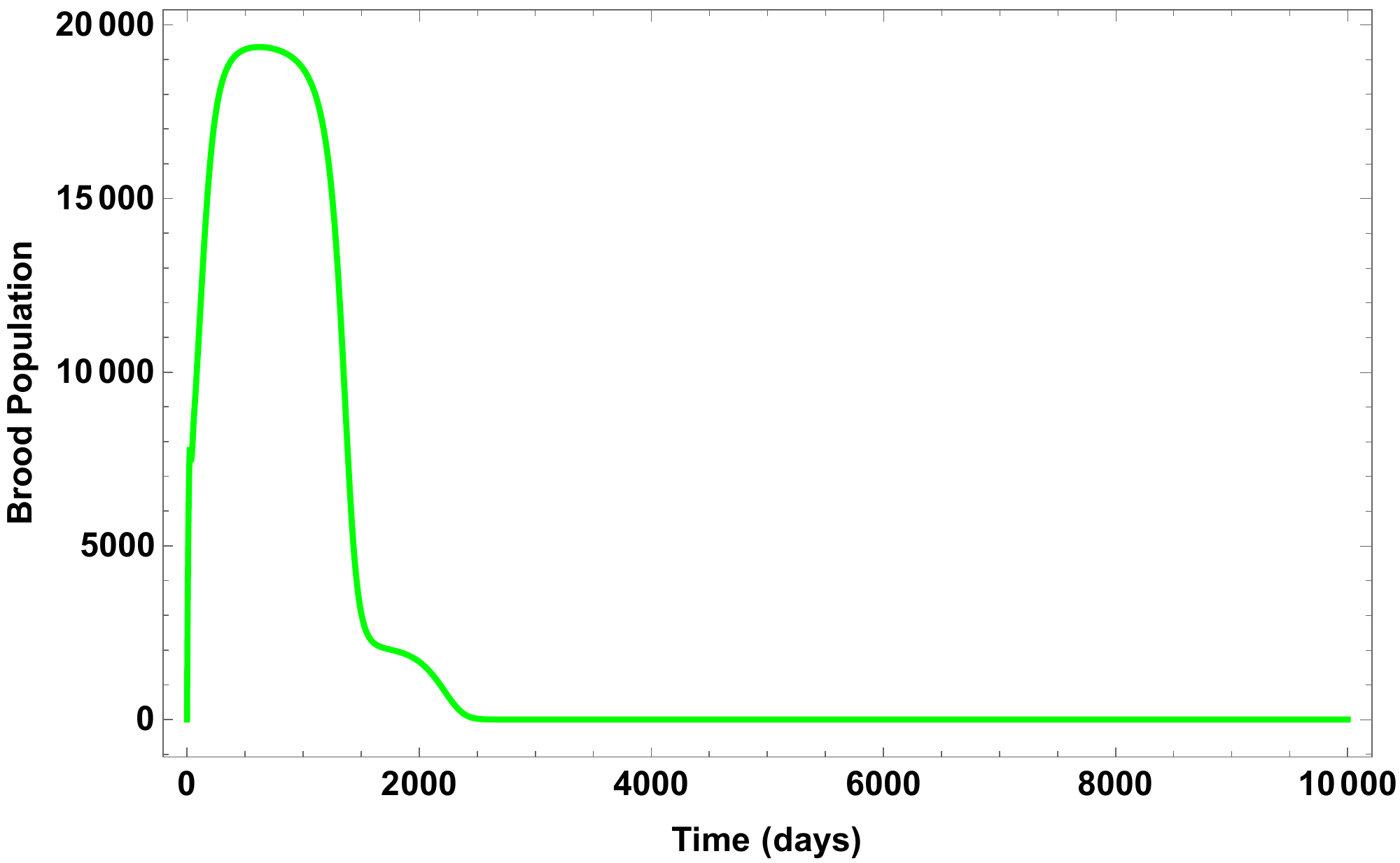}}
\subfigure[$\tau=26$]{\includegraphics[height = 40mm, width = 45mm]{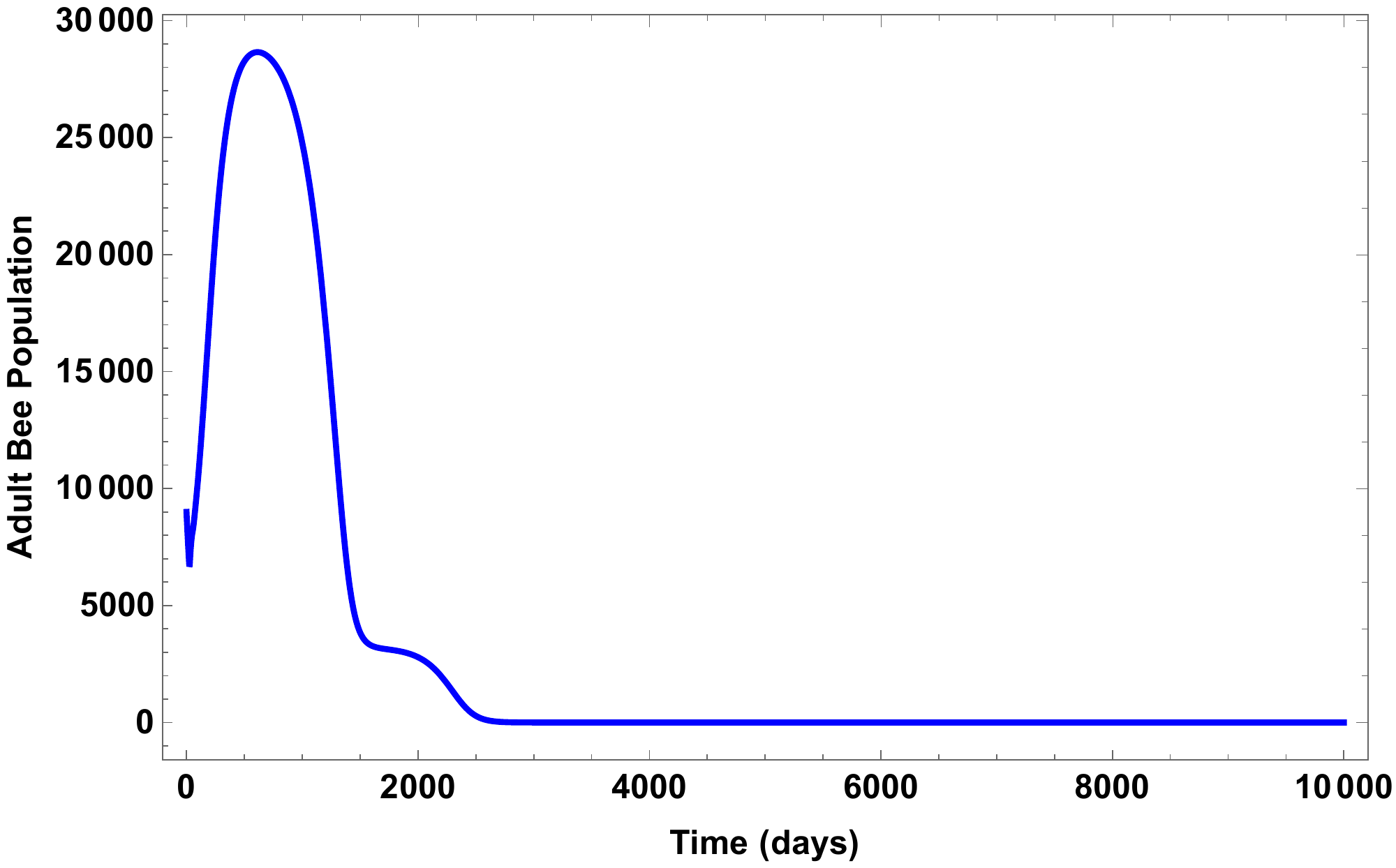}}
\subfigure[$\tau=26$]{\includegraphics[height = 40mm, width = 45mm]{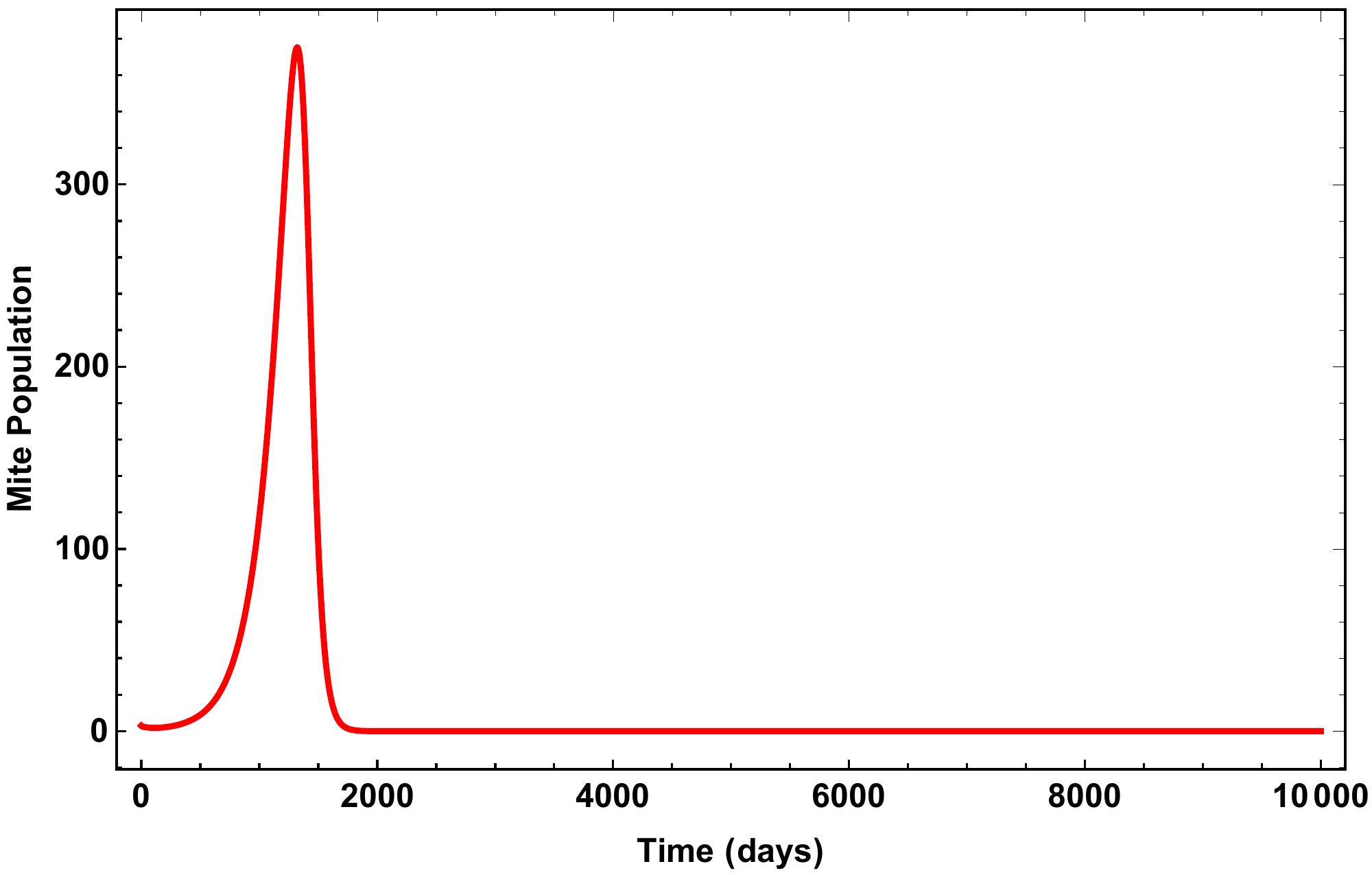}\label{fig:3D_taudynamics_i}}
\caption[Time Series: Brood, Adult Bee, and Mite with $\alpha_b=0.024$.]{{\small  Time series of the brood, adult bee, and mite population using $r=1500$, $K=95000000$, $d_b =0.051$, $d_h= 0.0121$, $d_m=0.027$, $\alpha_b =0.024$, $\alpha_h=0.8$, $c=1.9$, $a=8050$, with I.C. $B(0)=0$, $H(0)= 9000$, and $M(0) = 3$.}}\label{fig:3D_taudynamics}
\end{figure}

\subsection{Seasonality and Parameter Estimation}\label{c4Material}
The number of eggs laid by the queen bee can be predicted as a function of the ambient temperature, photoperiod, and adult population in the colony \citep{degrandi1989beepop}. In addition, it has also been shown that the total number of eggs laid daily by the queen is a decreasing function of the number of days the queen has been laying eggs \citep{degrandi1989beepop}, i.e., older queens lay fewer eggs than younger queens. Given that the number of eggs laid by the queen is temperature and photoperiod dependent (i.e., changed seasonally), the egg-laying rate must hence be described by a periodic function. It is well known that any periodic function can be represented as an infinite sum of sines and cosines \citep{feng2011modeling}. In order to keep our model simple and tractable, we combined these factors (i.e. temperature, photoperiod, etc.) and adapted the first order harmonic function presented in \citep{feng2011modeling} to the egg-laying rate $r_1$ and $r_2$ in the first and second part, respectively, of Model \eqref{BHM1}-\eqref{BHM2} to obtain:
\begin{align}
r_1 = r\left[1+cos\left(\frac{2\pi(t-\Phi))}{365}\right)\right]\quad \mbox{  and  }\quad r_2 = r\left[1+cos\left(\frac{2\pi(t-\tau-\Phi))}{365}\right)\right]\label{EggRateFunc}
\end{align}
where $\Phi$ denote the day of the year with the maximum egg-laying rate, $r$ is the baseline egg-laying rate from \citep{sumpter2004dynamics, eberl2010importance}, and $t$ is the time measured in days. Model \eqref{BHM1}-\eqref{BHM2} with a constant egg-laying rate $r$ is hence a model without seasonality. We introduced seasonality by changing the egg-laying rate to the harmonic function in equation \eqref{EggRateFunc}. The parameters $\Phi$ and $r$ were estimated by fitting the equation \eqref{EggRateFunc} to a one year simulated data of the number of eggs laid per day from the BEEPOP model \citep{degrandi1989beepop} (see Figure \ref{fig_EggLayingRate}).\\

\begin{figure}[H]
\begin{center}
{\includegraphics[height = 55mm, width = 100mm]{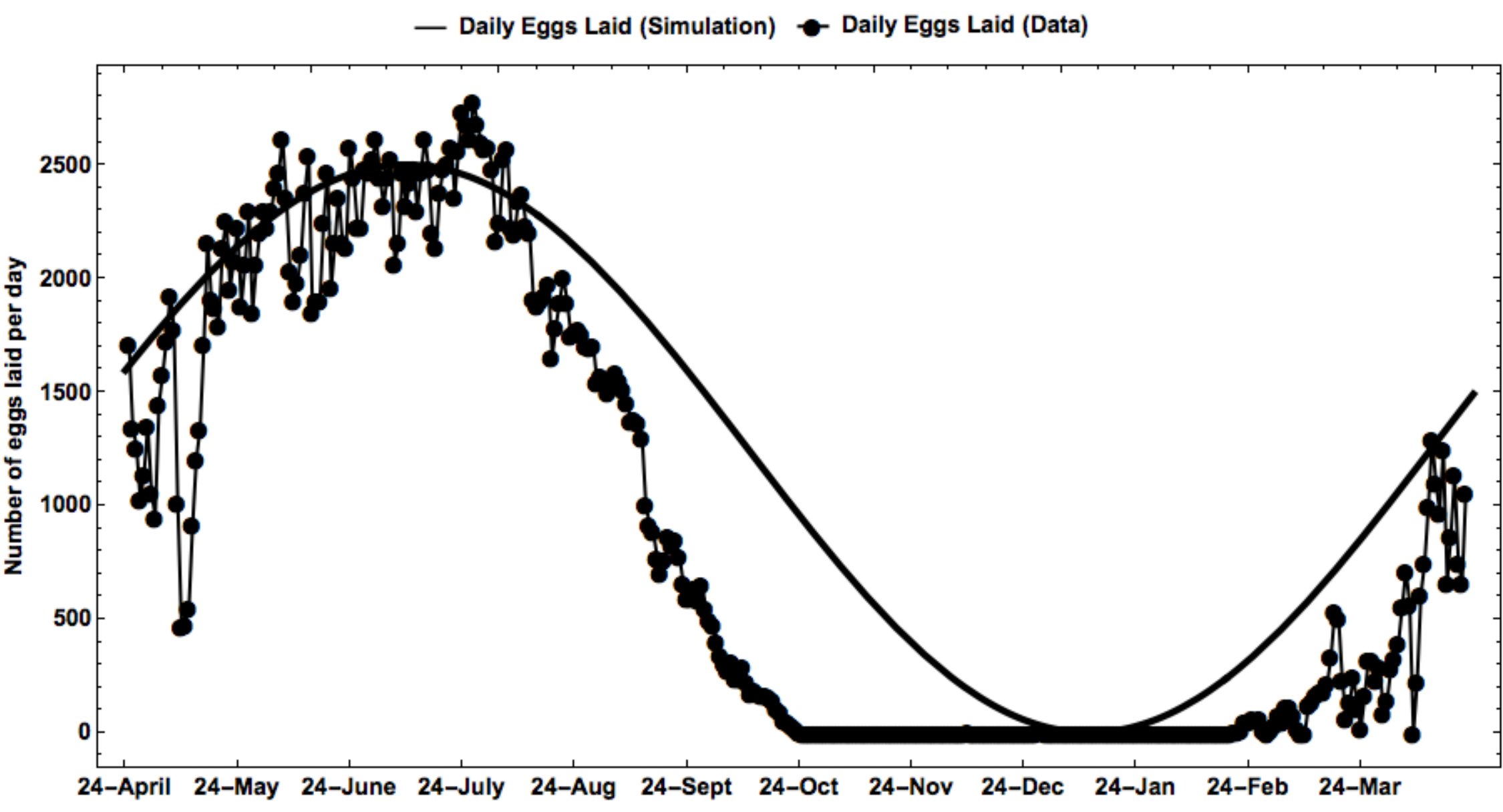}}
\end{center}
\vspace{-10pt}
\caption{Number of eggs laid by a strong full matted queen without a constraint over a period of one year following equation \eqref{EggRateFunc} with $r=1250$ and $\Phi = 75$ ($\approx$ July 8). The data was produced by using the BEEPOP model from \citep{degrandi1989beepop} by taking into account daily temperature, photoperiod, and adult population in the colony.}
\label{fig_EggLayingRate}
\end{figure}

In order to estimate parameters $\alpha_b$, $\alpha_h$, and $a$ in the model with and without seasonality, we first estimated the brood, adult honeybee, and mite population size per colony from field data provided by DeGrandi-Hoffman \cite{degrandi2016population}. The data were collected at the University of Arizona West Agricultural Facility, (20 colonies). The colonies were established in desert climate of Arizona where temperatures are favorable for bees foraging activity, especially, from April through November when the data were collected. All colonies initially were broodless, and had 9000 bees with a laying queen. A miticide treatment was applied to control the {\it Varroa} population at the beginning of the experiment (April of 2014). \\

In order to approximate the adult honeybee and brood population sizes in the colonies, frames of bees were measured monthly from May to November using a method from \citep{degrandi2008comparisons}. This method consist of estimating brood and adult bees on an area of the frames using a 5 cm $\times$ 5 cm grid which covers the entire side of the comb. Note that one frame of bees contains approximately 2506 bees and 5200 brood cells \citep{degrandi2008comparisons} and at most only 80\% of frames are cover with brood. Thus, each colony of adult bees is estimated by computing: the number of frames of adult bees $\times $ 2506, and a colony of brood is estimated as: the number of frames of brood $\times$ 0.8 $\times$ 5200. The {\it Varroa} mite population density in the colonies were also collected from May until November. During the experiment season (i.e. May to November), 300 bees were brushed into a jar then the number of mites on the 300 bees were counted monthly and these constitute the phoretic mites. The population of the reproductive mites was also estimated by counting the total number of mites per sampled cells. The total mite population in a colony is hence the sum of the phoretic and reproductive mite. We proceeded as follow to find the estimated mite population in colonies. Recall that the number of phoretic mites obtained is the mites per 300 bees. Then, the phoretic mite population size per colony was estimated by:  $\frac{\text{mites per 300 bees}}{300} \times$ population of bees per colony. We calculated the reproductive mites per colony by multiplying $\frac{\text{total number of mites}\times 5200}{\text{number of cells sampled}}$. In \citet{degrandi2016population}, the authors followed a similar approach to estimate the population size of brood, adult bees, and mites per colony. \\


Using {\it Varroa} mites and honeybee life history parameters in the ranges provided in Table \ref{C4ParamTab}, we estimated the parameters $\alpha_b$, $\alpha_h$, $a$ without considering seasonality ($\alpha_b=0.045$, $\alpha_h=0.49$, and $a=8500$) and in the presence of seasonality ($\alpha_b=0.0447$, $\alpha_h=0.8$, and $a=8050$) by fitting the model to the data when the egg-laying rate $r$ is constant and when $r$ is the harmonic function described in Equation \eqref{EggRateFunc}, respectively. To illustrate the importance of seasonality when modeling the dynamics of honeybee and mite populations, we present the best fit without seasonality in Figures \ref{fig:TimeSerieDataFit1} and the best fit with seasonality in Figures \ref{fig:TimeSerieDataFit2} using the egg-laying rate function in \eqref{EggRateFunc}.\\

\begin{table}[ht]
\centering
{\small\begin{tabular}{C{2cm}L{6.8cm}L{4.1cm}L{2cm}}
        \toprule
\textbf{Parameter} & \textbf{Description} & \textbf{Estimate/Units}  & \textbf{Reference}     \\   \midrule
$r$ & maximum egg-laying rate by the queen  & 0, 500, 1500 bees/day (season dependent)  & \cite{sumpter2004dynamics, eberl2010importance} \\
$d_{b}$ & average death rate of brood (larvae and pupae stage) & $\dagger$ 0.00602-0.036 (unsealed brood); 0.00303 (sealed brood) day$^{-1}$ & \cite{fukuda1968worker} \\
$d_{h}$ & average death rate of adult honeybee & 0-0.17 (hive bees); 0-0.8 (foragers) day$^{-1}$& \cite{rueppell2007regulation} \\
$d_{m}$ & average death rate of phoretic mite & (0.016-0.45) or 0.002 (winter), 0.006 (summer) day$^{-1}$ & \cite{branco2006comparative, martin1998population} \\
$c$ & conversion rate from mite feeding on honeybee to mite reproduction & 0-4.5 & \cite{huang2012varroa}  \\
$\sqrt{K}$ & colony size at which brood survivability is half maximal &$\leq$ 22007 (fall, spring), and  $\leq$ 37500 (summer) bees/day \hspace{1.5in}(upper bound values) & \cite{ratti2012mathematical}\\
$\alpha_b$ & parasitism rate on brood & 0.0447  day$^{-1}$ & Estimated (see Sec. \ref{c4Material})    \\
$\alpha_h$ & parasitism rate on adult bee & 0.8 day$^{-1}$ & Estimated (see Sec. \ref{c4Material})  \\
$a$ & size of honeybee population at which rate of attachment is half maximal& 8050 bees & Estimated (see Sec. \ref{c4Material})   \\
$\tau$ & Brood development time from egg to adult bee & 21 (workers) days & p. 83 in \cite{graham1992hive} \\\bottomrule
        \end{tabular}}
\caption{Standard parameters values used for simulation of honeybee and mite population of Model \eqref{BHM1}-\eqref{BHM2}. $\dagger$ calculated from the daily mortality $([1-\frac{330}{332}],[1-\frac{347}{360}])$ for unsealed brood and $(1-\frac{329}{330})$ for sealed brood.}
\label{C4ParamTab}
\end{table}
\begin{center}
{\bf Best fit without seasonality}
\end{center}

\begin{figure}[H]
\begin{center}
\subfigure[Mean brood population in colonies]{\includegraphics[height = 50mm, width = 40mm]{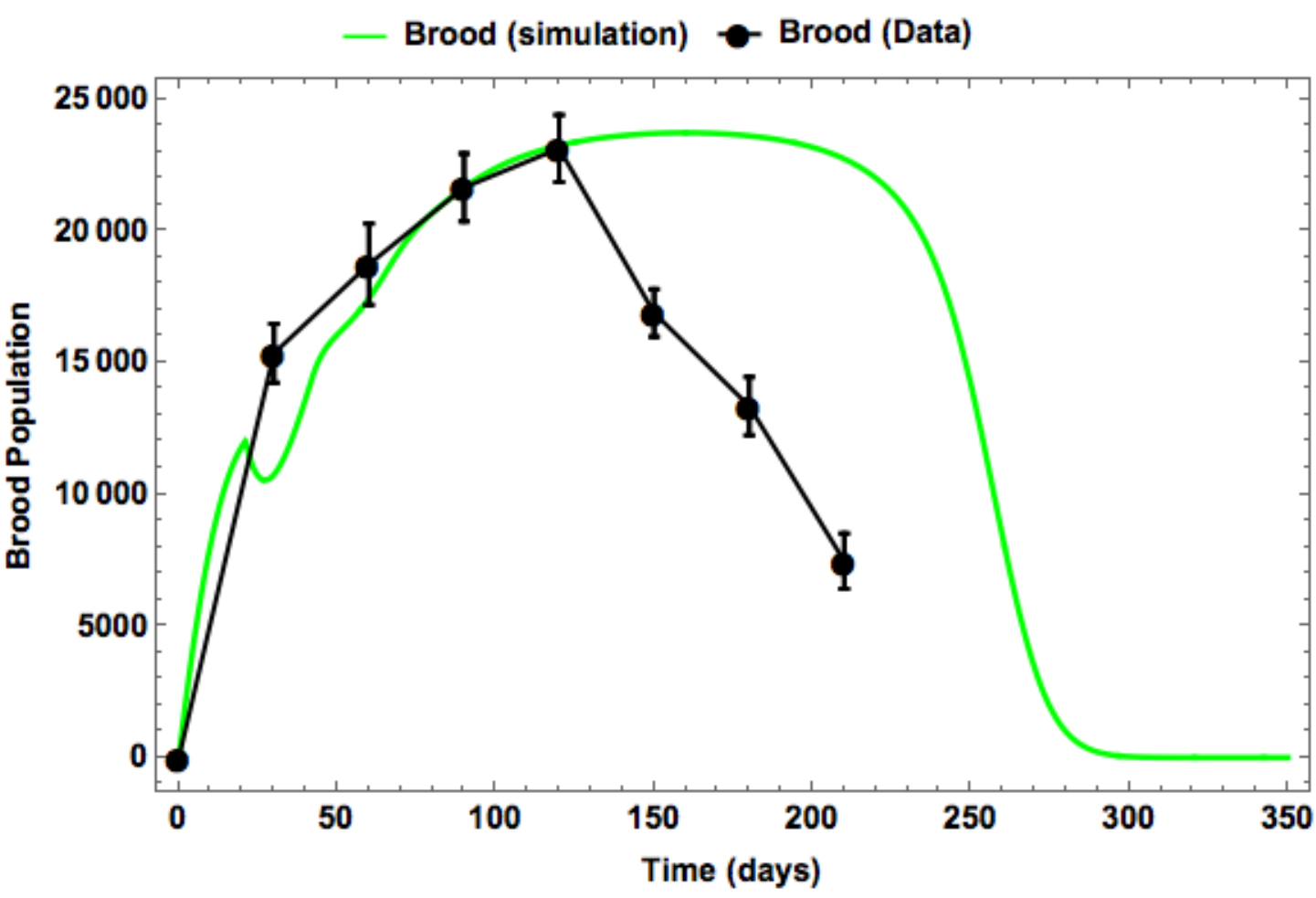}\label{fig_Model_Simulation_BroodS1}}\hspace{5mm}
\subfigure[Mean bee population in colonies]{\includegraphics[height = 50mm, width = 40mm]{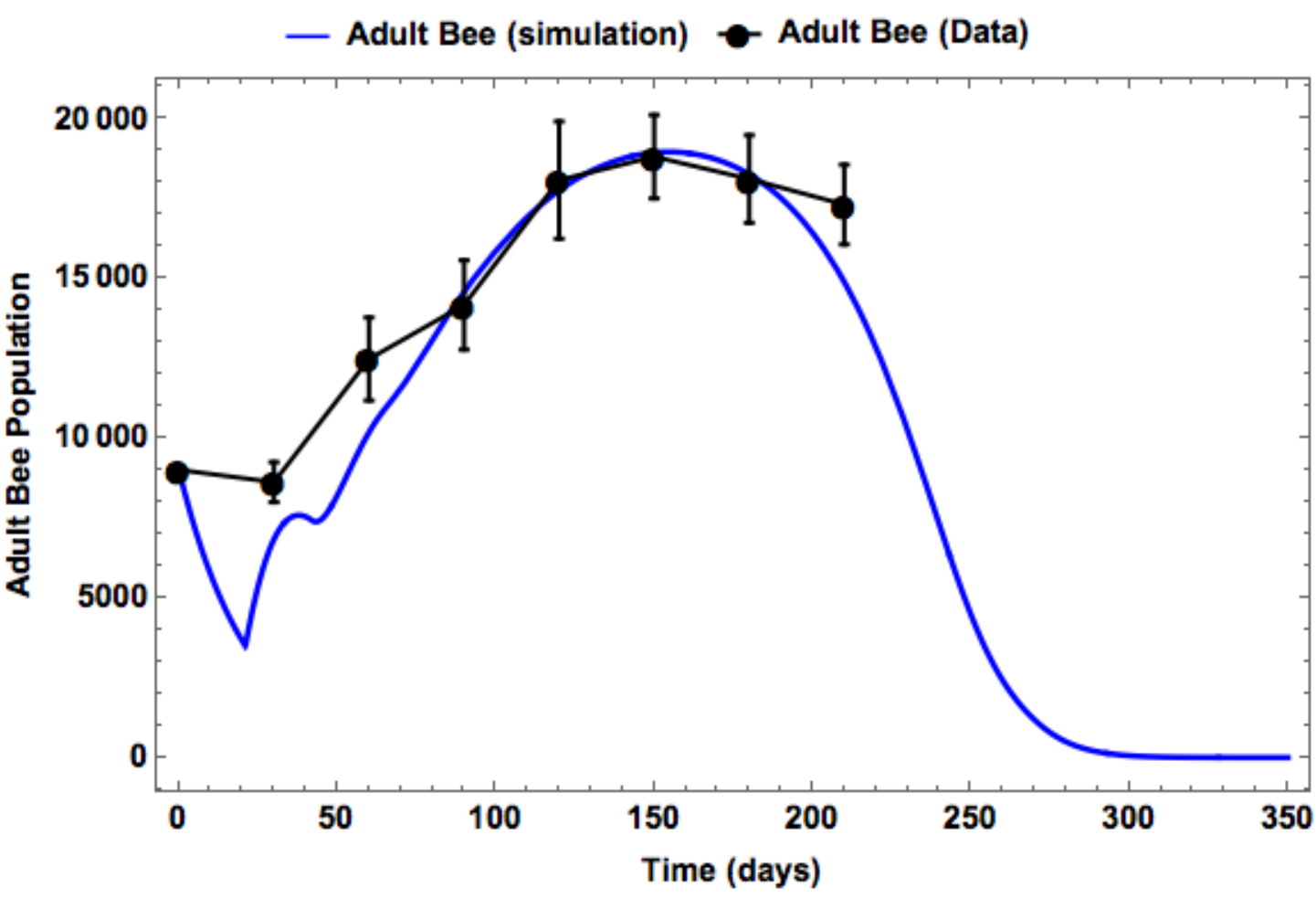}\label{fig_Model_Simulation_BeeS1}}\hspace{5mm}
\subfigure[Mean mite population in colonies]{\includegraphics[height = 50mm, width = 40mm]{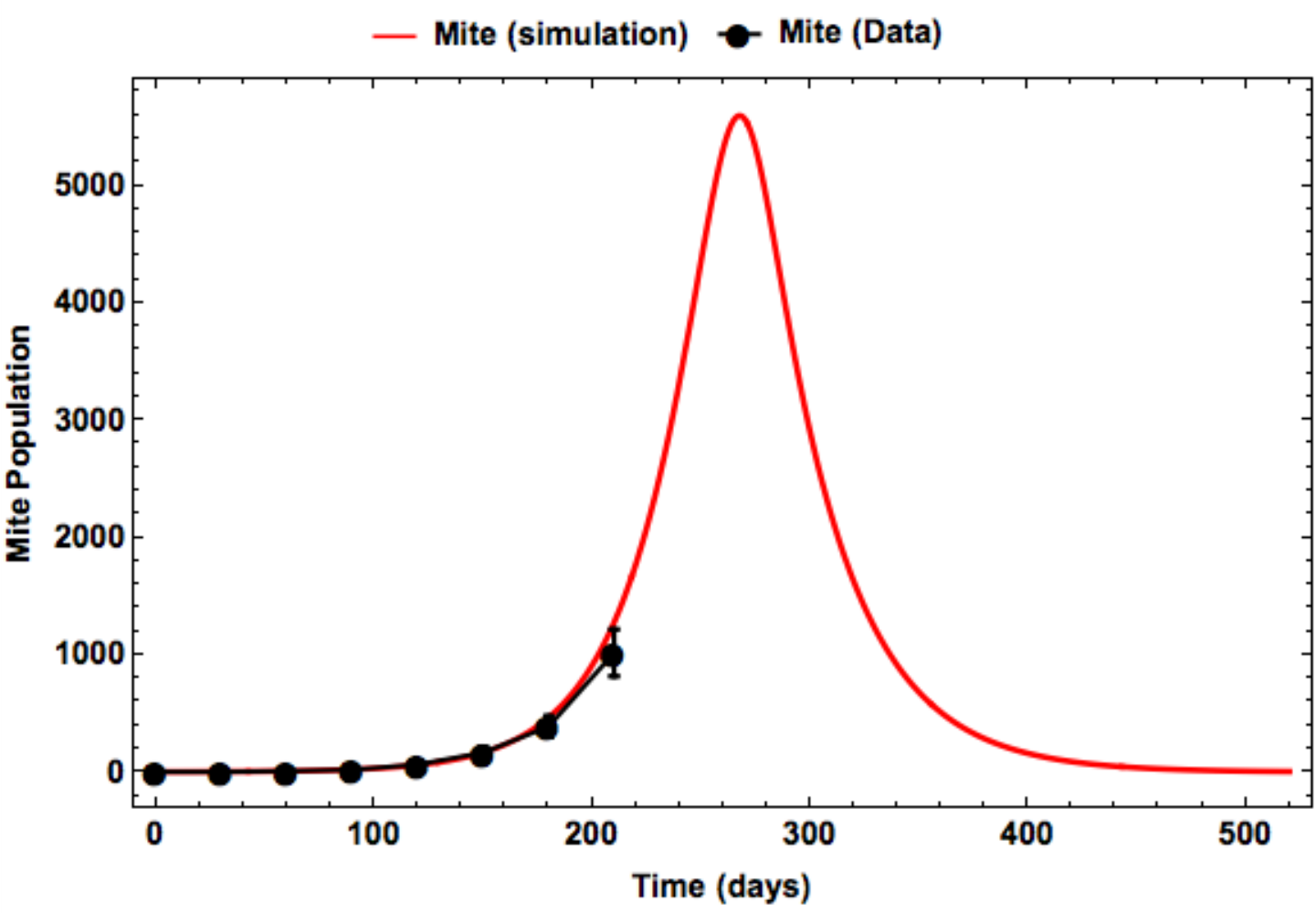}\label{fig_Model_Simulation_MiteS1}}\hspace{5mm}
\end{center}
\vspace{-15pt}
\caption{Time series of the brood, adult bee, and mite model simulation together with the average population data from the University of Arizona - West Campus Agricultural Facility. These figures represent respectively the average brood, adult bee, and mite population of 20 colonies with its standard error. The simulation is performed using $r=1500$, $K=35000000$, $d_b =0.0185$, $d_h= 0.045$, $d_m=0.029$, $\alpha_b=0.045$, $\alpha_h=0.49$, $c=1.9$, $a=8500$, $\tau=21$, $B_0(t)=B(0)=0$, $H(0)= 9000$, and $M(0) = 3$. Time $t=0$ corresponds to April 24.}
\label{fig:TimeSerieDataFit1}
\end{figure}

\begin{center}
    {\bf Best fit with seasonality}
\end{center} 


\begin{figure}[H]
\begin{center}
\subfigure[Mean brood population in colonies]{\includegraphics[height = 50mm, width = 40mm]{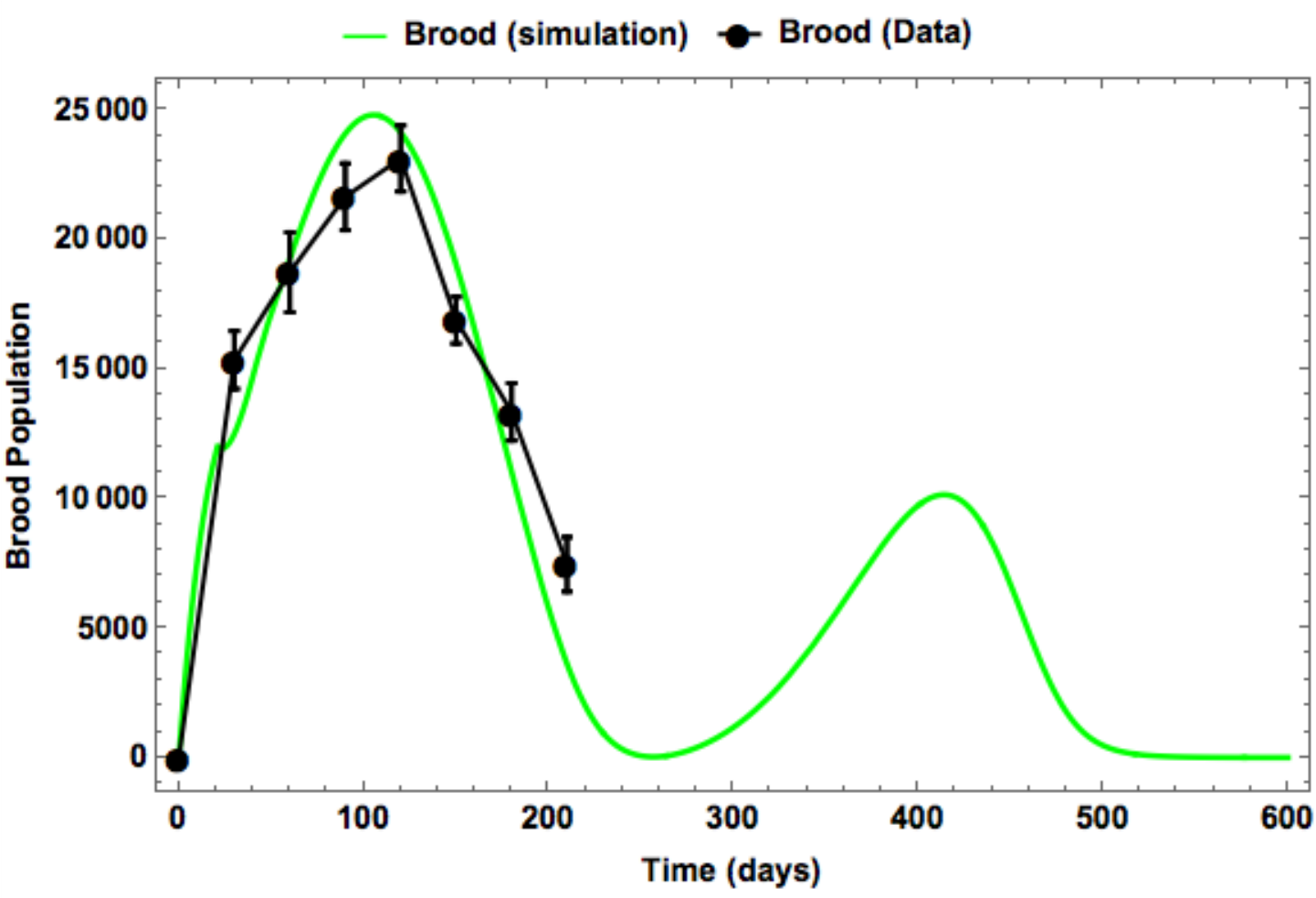}\label{fig_Model_Simulation_BroodS12}}\hspace{5mm}
\subfigure[Mean bee population in colonies]{\includegraphics[height = 50mm, width = 40mm]{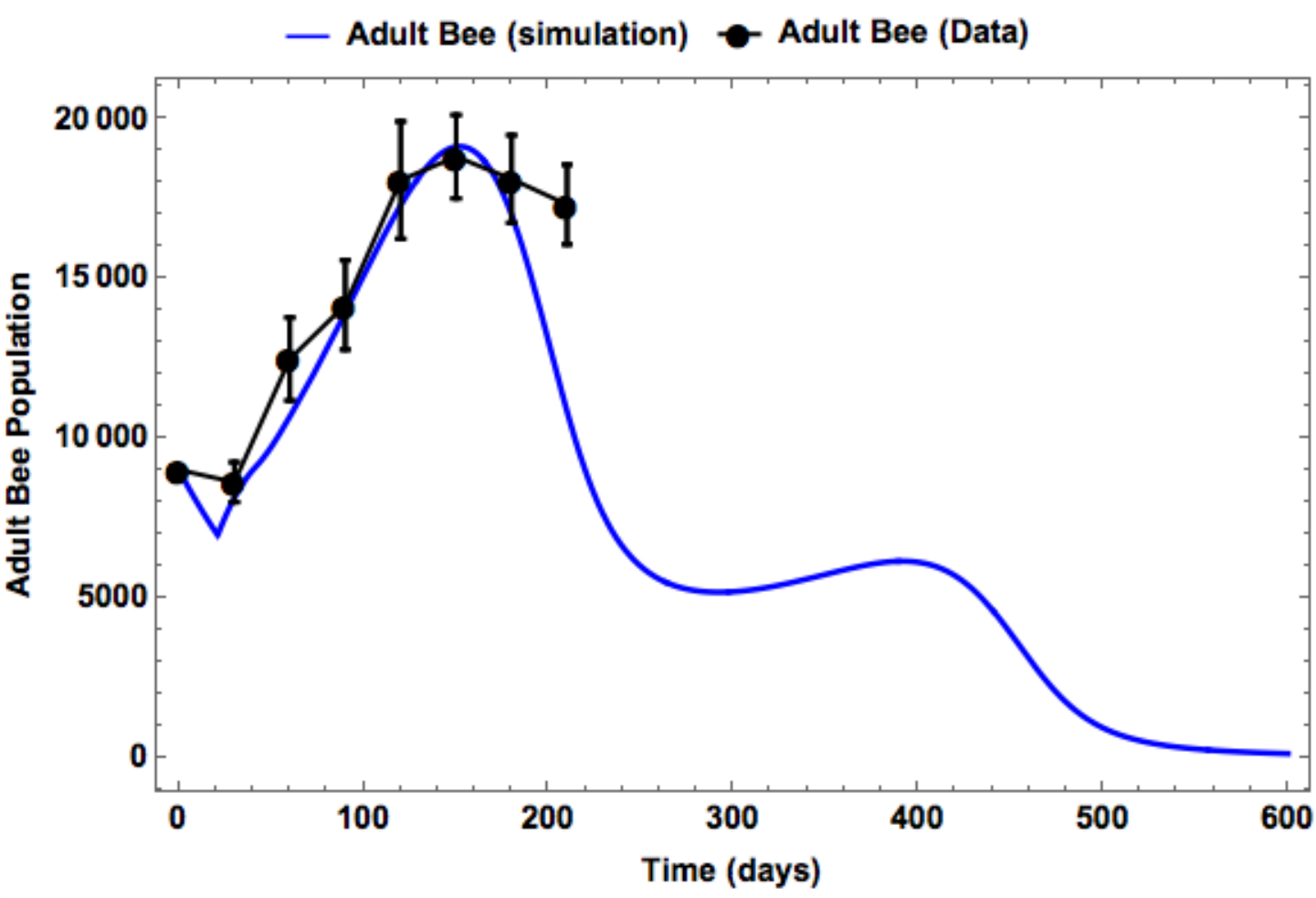}\label{fig_Model_Simulation_BeeS12}}\hspace{5mm}
\subfigure[Mean mite population in colonies]{\includegraphics[height = 50mm, width = 40mm]{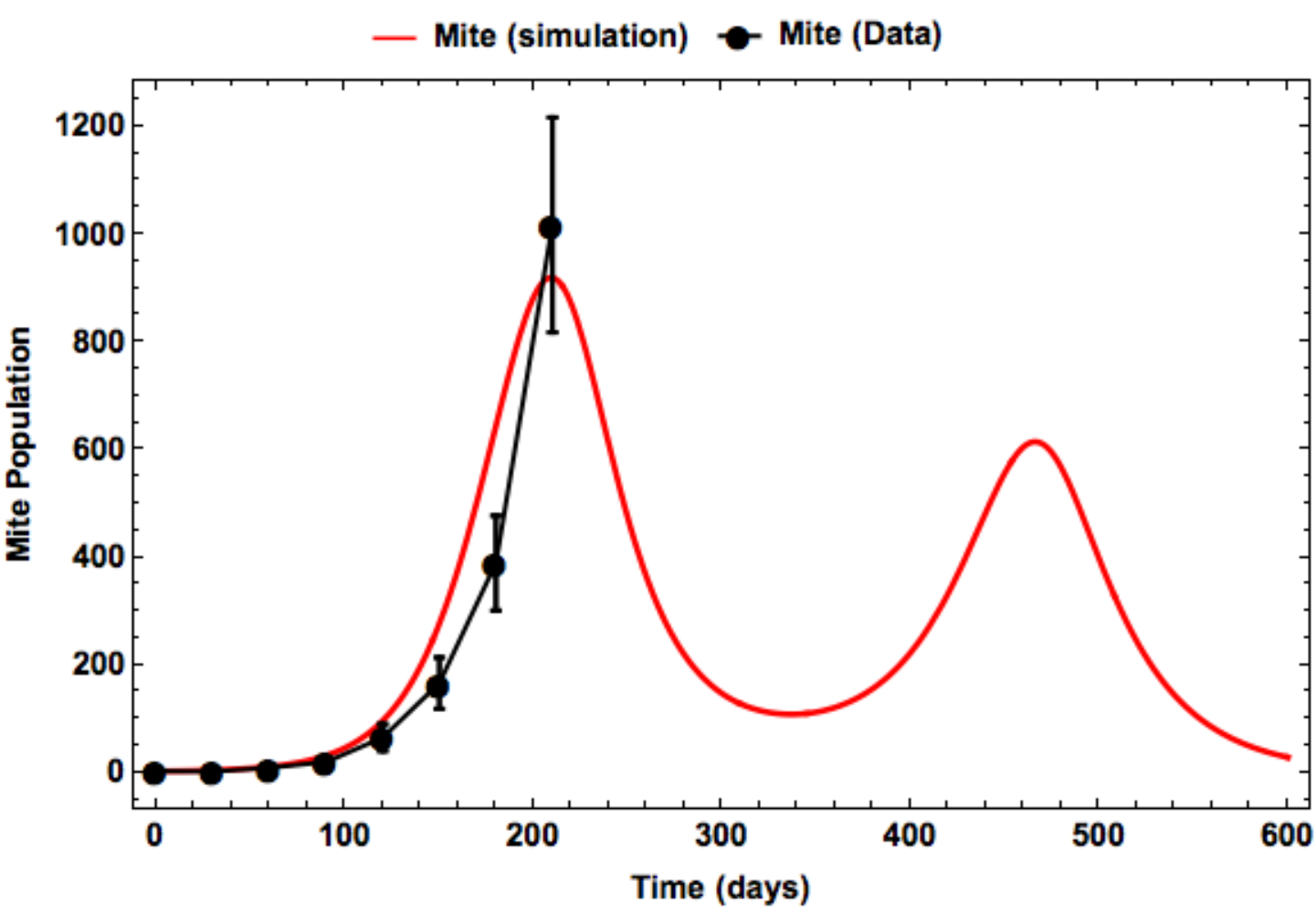}\label{fig_Model_Simulation_MiteS12}}
\end{center}
\vspace{-15pt}
\caption{Time series of the brood, adult bee, and mite model simulation together with the average population data from the University of Arizona - West Campus Agricultural Facility. These figures  represent respectively the average brood, adult bee, and mite population of 20 colonies with its standard error. The simulation is performed using using $r =1500$, $K=95000000$, $d_b =0.051$, $d_h= 0.0121$, $d_m=0.027$, $\alpha_b=0.0447$, $\alpha_h=0.8$, $c=1.9$, $a=8050$, $\Phi = 65$, $\tau=21$, $B_0(t)=B(0)=0$, $H(0)= 9000$, and $M(0) = 3$. Time $t=0$ corresponds to April 24.}
\label{fig:TimeSerieDataFit2}
\end{figure}


\textbf{Comparisons between the best fit of the model with and without seasonality (Figures \ref{fig:TimeSerieDataFit1} and \ref{fig:TimeSerieDataFit2}):} 
\begin{enumerate}
\item A better fit of the model simulation to the data is obtained when seasonality is taken into account in Figure \ref{fig:TimeSerieDataFit2} as oppose to Figure \ref{fig:TimeSerieDataFit1}. This reflects a more realistic life history parameters of honeybees and mites as presented in Figure \ref{fig:TimeSerieDataFit2}.
\item The parameter values used for both with and without seasonality Model \eqref{BHM1}-\eqref{BHM2} produce the following equilibrium points:
\[E_{000}=(0,0,0),\quad E_{B^*_1H^*_10}=(24171, 20930.2,0),\quad E_{B^*_2H^*_20}=(1931.15, 1672.22,0)\]
We highlight that under these parameter values, the sufficient conditions in Theorem \ref{pr1:IntEq} for the existence of interior equilibrium are not satisfied, thus, there is no interior point. According to Theorem \ref{th2:bq}, $E_{000}$ and $E_{B^*_1H^*_10}$ are asymptotically stable while $E_{B^*_2H^*_20}$ is unstable. Given initial conditions in the simulations shown in Figures \ref{fig_Model_Simulation_BroodS1}-\ref{fig_Model_Simulation_MiteS1}, honeybee and mite populations go extinct as $t\rightarrow\infty$. However, for other initial conditions, we could have the survival of only the honeybee. 
\item The population of brood and adult bees are driven extinct without seasonality in approximately 350 days, while mite population in approximately 500 days. This is an unrealistic situation as the mite population cannot outlive the colony population but rather should die with the colonie. With seasonality all populations went extinct in approximately 600 days. This result first indicates that including seasonality provides a more realistic scenario on the modeling of brood-bee-mite interaction. In addition, the environmental changes due to seasonality could promote a longer survival of honeybee colony infested by the {\it Varroa} mites. This highlights the effects of seasonal fluctuation on survivability of species.
\item In the presence of seasonality, the population of brood, adult bee, and mites tend to have a second rise after approximately one year and this is due to the resumption of egg-laying rate by the queen in the late winter as illustrated in Figure \ref{fig_EggLayingRate} (see the second peak starting in February 24).\\
\end{enumerate}

\noindent\textbf{Effects of infestation rate on brood population ($\alpha_b$) and seasonality:} In Appendix \ref{AppendixB}, we provide comparison on the role of $\alpha_b$ on the population dynamics of brood, adult bee, and mite in Figures \ref{fig:TimeSerieStable}, \ref{fig:TimeSerieOscillation}, \ref{fig:TimeSerieNonperiodic}, and \ref{fig:TimeSerieExtinction}. Those simulations show time series simulations and comparison of the dynamics with and without seasonality under different $\alpha_b$ values. For these simulations, parameters were chosen such that a unique interior equilibrium exists and is locally stable. For the smallest value of $\alpha_b$, (i.e. when the unique interior equilibrium is stable without seasonality and $\alpha_b=0.022$), taking seasonality into account causes the mite population to die out while the adult bee population stabilizes and the brood population fluctuates (Figure \ref{fig:TimeSerieStable}). An intermediate value of $\alpha_b = 0.024$ has the potential to generate fluctuating dynamics without seasonality while only the mite population dies out when seasonality is considered (Figure \ref{fig:TimeSerieOscillation}). In the presence of seasonality, a larger $\alpha_b$ (i.e. 0.027) has the potential to drive the brood, adult bee, and mite through non-periodic dynamics while all populations die out without seasonality (Figure \ref{fig:TimeSerieNonperiodic}). A large value of $\alpha_b=0.028$ has the ability to drive colonies to collapse irrespective of seasonality (Figure \ref{fig:TimeSerieExtinction}). While colonies can collapse under large $\alpha_b$ and in the absence of control measure, the results presented in Figure \ref{fig:TimeSerieExtinction} show that all populations die out before before the third year (1000$^{th}$ day) when seasonality is not taken into account. Populations persist over four years when seasonality is included. Such result highlights the importance of considering seasonality when modeling the population dynamics of honeybees colonies infested with \emph{Varroa} mite. \\

\subsection{Sensitivity Analysis }\label{c4sensitivityAnal}

Data fitting and parameter estimation with and without seasonality of $\alpha_b$, $\alpha_h$, and $a$ are provided in Section \ref{c4Material} using colonies' data from \cite{degrandi2008comparisons} (see Figure \ref{fig:TimeSerieDataFit1} for fitting without seasonality and Figure \ref{fig:TimeSerieDataFit2} for fitting with seasonality) and all other parameters sources are listed in Table \ref{C4ParamTab}. It is often noted in mathematical biology that natural variation, error in measurements may cause a variation in the parameter of the system \citep{marino2008methodology}. Thus, identifying critical input parameters of a model and quantifying how the uncertainty of such parameters impact model outcome is paramount. This section measures and quantifies the effect of parameter sensitivity on the population size of brood, adult bee, and mite, respectively, through global sensitivity analysis (SA). As noted by \cite{marino2008methodology}, different SA techniques will perform better for specific types of mathematical and computational models. There have been numerous global sensitivity methods discussed in the literature. For a detailed review on Monte Carlo analysis and variance decomposition methods, see \cite{iman1988investigation, cacuci2004comparative, ionescu2004comparative, saltelli1998alternative}. However, in order to obtain a holistic view regarding the sensitivity of the input parameters on the model outcome, two different SA methods were employed: (1) the Partial Rank Correlation Coefficient (PRCC) SA with Latin Hypercube Sampling (or LHS first introduced by \cite{mckay1979comparison})  as the sampling technique; (2) and the Extended Fourier Amplitude Sensitivity Test (eFAST). We followed the methodology discussed in \cite{marino2008methodology} for both SA methods (i.e. LHS/PRCC and eFAST). \\

\begin{figure}[H]
\begin{center}
\subfigure
{\includegraphics[height = 55mm, width = 60mm]{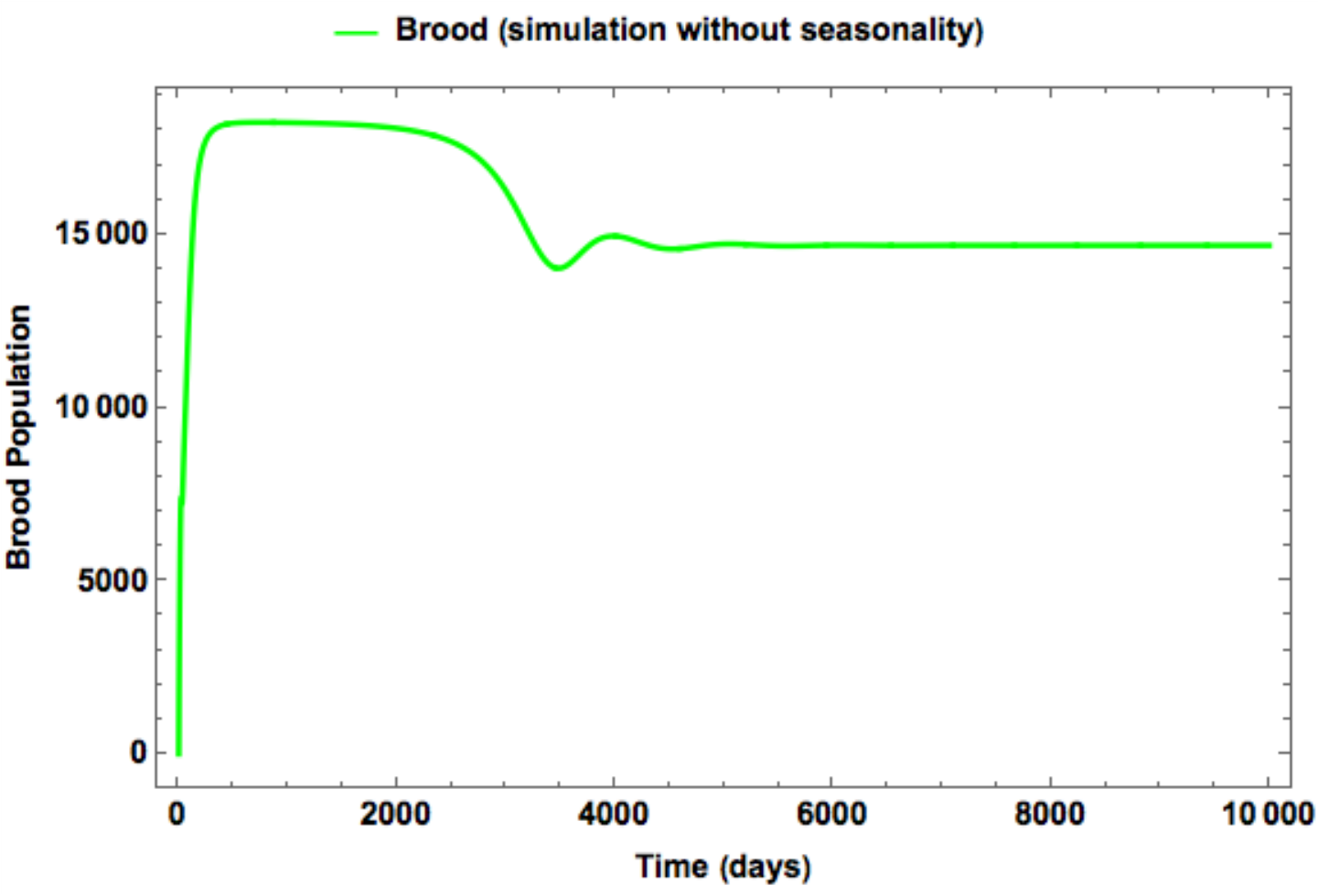}\label{fig_Brood_Stable_NoS}}\hspace{5mm}
\subfigure
{\includegraphics[height = 55mm, width = 60mm]{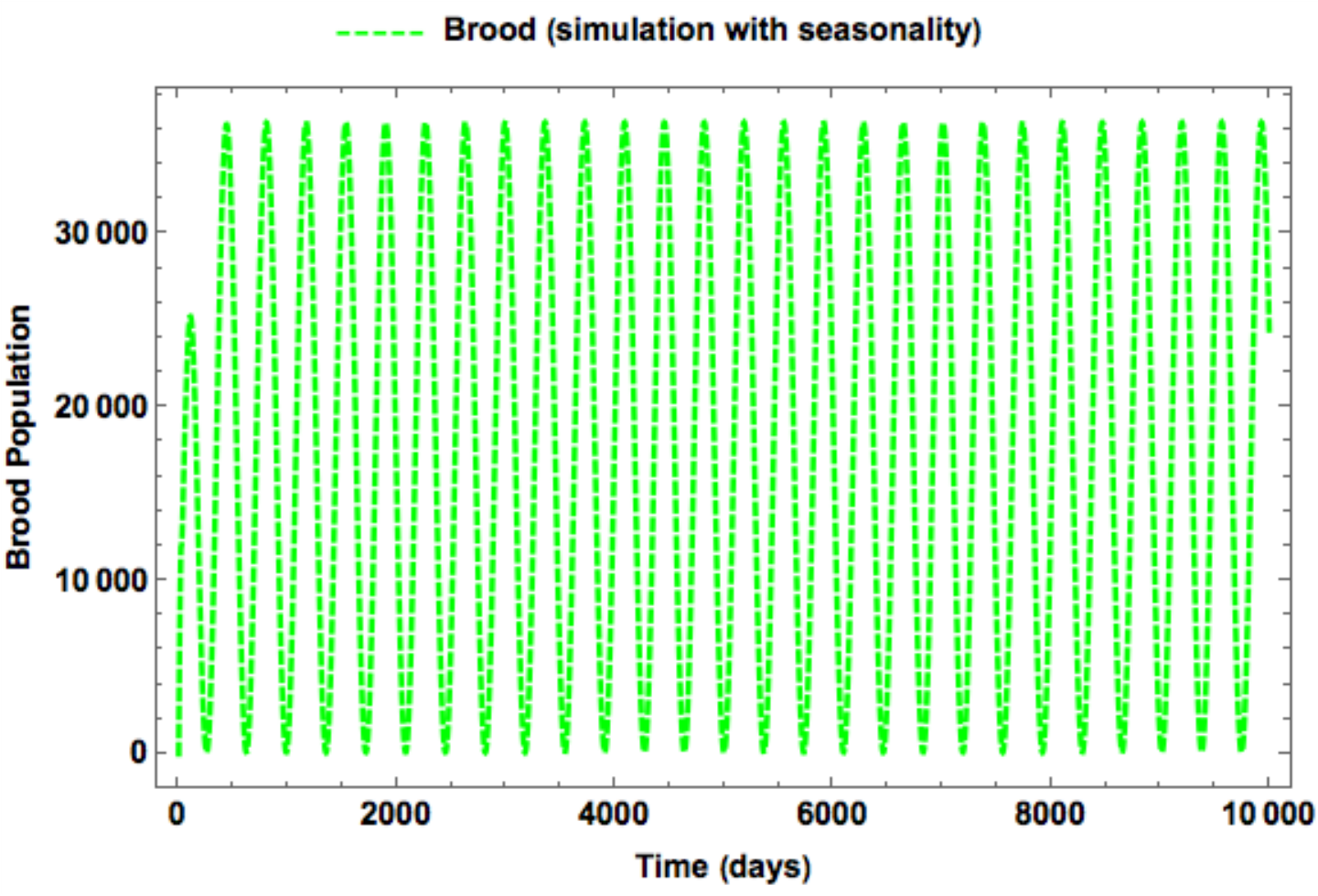}\label{fig_Brood_Stable_S}}\\
\subfigure
{\includegraphics[height = 55mm, width = 60mm]{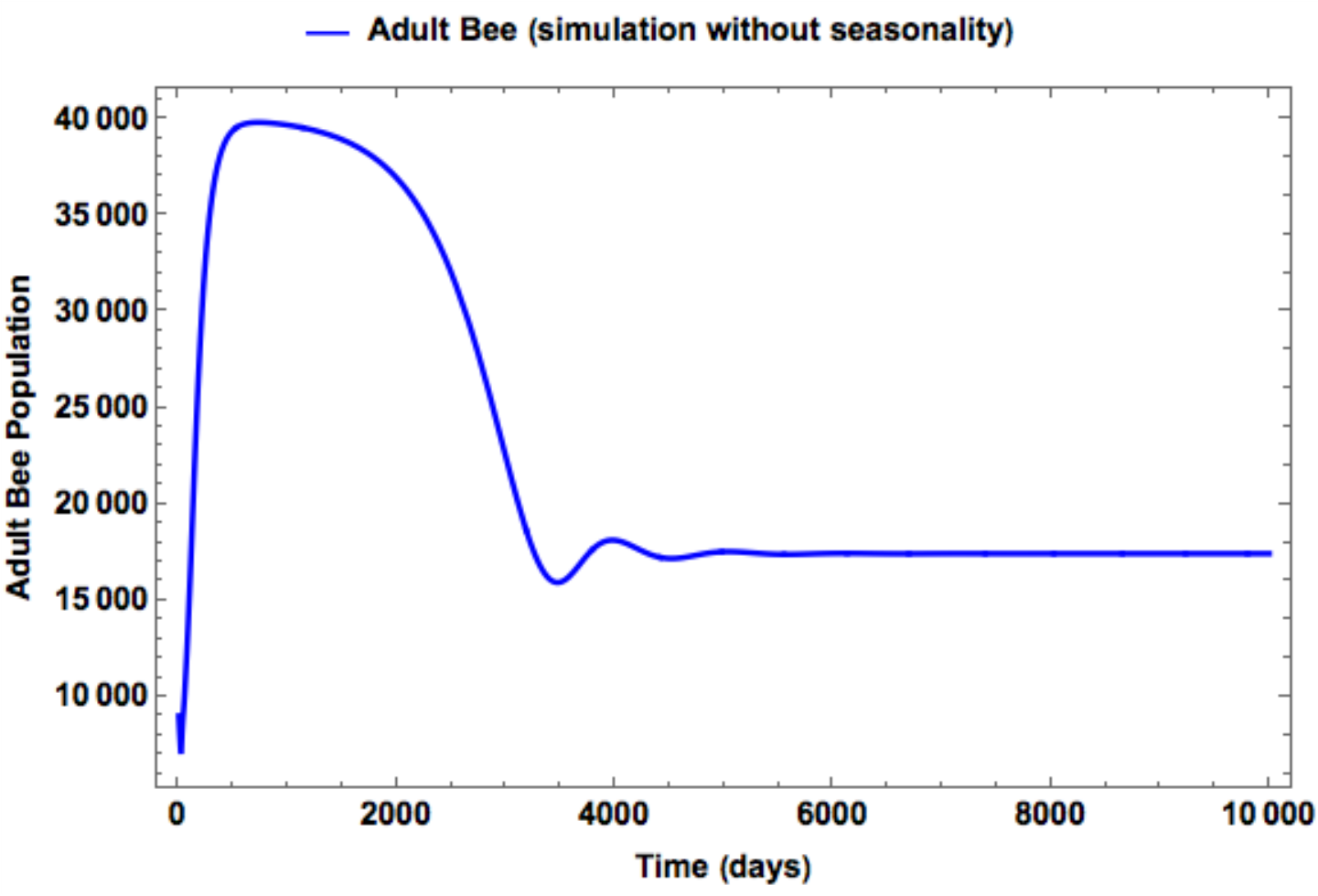}\label{fig_Bee_Stable_NoS}}\hspace{5mm}
\subfigure
{\includegraphics[height = 55mm, width = 60mm]{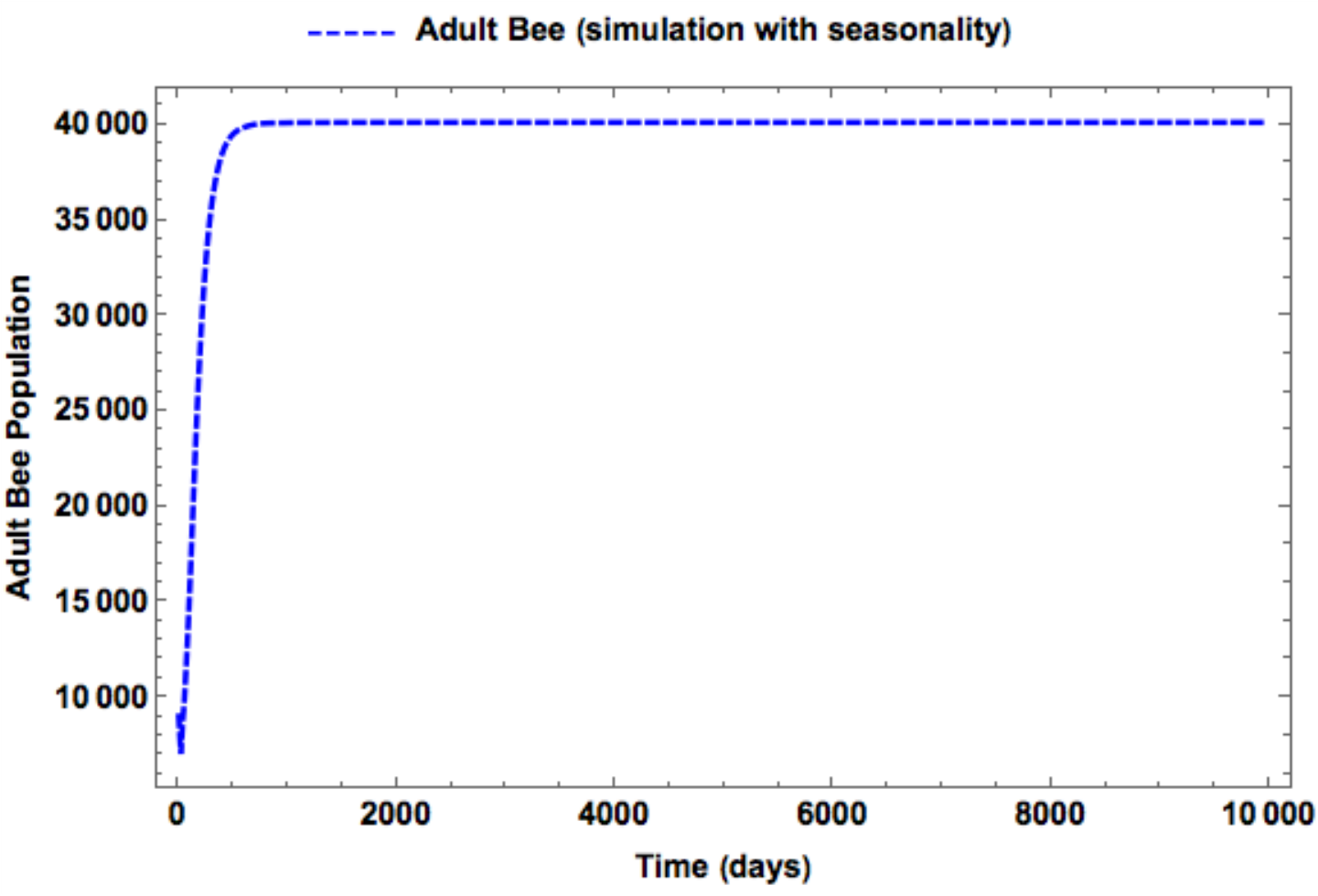}\label{fig_Bee_Stable_S}}\\
\subfigure
{\includegraphics[height = 55mm, width = 60mm]{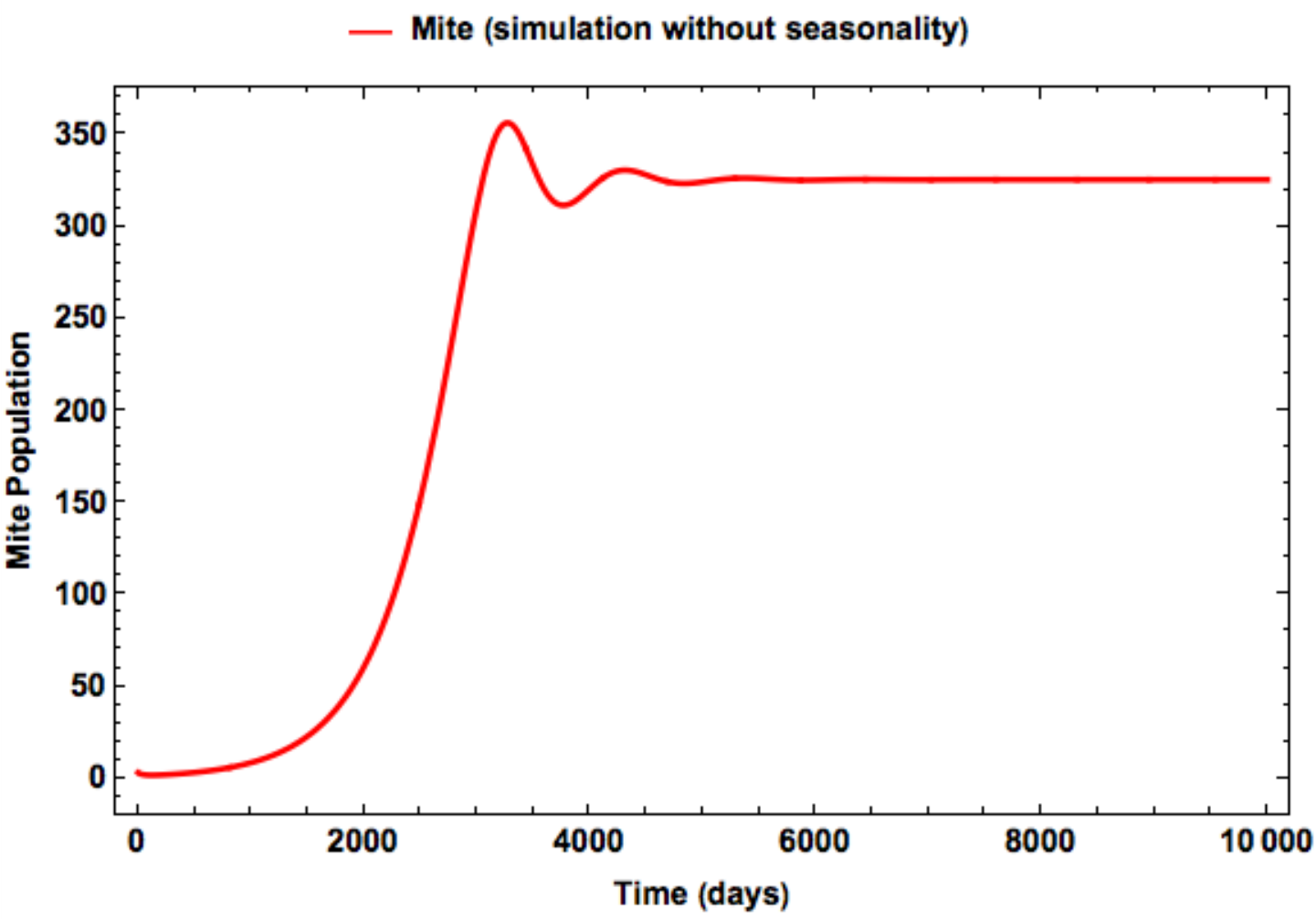}\label{fig_Mite_Stable_NoS}}\hspace{5mm}
\subfigure
{\includegraphics[height = 55mm, width = 60mm]{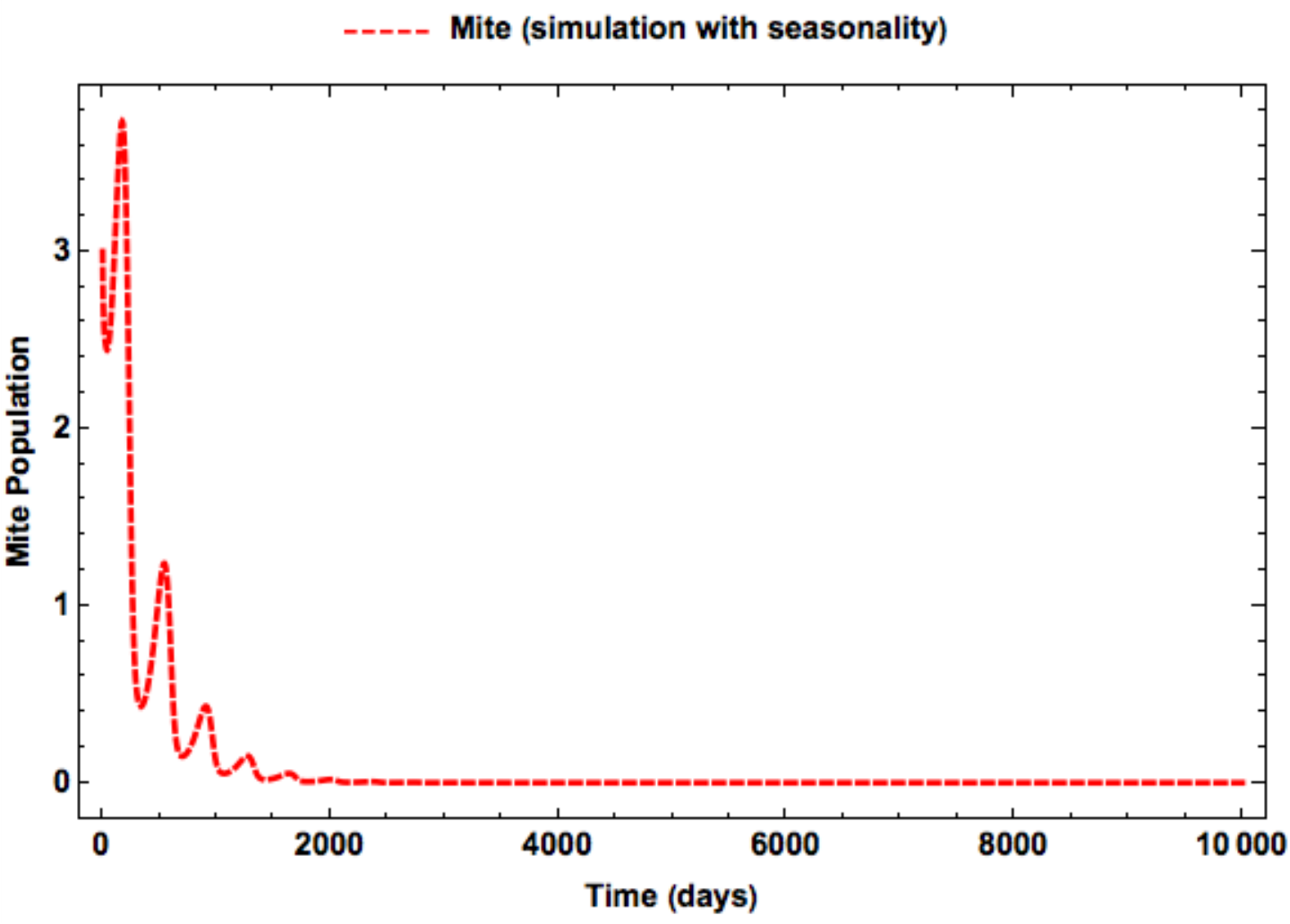}\label{fig_Mite_Stable_S}}\\
$\alpha_b=0.022$
\end{center}
\vspace{-10pt}
\caption[Time Series: Brood, Adult Bee, and Mite Simulation When the Queen's Eggs Laying Rate is Constant and $\alpha_b=0.022$.]{{\small  Time series of the brood, adult bee, and mites simulation using $r=1500$, $K=95000000$, $d_b =0.051$, $d_h= 0.0121$, $d_m=0.027$, $\alpha_h=0.8$, $c=1.9$, $a=8050$, $\tau=21$, $\Phi=65$, $B_0(t)=B(0)=0$, $H(0)= 9000$, and $M(0) = 3$ when the queen's eggs laying rate is constant in figures on the left column 
(i.e. no seasonality) and when the queen's eggs laying rate has seasonality in figures on the right column with $\alpha_b =0.022$.}} 
\label{fig:TimeSerieStable}
\end{figure}
For our analysis, SA was conducted on the time corresponding to the largest population size in Figure \ref{fig:TimeSerieDataFit2} as output and the eleven parameters of Model \eqref{BHM1}-\eqref{BHM2} when seasonality is taken into account using the egg laying rate formula \eqref{EggRateFunc}. The time corresponding to the largest population size was selected to determine how the input parameters might affect the brood, adult bee, and mite population at their peak thus maintaining or causing the collapse of the colony.
 The results of the SA are presented in Figures \ref{fig:SensitivityBrood}, \ref{fig:SensitivityBee}, and \ref{fig:SensitivityMite} for the brood, adult honeybee and mite populations.\\


\begin{figure}[ht]
\begin{center}
\subfigure[eFAST sensitivity at time = 96]{\includegraphics[height = 65mm, width = 70mm]{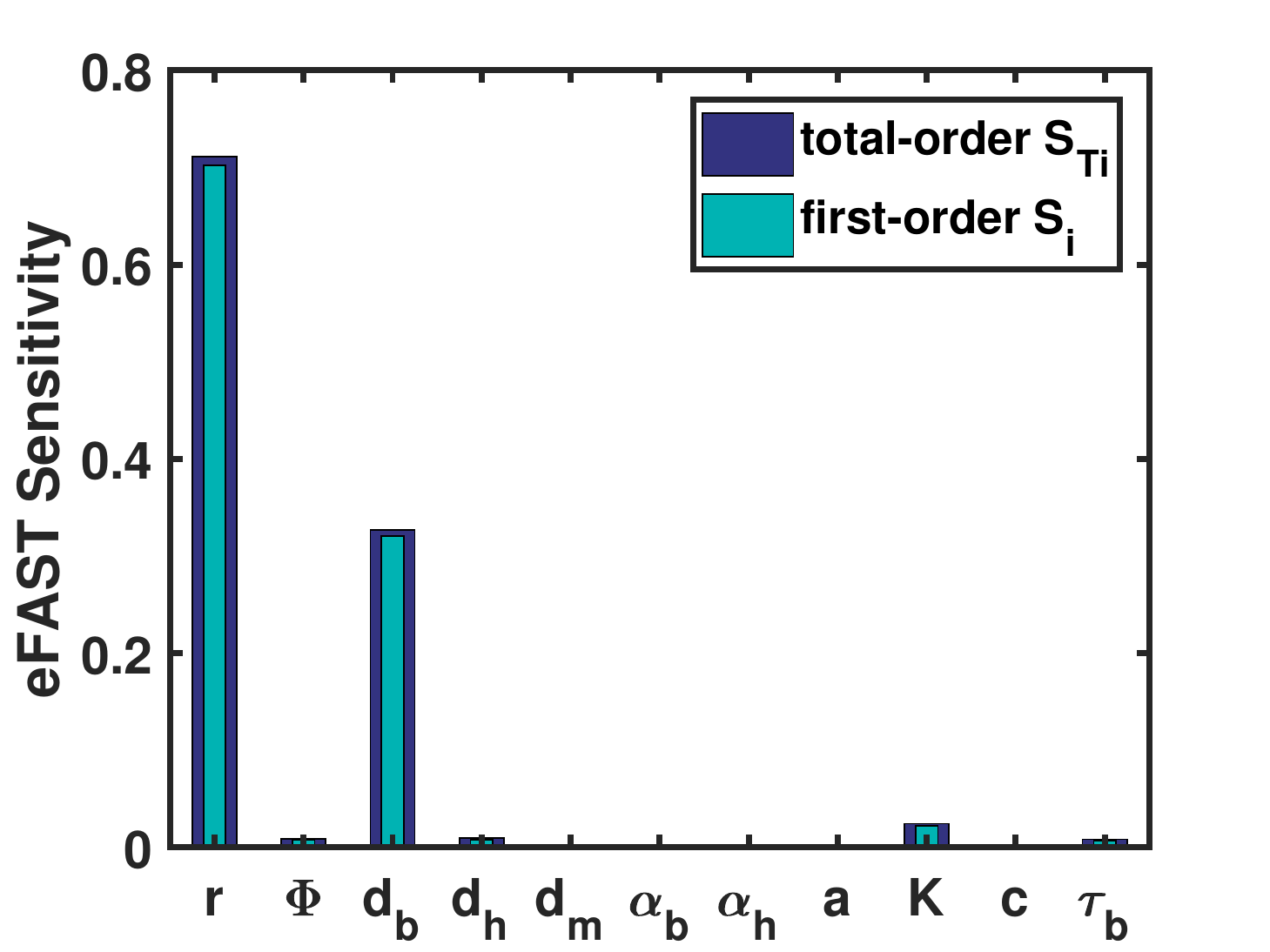}\label{fig_Sensitivity_Brood_efast}}
\subfigure[PRCC sensitivity at time = 96]{\includegraphics[height = 65mm, width = 70mm]{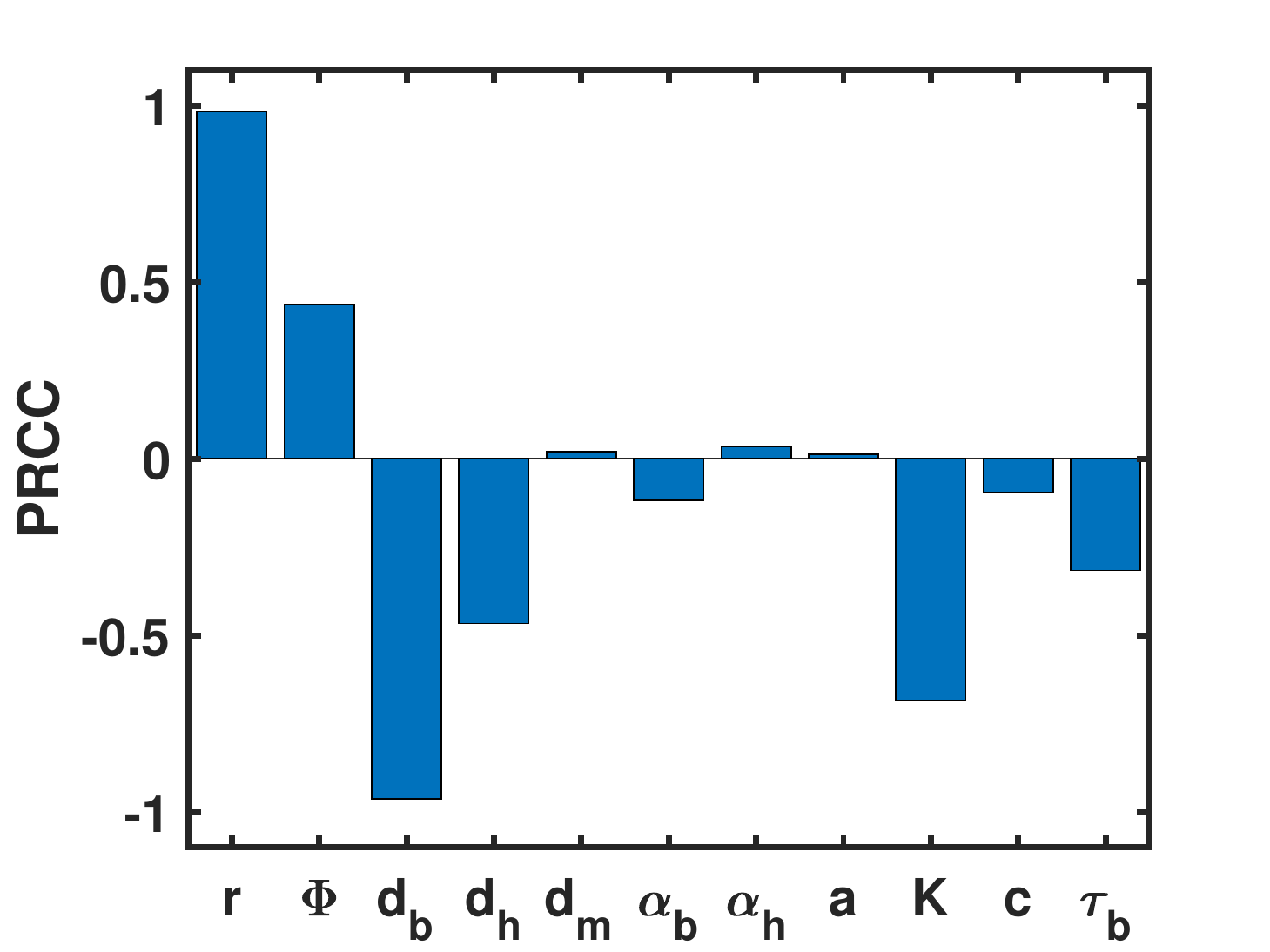}\label{fig_Sensitivity_Brood_PRCC}}
\end{center}
\vspace{-15pt}
\caption{eFAST and PRCC Sensitivity analysis on Model \eqref{BHM1} and \eqref{BHM2} using parameter from Figures \ref{fig:TimeSerieDataFit2} where time point chosen correspond to the highest population point from the brood population in Figure \ref{fig_Model_Simulation_BroodS12}. Figures \ref{fig_Sensitivity_Brood_efast} shows the eFAST results with resampling and search curves were resampled five times $(N_R = 5)$, for a total of 3575 model evaluations $(N_S = 65)$. First-order $S_i$ and total-order $S_{Ti}$ are shown for each parameter as shown in the legend. Figures \ref{fig_Sensitivity_Brood_PRCC} illustrates the result of the PRCC results with $N = 1000$.}
\label{fig:SensitivityBrood}
\end{figure}

Both PRCC and eFAST values in Figure \ref{fig:SensitivityBrood} indicate that $r$, $d_b$, and $K$ are the most sensitive parameters affecting the brood population size with $r$ (the maximum queen's egg-laying rate) being the most sensitive of the three. The PRCC values of $\Phi$ and $d_h$ in Figure \ref{fig_Sensitivity_Brood_PRCC} appear to be at the intermediate level ($\approx$ 0.5  as shown in Table \ref{fig:PRCCeFAST96} in Appendix \ref{AppendixB}). The parameters $d_m$, $\alpha_b$, $\alpha_h$, $a$, and $c$ are shown not to have a high sensitivity value in both Figures \ref{fig_Sensitivity_Brood_efast} and \ref{fig_Sensitivity_Brood_PRCC} suggesting that the largest brood population size is not sensitive to mite infestation rate but rather the queen's ability to lay eggs, which is a function of seasonality, age of the queen, colony size, nutrition, etc. \cite{degrandi1989beepop}. It is observable in Figure \ref{fig:SensitivityBee} and \ref{fig:SensitivityMite} that the most sensitive parameters affecting the adult bee and mite populations size from both the PRCC and eFAST indexes are $r$, $\alpha_b$, and $c$. Moreover, the input parameters $d_b$ and $d_m$ have a high sensitivity to the mite population from both the PRCC and eFAST indexes (Figure \ref{fig:SensitivityMite}) indicating that reduction of the mite mortality through proper control measure may release parasitic pressure on the colony. It is significant to point out that the maximum queen's egg-laying rate, $r$, appears to be the most sensitive parameter affecting the population size of the brood, adult honeybees (i.e. the entire colony of honeybees), and mites under both PRCC and eFAST SA methods (Figures \ref{fig:SensitivityBrood}, \ref{fig:SensitivityBee}, and \ref{fig:SensitivityMite}). These results have been confirmed by \cite{degrandi1989beepop} where the authors stated that the queen's egg-laying potential has the greatest effect on colony population size. It is also noticeable that the natural mortality of adult bees (i.e. $\alpha_h$) is not very sensitive to the brood, adult bee, and mite populations. The SA also reveals that the infestation rate on the brood (i.e. $\alpha_b$) may be another important parameter affecting the population size of the colony (see high PRCC and eFAST indexes values in Table \ref{fig:PRCCeFAST132} and \ref{fig:PRCCeFAST183} in Appendix \ref{AppendixB}).\\


\begin{figure}[H]
\begin{center}
\subfigure[eFast sensitivity at time = 132]{\includegraphics[height = 65mm, width = 70mm]{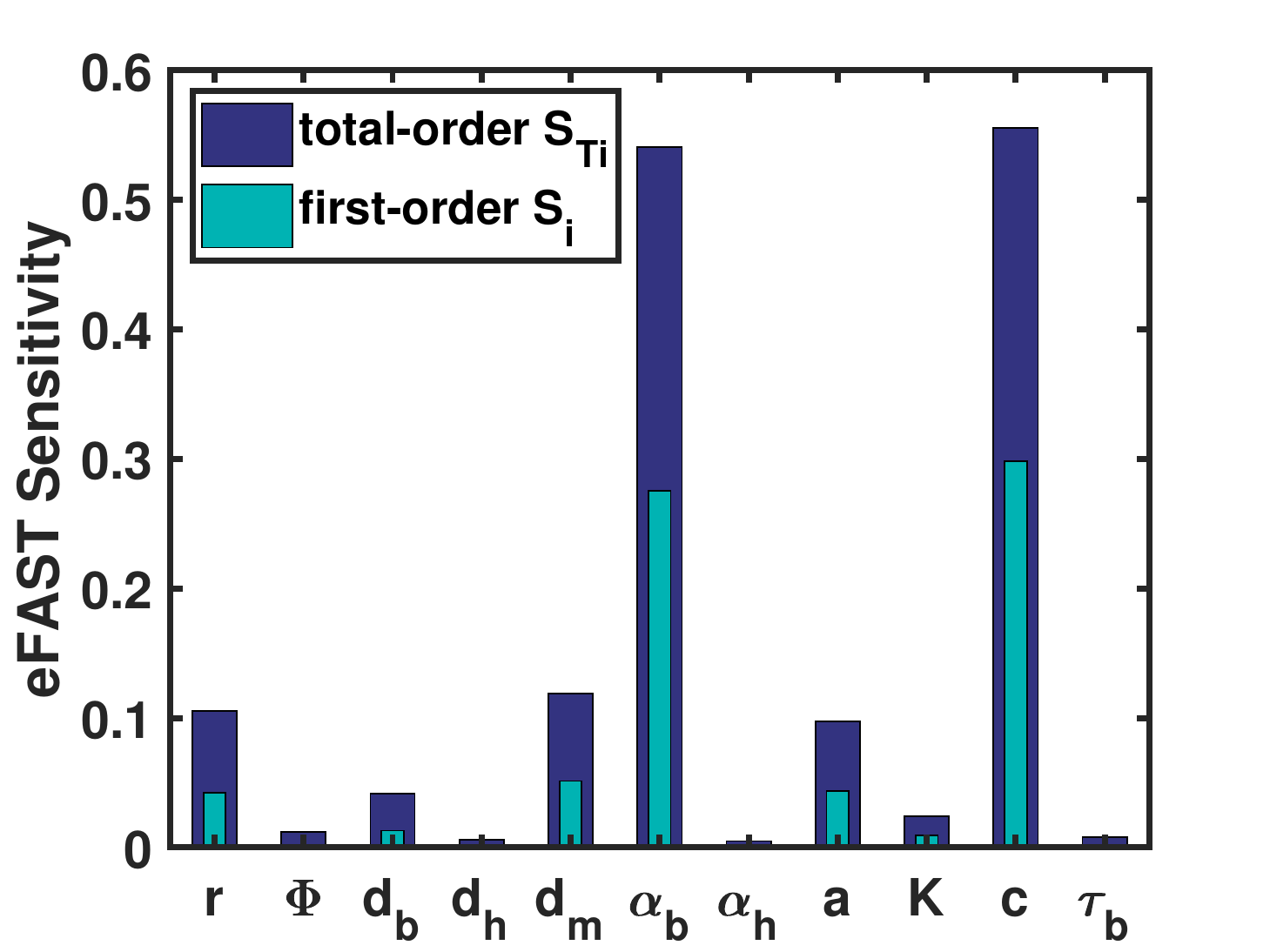}\label{fig_Sensitivity_Bee_efast}}
\subfigure[PRCC sensitivity at time = 132]{\includegraphics[height = 65mm, width = 70mm]{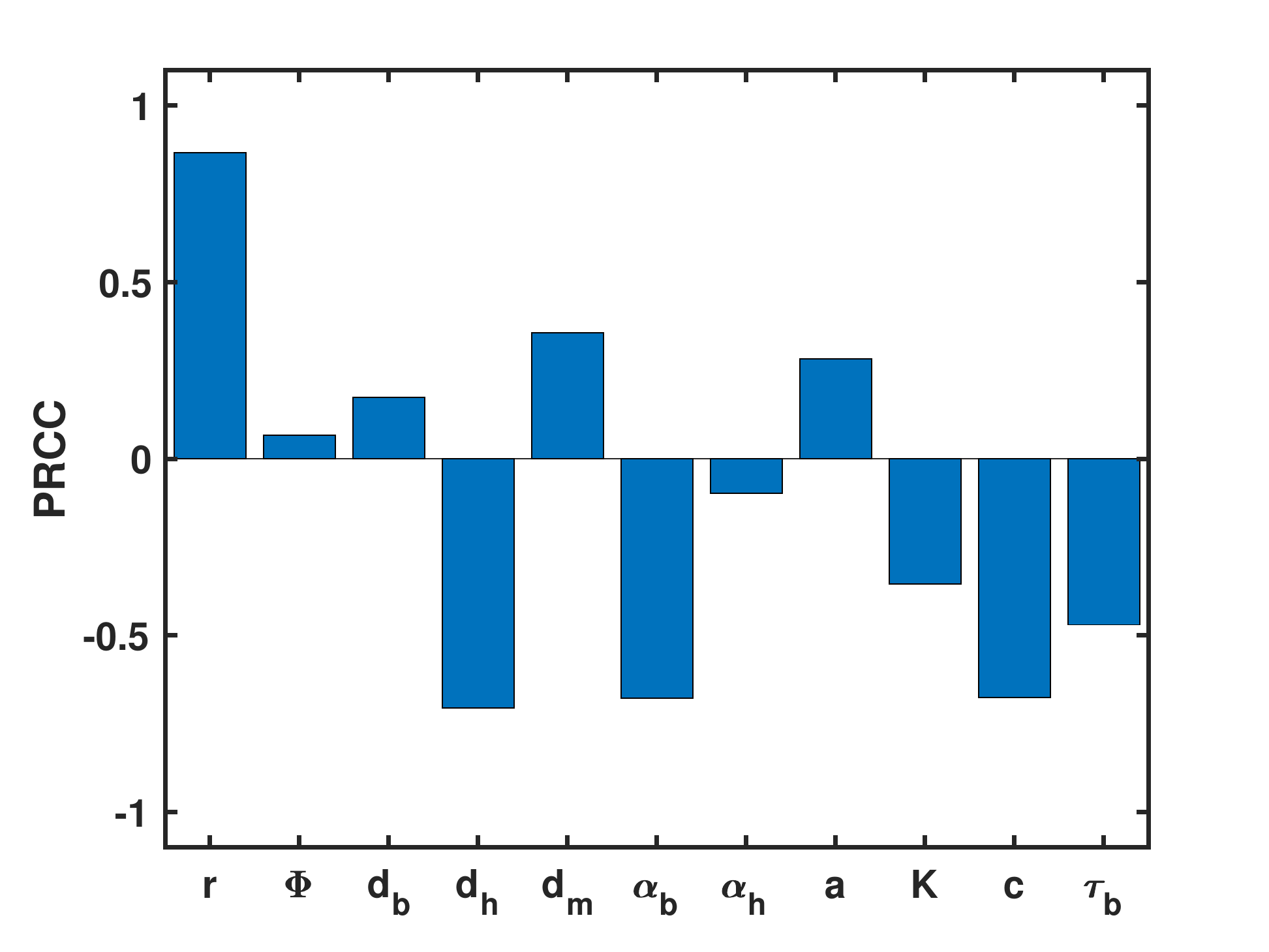}\label{fig_Sensitivity_Bee_PRCC}}
\end{center}
\vspace{-15pt}
\caption{eFAST and PRCC Sensitivity analysis on Model \eqref{BHM1} and \eqref{BHM2} using parameter from Figures \ref{fig:TimeSerieDataFit2} where time point chosen correspond to the highest population point from the adult bee population in Figure \ref{fig_Model_Simulation_BeeS12}. Figures \ref{fig_Sensitivity_Bee_efast} shows the eFAST results with resampling and search curves were resampled five times $(N_R = 5)$, for a total of 3575 model evaluations $(N_S = 65)$. First-order $S_i$ and total-order $S_{Ti}$ are shown for each parameter as shown in the legend. Figures \ref{fig_Sensitivity_Bee_PRCC} illustrates the result of the PRCC results with $N = 1000$.}
\label{fig:SensitivityBee}
\end{figure}


\begin{figure}[H]
\begin{center}
\subfigure[eFast sensitivity at time = 183]{\includegraphics[height = 65mm, width = 70mm]{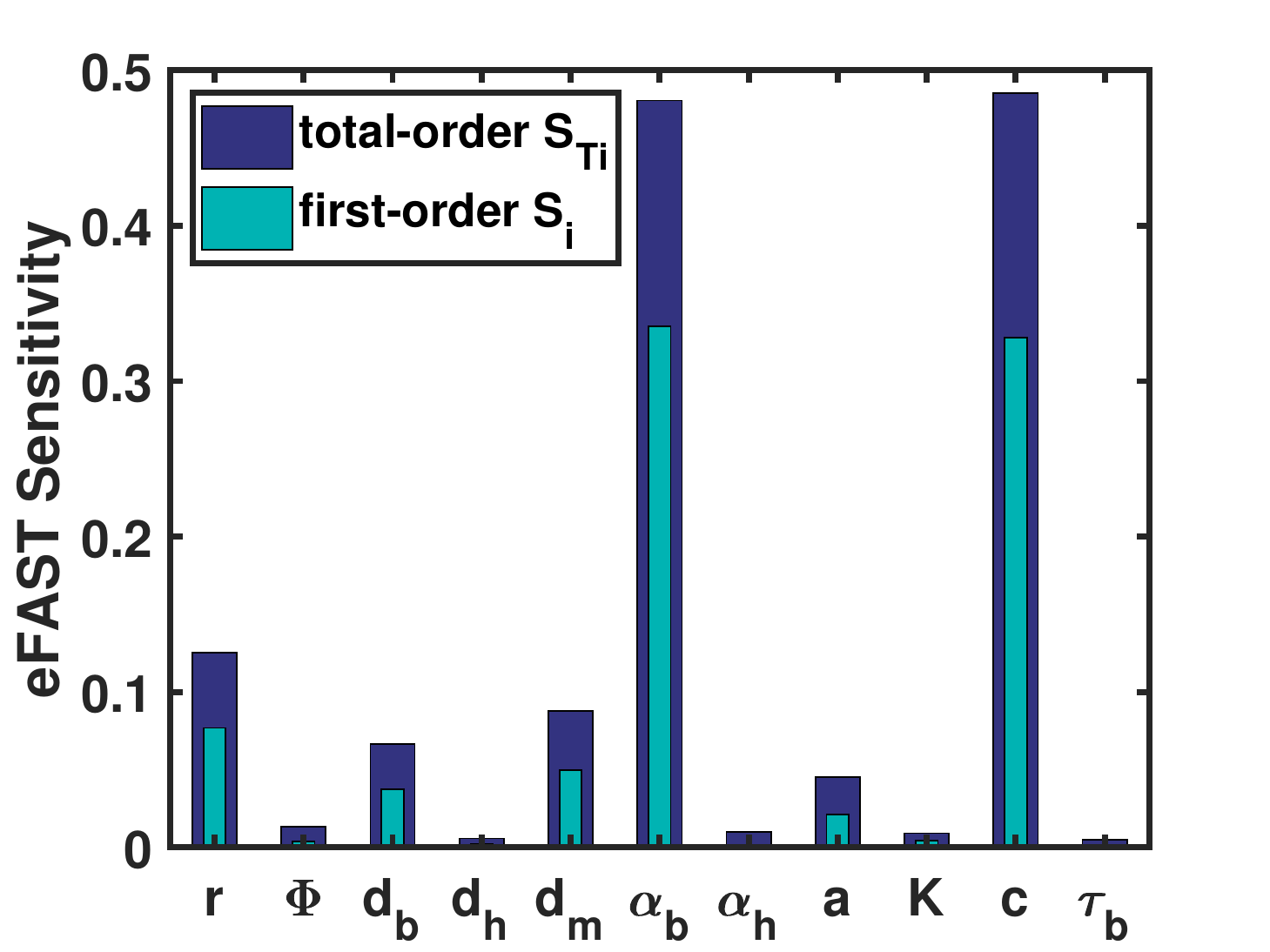}\label{fig_Sensitivity_Mite_efast}}
\subfigure[PRCC sensitivity at time = 183]{\includegraphics[height = 65mm, width = 70mm]{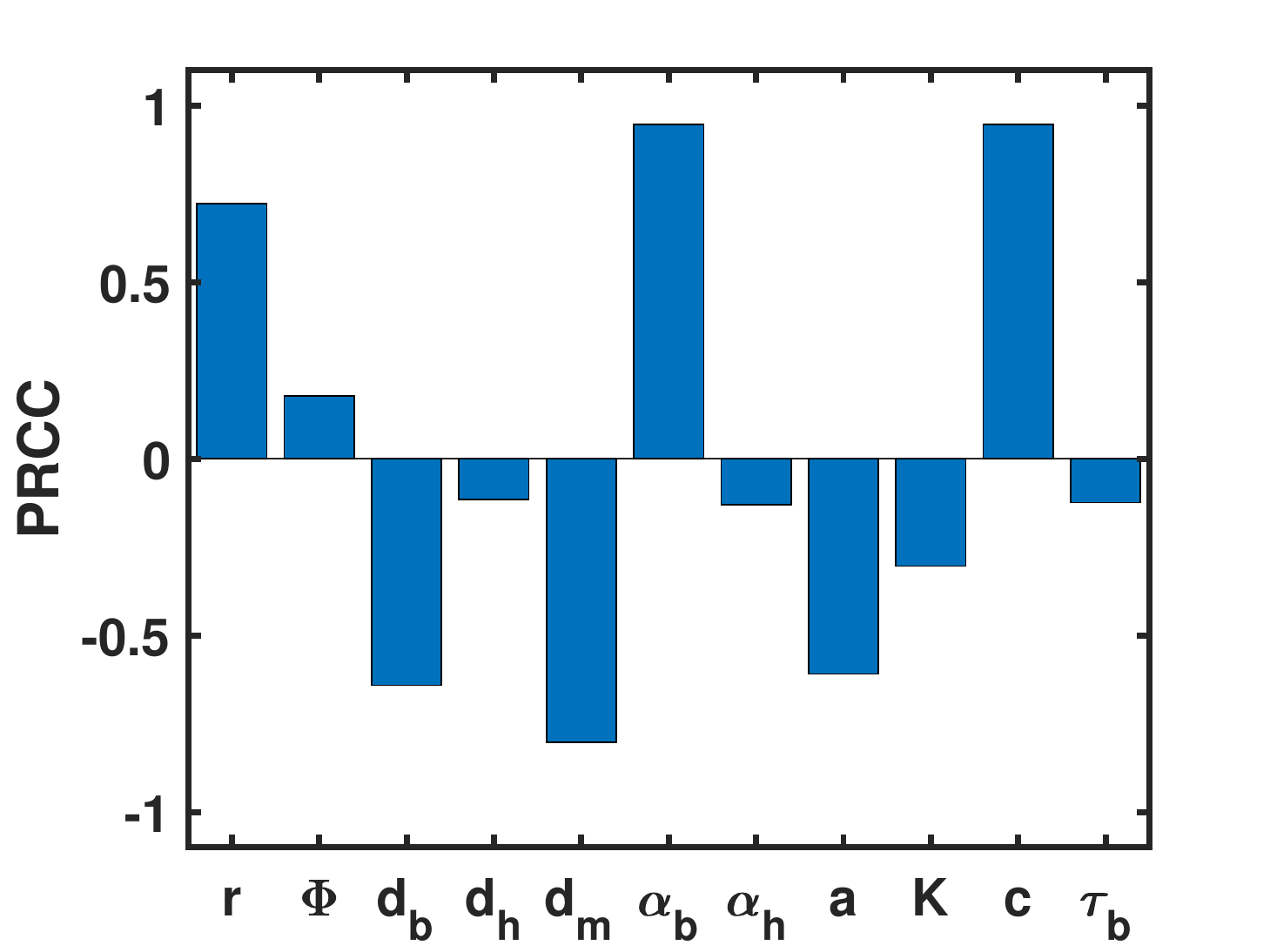}\label{fig_Sensitivity_Mite_PRCC}}
\end{center}
\vspace{-15pt}
\caption{eFAST and PRCC Sensitivity analysis on Model \eqref{BHM1} and \eqref{BHM2} using parameter from Figures \ref{fig:TimeSerieDataFit2} where time point chosen correspond to the highest population point from the mite population in Figure \ref{fig_Model_Simulation_MiteS12}. Figures \ref{fig_Sensitivity_Mite_efast} shows the eFAST results with resampling and search curves were resampled five times $(N_R = 5)$, for a total of 3575 model evaluations $(N_S = 65)$. First-order $S_i$ and total-order $S_{Ti}$ are shown for each parameter as shown in the legend. Figures \ref{fig_Sensitivity_Mite_PRCC} illustrates the result of the PRCC results with $N = 1000$.}
\label{fig:SensitivityMite}
\end{figure}



\section{Discussion}\label{C4Conclusion}
Colonies of honeybees have been declining for over a decade \citep{le2010varroa}. Colony losses are due to a combination of stressors \citep{le2010varroa, hayes2008survey}, but the presence of {\it Varroa} mites has been considered as one of the most important factors  \citep{kang2016disease, degrandi2004mathematical, messan2017migration}. In this study, we proposed a nonlinear stage-structure delay differential equations model that describes the interactions between brood, adult honeybees, and mites in a single patch framework where the maturation from brood to adult honeybees was taken into account. Moreover, noting that the temporal dynamics of honeybee colonies vary with respect to time of year (e.g. temperature, photoperiod, etc.) \citep{degrandi1989beepop} and the effects on the egg laying rate of the queen, seasonality was incorporated into our model. Comparisons are made between simulation predictions with and without seasonality. The theoretical results provide insights on how the presence of mites affect population dynamics of the adult honeybee and brood population. \\

It follows from our results that initial population size plays an important role in sustaining a healthy bee colony. Colony size affects egg lay rates, and larger colony sizes can rear more brood enabling the population to grow. A low natural death rate also contributes to a relatively rapid increase of the bee population. The introduction of mites into a healthy colony was shown to have dire effects on the colony growth. Mites can generate fluctuating dynamics of the colony population (i.e. brood, adult bees, and mites). Though the process behind the population oscillation is not completely clear, this phenomenon can result in the extinction of the colony \cite{kang2016disease}}. \\

The delay parameter (i.e. the development time from brood to adult bee) can stabilize, destabilize, or even promote fluctuating dynamics, leading to the coexistence or death of all species. In this study, larger development time combined with mites can destabilize population dynamics that could drive the colony collapsing. It was pointed out in \cite{puvskadija2017late} that unfavourable weather conditions after the queen bee starts with intensive oviposition during the early spring may cause an imbalance in the division of tasks among worker bees leading to slow spring development. Moreover, the development time is known to be influenced by climatic conditions and food availability \cite{nunes2006rate}. Possible explanation of the destabilization effect shown in this work due to large development time could hence be a detrimental weather condition.  \\

Seasonality has huge impacts on population dynamics. More specifically,  seasonality in the queen egg-laying rate could either promote or suppress the colony survival. For example, some our simulations showed that our honeybee colony could have stable equilibrium dynamics that both honeybee and mites coexist at the unique interior equilibrium when seasonality is not considered (i.e. when brood rearing does not depend on seasonality) while the colony collapses when seasonality is considered. In general, our simulations of our model without incorporating seasonality provided unrealistic scenarios where mite population outlives the colony population. This unrealistic dynamic was expected as mite population size varies from fall to spring as shown in the work of \cite{degrandi2004mathematical, degrandi2017dispersal} and thus illustrating the importance of year-round seasonality on predicting the mite population size and its effect on a colony. When the brood cycle was assumed to be seasonal through the incorporation of a time periodic dependent parameter in the egg-laying ability of the queen, the simulated model aligned well with the data. This exemplified a more realistic scenario of the brood-adult bee-mite system and the capabilities of our model to predict future population cycles. Under this scenario, the mite population dies out and the healthy colony is left with brood and adult bees. This result points to the importance of seasonality in honeybee colony survival. \\

The results of our sensitivity analysis from our PRCC and eFAST showed that the queen's egg-laying rate has the greatest impact on the colony's population size, which supports the work of \citep{degrandi1989beepop}. Colony size also was sensitive to the mite infestation rate of the brood population. This was expected as the infestation by the \emph{Varroa} destructor has been noted to be one of the major stressors of colony decline \citep{koleoglu2017effect}. Our results also illustrated the dynamics generated by the mite to brood infestation rate on the population size with and without seasonality. When seasonality is not taken into account, small infestation on the brood population could promote coexistence of all species at the interior equilibrium, intermediate infestation rate could yield the coexistence of all species through fluctuating dynamics, and large infestation rate could drive the colony to collapse. However, by incorporating seasonality, our results showed that mite population could die out under small and large infestation rate, but all the populations may go through non-periodic dynamics under an intermediate infestation rate on the brood population. Moreover, by comparing the seasonal and non-seasonal dynamics, our results indicated that in the environment where high seasonal fluctuation is present, infested colonies may survive longer than in an environment when small or no fluctuation is observed. These findings highlight the importance of seasonality in the honeybee interaction with {\it Varroa} mite and some of the parameters that may promote a mite-free colony or drive the colony to collapse. 


\section{Conclusion}\label{C5Conclusion}

As honeybee population continues to decline, understanding the related causes  is critical to alleviate this ecological disturbance. This study is the first to explicitly model brood-adult bee-mite interaction and incorporate the development time from brood to adult bee by using a distributed delay. Our findings revealed the catastrophic effects that mites can have on a healthy colony. By illustrating that the large development time from brood to adult bee can have the destabilizing effects on an infested colony (i.e. brood, adult bee, and mite population), this study demonstrates that while colony survival can be threaten by the availability of mites, favorable environmental conditions (e.g. weather, food resources) may promote species coexistence and thus represent complex intertwined ecological processes. Moreover, we elucidated how higher climatic fluctuations could promote longer survivability of brood and adult bees in an infested colony. Comparing the model with and without seasonality illustrated that incorporating seasonality is viable to simulate realistic honeybee population dynamics. Seasonality was shown to play a crucial role in the honeybee population cycles and it represents an important component in mathematical models describing the interaction of honeybee and its parasitic \emph{Varroa} mite.  It will be interesting to study similar dynamics when honeybee population is prone to use a defensive mechanism such as grooming behavior. This will be a subject for future study.     


\begin{appendices}
\section{Proofs}\label{proofs}
%

\subsection*{Proof of Theorem \ref{th1:pb}}

\begin{proof}
\begin{enumerate}
\item  We will proceed by first showing the positivity of our system. First, we prove that $B(t)>0, H(t)>0, M(t)>0$ for all $t\in[0, \tau]$. On the contrary, we assume that there exists $t_0\in (0,\tau]$ such that $B(t_0)=0$ and $B(t)>0$ for $t\in (0,t_0)$. Then we have from the first equation of \eqref{BHM1} that
 \begin{eqnarray*}
\frac{dB}{dt} \ge -\alpha_b  \frac{B}{a+B}M-d_b B-e^{-\int_{t-\tau}^t \left[d_b+\frac{\alpha_b M(s)}{a+B(s)}\right] ds}B_0(t-\tau),\ t\in(0, t_0).
\end{eqnarray*}
Integrating from $0$ to $t_0$, we have
\begin{eqnarray}\label{inqu}
B(t_0) e^{\int_{0}^{t_0} \left[d_b+\frac{\alpha_b M(s)}{a+B(s)}\right] ds} -B(0)\ge -\int_0^{t_0}e^{-\int_{t-\tau}^0 \left[d_b+\frac{\alpha_b M(s)}{a+B(s)}\right] ds}B_0(t-\tau)dt.
\end{eqnarray}
Substituting \eqref{initial1} into \eqref{inqu}, we get
\begin{eqnarray*}
\begin{array}{ll}
\int_{-\tau}^{0}B(t_0) e^{-\int_{t}^0 \left[d_b+\frac{\alpha_b M_0(s)}{a+B_0(s)}\right] ds}dt&\le
\int_0^{t_0}e^{-\int_{t-\tau}^0 \left[d_b+\frac{\alpha_b M(s)}{a+B(s)}\right] ds}B_0(t-\tau)dt\\
&= \int_{-\tau}^{t_0-\tau}e^{-\int_{t}^{0} \left[d_b+\frac{\alpha_b M_0(s)}{a+B_0(s)}\right] ds}B_0(t)dt,
\end{array}
\end{eqnarray*}
which is a contradiction since $t_0-\tau<0$ and $B_0(t)>0, t\in[-\tau, 0]$. Therefore, $B(t)>0$ for all $t\in [0,\tau]$.

If there exists $t_0\in (0,\tau]$ such that $H(t_0)=0$ and $H(t)>0$ for $t\in (0,t_0)$. Then $H'(t_0)\leq 0$. By the second equation of  \eqref{BHM1}, we get a contradiction that
\begin{equation*}
0\geq H'(t_0)=e^{-\int_{t_0-\tau}^{t_0} \left[d_b+\frac{\alpha_b M(s)}{a+B(s)}\right] ds}B_0(t_0-\tau)>0
\end{equation*}
since $B_0(t)>0, t\in[-\tau, 0]$. Therefore, $H(t)>0$ for all $t\in[0, \tau]$. Furthermore, the third equation of  \eqref{BHM1} implies that
\begin{equation*}
M(t)=M(0)e^{-\int_{0}^{t} \left[d_m-\frac{c\alpha_b B(s)}{a+B(s)}\right] ds}>0,\ t\in [0,\tau].
\end{equation*}

Now, we show by induction that both $B(t), H(t)$ and $M(t)$ are positive on $n\tau\leq t\le (n+1)\tau, n=0, 1, \cdots$. We have proved that it is valid for $n=0$. We only show that it is also valid for the case $n=1$. For $n\ge 2$, it can be dealt with similarly. On the contrary, we assume that there exists $t_0\in (\tau, 2\tau]$ such that $B(t_0)=0$ and $B(t)>0$ for $t\in(0,t_0)$. Then by the first equation of  \eqref{BHM2}, we have
 \begin{eqnarray*}
\frac{dB}{dt} \ge -\alpha_b  \frac{B}{a+B}M-d_b B-e^{-\int_{t-\tau}^t \left[d_b+\frac{\alpha_b M(s)}{a+B(s)}\right] ds}\frac{rH^2(t-\tau)}{K+ H^2(t-\tau)},\  t\in(0, t_0).
\end{eqnarray*}
It follows that
\begin{eqnarray*}
\frac{d}{dt}\left(B(t)e^{\int_{\tau}^t \left[d_b+\frac{\alpha_b M(s)}{a+B(s)}\right] ds}\right)\ge -e^{-\int_{t-\tau}^\tau \left[d_b+\frac{\alpha_b M(s)}{a+B(s)}\right] ds}\frac{rH^2(t-\tau)}{K+ H^2(t-\tau)}.
\end{eqnarray*}
Integrating from $\tau$ to $t_0$, we get
\begin{eqnarray}\label{inqu2}
B(t_0) e^{\int_{\tau}^{t_0} \left[d_b+\frac{\alpha_b M(s)}{a+B(s)}\right] ds} -B(\tau)\ge -\int_{\tau}^{t_0}e^{-\int_{t-\tau}^\tau \left[d_b+\frac{\alpha_b M(s)}{a+B(s)}\right] ds}\frac{rH^2(t-\tau)}{K+ H^2(t-\tau)}dt.
\end{eqnarray}
From the first equation of  \eqref{BHM1} and \eqref{initial1}, we have
\begin{eqnarray}\label{inqu3}
B(\tau) = \int_{0}^{\tau}e^{-\int_{s}^\tau \left[d_b+\frac{\alpha_b M(s)}{a+B(s)}\right] ds}\frac{rH^2(t)}{K+ H^2(t)}dt.
\end{eqnarray}
Substituting \eqref{inqu3} into \eqref{inqu2}, we obtain
\begin{eqnarray*}
\begin{array}{ll}
\int_{0}^{\tau}e^{-\int_{s}^\tau \left[d_b+\frac{\alpha_b M(s)}{a+B(s)}\right] ds}\frac{rH^2(t)}{K+ H^2(t)}dt&<\int_{\tau}^{t_0}e^{-\int_{t-\tau}^\tau \left[d_b+\frac{\alpha_b M(s)}{a+B(s)}\right] ds}\frac{rH^2(t-\tau)}{K+ H^2(t-\tau)}dt\\
&=\int_{0}^{t_0-\tau} e^{-\int_{s}^\tau \left[d_b+\frac{\alpha_b M(s)}{a+B(s)}\right] ds}\frac{rH^2(s)}{K+ H^2(s)}ds,
\end{array}
\end{eqnarray*}
which is contradiction since $t_0-\tau<\tau$ and $H(t)>0, t\in[0, \tau]$. Therefore, $B(t)>0$ for all $t\in [0,2\tau]$.

Similar to the arguments for case $t\in(0, \tau]$, it is easy to verify that $H(t)$ and $M(t)$ are positive on $\tau\leq t\le 2\tau$. Furthermore, we can get by induction that $B(t)>0, H(t)>0, M(t)>0$ for all $t>0$.

\item We now proceed with the boundedness of our system in below.
Define $W= cB+cH+M$, then we have
$$\begin{array}{lcl}
\frac{dW}{dt}&=& c\frac{dB}{dt}+c\frac{dH}{dt}+\frac{dM}{dt}\\
&=& \frac{crH^2}{K+H^2}-\frac{c\alpha_hHM}{a+H}-cd_bB-cd_hH-d_mM\\
&\leq& \frac{crH^2}{K+H^2}-cd_bB-cd_hH-d_mM\\
&\leq& cr - \min\{d_b,d_h,d_m\}(cB+cH+M)=cr-\min\{d_b,d_h,d_m\} W.
\end{array}$$
Therefore, we have
$$\limsup_{t\rightarrow\infty} W(t)=\limsup_{t\rightarrow\infty}(cB(t)+cH(t)+M(t))\leq \frac{cr}{\min\{d_b,d_h,d_m\}}.$$
\end{enumerate}
\end{proof}


\subsection*{Proof of Theorem \ref{th2:bq}}

\begin{proof}
As $M=0$, model (\ref{BHM2}) reduces to the model of Chen et al., then the existence of the boundary equilibria can be obtained directly by Proportion 3.1 in their paper \cite{chen2020model}.
We proceed with the stability of the boundary equilibria $E_{000},~ E_{B^{*}_1H^{*}_10}, \mbox{  and  }E_{B^{*}_2H^{*}_20}$ by linearizing our system. First, we note that $E_{B^{*}_1H^{*}_10}$ is unstable since $E_{B_1^*,H_1^*}$ is unstable in the model of Chen et al. by Theorem 3.3 \cite{chen2020model}. So, we only consider the stability of $E_{000}$ and $E_{B^{*}_2H^{*}_20}$.

To facilitate our analysis, we introduce the variable $P(t) = e^{-\int_{t-\tau}^{t}\left(d_b+\frac{\alpha_bM(s)}{a+B(s)}\right)ds}$ and Model \eqref{BHM2} becomes:

{\footnotesize
\begin{equation}
\begin{aligned}
\frac{dB}{dt} &= \frac{rH^2}{K+ H^2}- \alpha_b  \frac{B}{a+B} M - d_b B
- \frac{rPH(t-\tau)^2}{K+ H(t-\tau)^2} \\
\frac{dH}{dt} &=\frac{rPH(t-\tau)^2}{K+ H(t-\tau)^2}-\alpha_h \frac{H}{a+H} M - d_h H\\
\frac{dM}{dt} &=c\alpha_b \frac{B}{a+B}M- d_m M\\
\frac{dP}{dt} &= \frac{\alpha_bPM(t-\tau)}{a+B(t-\tau)}-\frac{\alpha_bPM}{a+B}
 \label{BHM2_Modify}
\end{aligned}
\end{equation}.
}

 Let $(B^*, H^*, M^*, P^*)$ be the equilibrium of the system \eqref{BHM2_Modify} where $P^* = e^{-\left(d_b+\frac{\alpha_bM^*}{a+B^*}\right)\tau}$.  The linearization matrix of Model \eqref{BHM2_Modify} at the equilibrium $(B^*, H^*, M^*, P^*)$ can be represented as follows:
{\footnotesize
\begin{equation}
\begin{aligned}
D  \left(  \left[ \begin{array}{c}
                 \dot{B}(t)\\
                 \dot{H}(t)\\
                 \dot{M}(t) \\
                 \dot{P}(t)
                \end{array}
            \right] \right)
            \Bigg\vert_{(B^*,H^*,M^*,P^*)}
                       &= \left[ \begin{array}{cccc}
               \frac{-a\alpha_bM^*}{(a+B^*)^2}-d_b & \frac{2rKH^*}{(K+(H^*)^2)^2}&-\frac{\alpha_b B^*}{a+B^*} & -\frac{r(H^*)^2}{K+(H^*)^2}\\
               0 &  \frac{-a\alpha_hM^*}{(a+H^*)^2}-d_h &-\frac{\alpha_h H^*}{a+H^*} & \frac{r(H^*)^2}{K+(H^*)^2}\\
               \frac{ac\alpha_bM^*}{(a+B^*)^2} &0& \frac{c\alpha_b B^*}{a+B^*}-d_m & 0\\
               \frac{\alpha_bP^*M^*}{(a+B^*)^2} & 0 & -\frac{\alpha_bP^*}{a+B^*} & 0
              \end{array} \right]
	      \left[ \begin{array}{c}
	              B(t)\\
	              H(t)\\
	              M(t)\\
	              P(t)
	             \end{array} \right]\\
	   & +  \left[ \begin{array}{cccc}
		  0 & - \frac{2rKP^*H^*}{(K+(H^*)^2)^2}&0 & 0\\
		    0& \frac{2rKP^*H^*}{(K+(H^*)^2)^2}&0 & 0\\
		    0 & 0 & 0 & 0\\
		    - \frac{\alpha_bP^*M^*}{(a+B^*)^2} & 0 & \frac{\alpha_bP^*}{a+B^*} & 0
		  \end{array} \right]
		  \left[ \begin{array}{c}
		     B(t-\tau)\\
	             H(t-\tau)\\
	              M(t-\tau)\\
	              P(t-\tau)
			\end{array} \right].\\
&:=U\Phi(t)+V\Phi(t-\tau).
			\label{JE}
	   \end{aligned}
	   \end{equation}
}
 The characteristic equation of \eqref{JE} is given by
\begin{eqnarray*}
C(\lambda)& = \bigg| \lambda I - U - e^{-\lambda \tau} V    \bigg|=0.
\end{eqnarray*}
Notice $P^*=e^{-d_b\tau}$ when $M^*=0$. By a direct computation, we get
\begin{eqnarray*}
C(\lambda)=\lambda(\lambda+d_b)\left(\lambda-\frac{c\alpha_b B^*}{a+B^*}+d_m\right)\left(\lambda+d_h-\frac{2rKH^*}{(K+(H^*)^2)^2}e^{-(\lambda+d_b)\tau}\right).
\end{eqnarray*}
which always has eigenvalues $\lambda_0=0$, which is in the direction $P$, $\lambda_1=-d_b<0$ and $\lambda_2=\frac{c\alpha_b B^*}{a+B^*}-d_m$. The other eigenvalues satisfy the following algebraic equation
\begin{eqnarray}\label{Plambda}
L(\lambda):=\lambda+d_h-\frac{2rKH^*}{(K+(H^*)^2)^2}e^{-(\lambda+d_b)\tau}=0.
\end{eqnarray}
 Therefore, the stability of $E_{000}$ and $E_{B^{*}_2H^{*}_20}$ is determined by the signs of $\lambda_2$ and of the roots of $L(\lambda)=0$.

At extinction equilibrium $E_{000}=(0,0,0)$, $\lambda_2=-d_m$ and $L(\lambda) = \lambda+d_h$,
therefore, $E_{0}=(0,0)$ is locally asymptotically stable for all $\tau>0$.

At $E_{B^{*}_2H^{*}_20}$, form the proof of Theorem 3.3 in the paper of Chen et al. \cite{chen2020model} (by Theorem 4.7 of Smith \cite{smith2011introduction}), we know that all roots of $L(\lambda)$ have negative real parts. Thus, we can conclude that if $d_m>\frac{c\alpha_b B^*}{a+B^*}$ then $E_{B^{*}_2H^{*}_20}$ is locally asymptotically stable, while unstable if $d_m<\frac{c\alpha_b B^*}{a+B^*}$.

\end{proof}

\subsection*{Proof of Theorem \ref{pr1:IntEq}}

\begin{proof}
Note that from Equation \eqref{Equilibrium1c}, $B^{*} = \frac{a}{\frac{c\alpha_b}{d_m}-1}$, and from Equation \eqref{Equilibrium1a} and \eqref{Equilibrium1b}, we obtain
\begin{align*}
\frac{rH^*{^2}}{K+H^*{^2}}e^{-\left(d_b+\frac{\alpha_bM^*}{a+B^*}\right)\tau} &= \frac{rH^*{^2}}{K+H^*{^2}} - \frac{\alpha_bB^*M^*}{a+B^*}-d_bB^* \\[.2cm]
 \frac{rH^*{^2}}{K+H^*{^2}}e^{-\left(d_b+\frac{\alpha_bM^*}{a+B^*}\right)\tau} &= \frac{\alpha_hH^*M^*}{a+H^*}+d_hH^*
\end{align*}
which gives
\begin{align}
 \frac{rH^*{^2}}{K+H^*{^2}} - \frac{\alpha_bB^*M^*}{a+B^*}-d_bB^* =  \frac{\alpha_hH^*M^*}{a+H^*}+d_hH^*\label{EqM1}.
\end{align}
Then
\begin{eqnarray}\label{M1}
M^*=\frac{\frac{rH^*{^2}}{K+H^*{^2}}-d_bB^*-d_hH^*}{\frac{\alpha_hH^*}{a+H^*}+\frac{\alpha_bB^*}{a+B^*}},
\end{eqnarray}

From Equation \eqref{Equilibrium1a} we have the following:
$$\frac{rH^*{^2}}{K+H^*{^2}}\left(1-e^{-\left(d_b+\frac{\alpha_bM^*}{a+B^*}\right)\tau}\right) =B^*\left( d_b+\frac{\alpha_bM^*}{a+B^*}\right)\quad \Leftrightarrow \quad \frac{rH^*{^2}}{K+H^*{^2}}=\frac{B^*\left(\frac{\alpha_b M^*}{a+B^*}+d_b\right)}{1-e^{-\left(\frac{\alpha_b M^*}{a+B^*}+d_b\right)\tau}}$$
Let
\begin{eqnarray*}
f_1(H^*)=\frac{rH^*{^2}}{K+H^*{^2}},\quad \mbox{and}\quad f_2(H^*)=\frac{B^*\left(\frac{\alpha_b M^*}{a+B^*}+d_b\right)}{1-e^{-\left(\frac{\alpha_b M^*}{a+B^*}+d_b\right)\tau}}.
\end{eqnarray*}
Thus, $(B^*, H^*, M^*)$ is a interior equilibria if and only if $B^*=\frac{a}{\frac{c \alpha_b}{d_m}-1}>0$, i.e. $\frac{c \alpha_b}{d_m}>1$, $H^*>0$ is a positive root of $f_1(H^*)=f_2(H^*)$ and $M^*$ defined in \eqref{M1} is positive.

In what follows, we assume $\frac{c \alpha_b}{d_m}>1$.

Regard $M^*$ defined in \eqref{M1} as a function on $H^*$, denoted as $M^*(H^*)$, we rewrite it as $M^*(H^*)=\frac{Q(H^*)}{P(H^*)}$, where
\begin{eqnarray*}
\begin{array}{ll}
Q(H^*)=-d_hH^*{^3}+(r-B^*d_b)H^{*^2}-d_hKH^*-d_bKB^*,\\
P(H^*)=(K+H^*{^2})\left(\frac{\alpha_hH^*}{a+H^*}+\frac{\alpha_bB^*}{a+B^*}\right).
\end{array}
\end{eqnarray*}
Clearly, $Q(0)=-d_bKB^*<0, P(0)=K\frac{\alpha_bB^*}{a+B^*}$, and $M(0)=-\frac{d_b}{\alpha_b}(a+B^*)$. Also, for all $H^*\ge 0$, $P(H^*)>0$. Thus, the sign of $M(H^*)$ is determined by $Q(H^*)$.

(i)\quad If $r-B^*d_b\le 0$, then for all $H^*\ge 0$, $Q(H^*)<0$. In this case,  Model \eqref{BHM2} has no interior equilibria.

(ii)\quad Let $r-B^*d_b> 0$. By $Q'(H^*)=-3d_hH^*{^2}+2(r-B^*d_b)H^{*}-d_hK$, we have two cases:\\
(1)\quad If $\Delta=4(r-B^*d_b)^2-12Kd_h^2\leq 0$, then for all $H^*\in\mathbb{R}$, $Q'(H^*)\le 0$. Notice $Q(0)<0$, we know that for all $H^*\ge 0$, $Q(H^*)<0$. This implies that Model \eqref{BHM2} has no interior equilibria for this case. \\
(2)\quad If $\Delta=4(r-B^*d_b)^2-12Kd_h^2> 0$, i.e., $d_h<\frac{r-B^*d_b}{\sqrt{3K}}$, then $Q'(H^*)$ has two positive roots $H_1^c<H_2^c$:
\begin{eqnarray*}
H_{1,2}^c=\frac{(r-d_b B^*)\pm\sqrt{(r-d_b B^*)^2-3Kd_h^2}}{3d_h},
\end{eqnarray*}
in which $H_1^c$ is the minimum point and $H_2^c$ is the maximum point of $Q(H^*)$.
\begin{itemize}
  \item  If $Q(H_2^c)\le 0$, then for all $H^*\ge 0$, $Q(H^*)\leq 0$ and Model \eqref{BHM2} has no interior equilibria.
  \item If $Q(H_2^c)> 0$, then $Q(H^*)$ has exact two positive roots, denoted as $H_1^r<H_2^r$, satisfying $Q(H^*)>0$ for $H^*\in (H_1^r, H_2^r)$ and  $Q(H^*)\leq 0$ for $H^*\in [0, H_1^r]\cup[H_2^r, \infty)$.
\end{itemize}
Thus, in order to show the existence of at least one interior equilibria, we only need to find a root of $f_1(H^*)=f_2(H^*)$ in $(H_1^r, H_2^r)$.

Note that $f_1(H^*)$ and $f_2(H^*)$ have the following properties.\\
(a)\quad $f_1(0)=0$, $\lim_{H\rightarrow\infty}f_1(H^*)=r$, and $f_1(H^*)$ is strictly increasing on $[0, \infty)$, which implies $f_1(H_1^r)<f_2(H_2^r)$.\\
(b)\quad
$\lim_{H^*\rightarrow0}f_2(H^*)=\frac{B^*}{\tau}>0$,
$\lim_{H^*\rightarrow\infty} f_2(H)=0$ since $\lim_{H^*\rightarrow\infty}M^*(H^*)= -\infty$.\\
(c)\quad $f_2(H_1^r)=f_2(H_2^r)=\frac{B^*d_b}{1-e^{-\tau d_b}}$ since $M(H_1^r)=M(H_2^r)=0$.\\
(d)\quad By \eqref{M1} and the fact $M(H_1^r)=M(H_2^r)=0$, $f_2(H_1^r)=B^*d_b+d_hH_1^r, f_2(H_2^r)=B^*d_b+d_hH_2^r$, which implies $f_2(H_1^r)>B^*d_b, f_2(H_2^r)>B^*d_b$.

From the properties (a) and (b), we can claim that if $f_1(H_1^r)<f_2(H_1^r)$ and $f_2(H_2^r)<f_1(H_2^r)$, illustrated in Figure \ref{proof_graph}, then Model \eqref{BHM2} has at least one interior equilibrium. Thus, by property (c), if $\tau>0$ satisfies the inequalities
\begin{eqnarray}\label{inequtau}
f_1(H_1^r)<\frac{B^*d_b}{1-e^{-\tau d_b}}<f_1(H_2^r),
\end{eqnarray}
then Model \eqref{BHM2} has at least one interior equilibrium. Noticing property (d) and solving \eqref{inequtau}, we get
\begin{eqnarray*}
\beta_1<\tau<\beta_2,
\end{eqnarray*}
where
\begin{eqnarray*}
\beta_1=\frac{1}{d_b}\ln\left(\frac{f_1(H_2^r)}{f_1(H_2^r)-B^*d_b}\right),\quad \beta_2=\frac{1}{d_b}\ln\left(\frac{f_1(H_1^r)}{f_1(H_1^r)-B^*d_b}\right).
\end{eqnarray*}
Therefore, if $\tau\in(\beta_1, \beta_2)$, then Model \eqref{BHM2} has at least one interior equilibria.

At last, we give a sufficient condition, which is easy to be verified, such that $Q(H_2^c)> 0$. From the property of cubic function $Q(H)$, we know that it has unique point of inflection $H_0=\frac{r-B^*d_b}{3d_h}$, and that if $Q(H_0)>0$ then $Q(H_2^c)> 0$. By a direct computation, we have
\begin{eqnarray*}
Q(H_0)=\frac{2(r-B^*d_b)^3}{27d_h^2}-\frac{K(r-B^*d_b)}{3}-d_bKB^*.
\end{eqnarray*}
Thus, if
\begin{eqnarray*}
d_h<\frac{(r-B^*d_b)\sqrt{2(r-B^*d_b)}}{3\sqrt{K(r-B^*d_b)+3d_bKB^*}}
\end{eqnarray*}
then $Q(H_0)>0$, and hence $Q(H_2^c)> 0$.
\begin{figure}[H]
\centering
\includegraphics[scale=0.8]{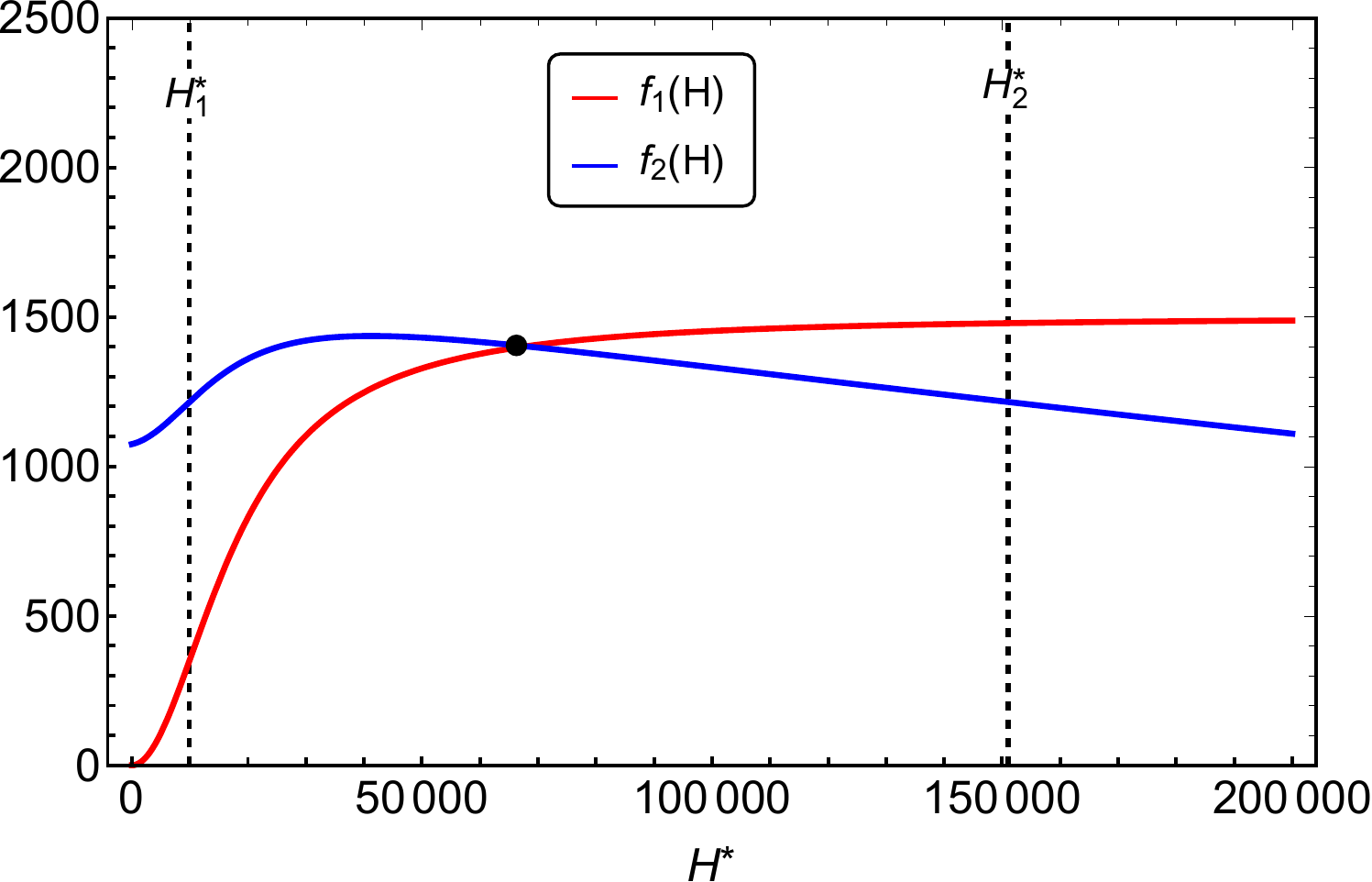}
\caption{Graph showing the existence of a unique interior equilibrium whenever $f_1(H^*)=f_2(H^*)$ occurring at the black dot with $r=1500$, $K=324000000$, $d_b = 0.012$, $d_h=0.008$, $d_m=0.028$, $\alpha_b=0.038$, $\alpha_h=0.022$, $c=1.23$, $a=15100$, and $\tau =21$. Vertical dashed lines, $H^*_1$ and $H^*_2$, are the positive solutions of $Q(H^*)$ (eq. \eqref{Q}).}\label{proof_graph}
\end{figure}
\end{proof}

\subsection*{Proof of Theorem \ref{globalstability-fullsystem}}

\begin{proof}
By the positivity of solutions of Model \eqref{BHM1}-\eqref{BHM2}, the third equation of \eqref{BHM2} implies that
\begin{eqnarray*}
\frac{dM}{dt}<(c\alpha_b-d_m)M(t)<0,
\end{eqnarray*}
since $\frac{c\alpha_b}{d_m}<1$. Thus, $\lim_{t\to\infty}M(t)=0$. Then, the model reduces to the model of Chen et al. \cite{chen2020model}, the global stability of $E_{00}$, we know that $\lim_{t\to\infty}B(t)=\lim_{t\to\infty}H(t)=0$ if  $d_h > \frac{re^{-d_b\tau}}{2\sqrt{K}}$. The proof is complete.
\end{proof}



\section{}\label{AppendixB}
\begin{figure}[H]
\begin{center}
\subfigure
{\includegraphics[height = 55mm, width = 60mm]{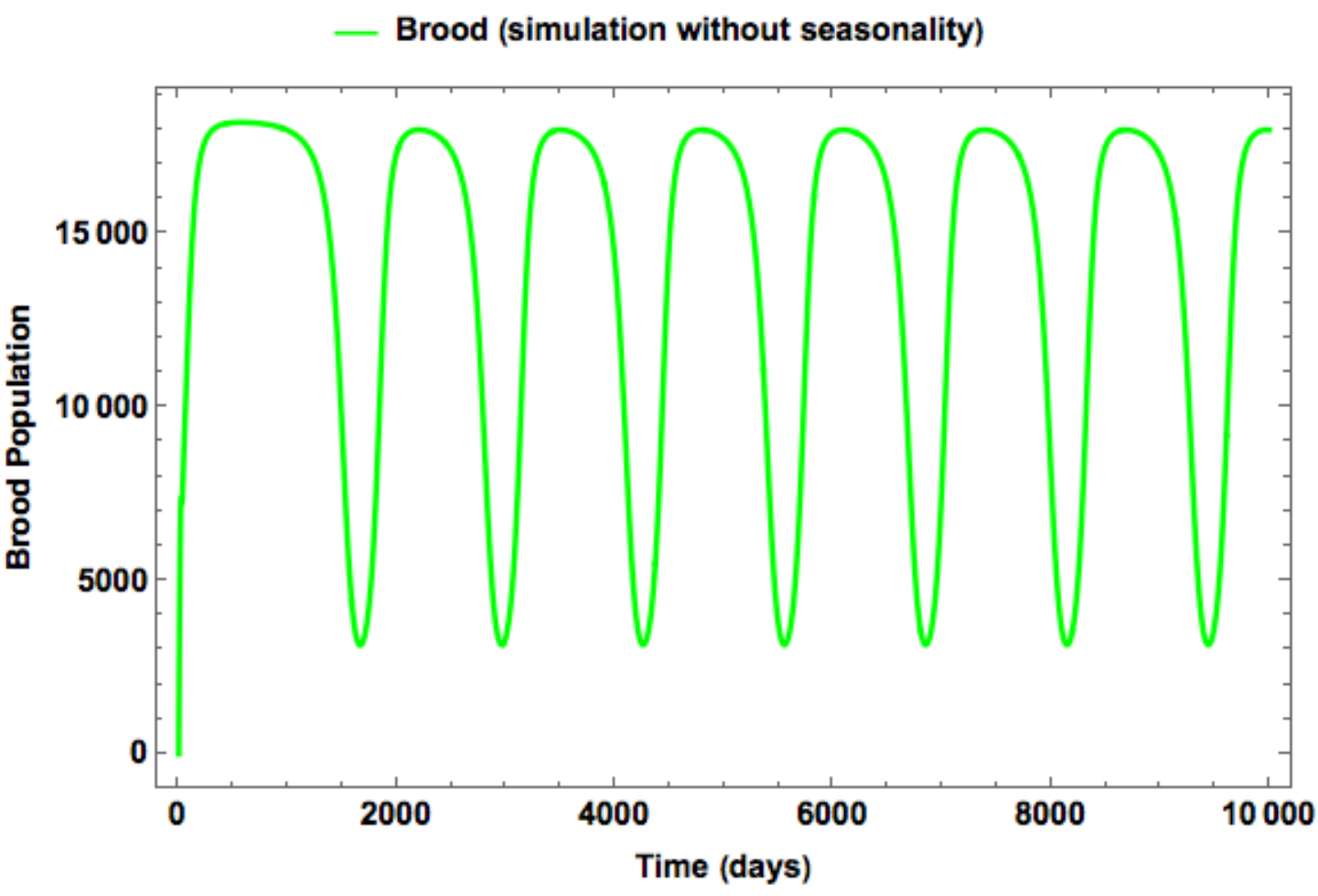}\label{fig_Brood_Oscillation_NoS}}\hspace{5mm}
\subfigure
{\includegraphics[height = 55mm, width = 60mm]{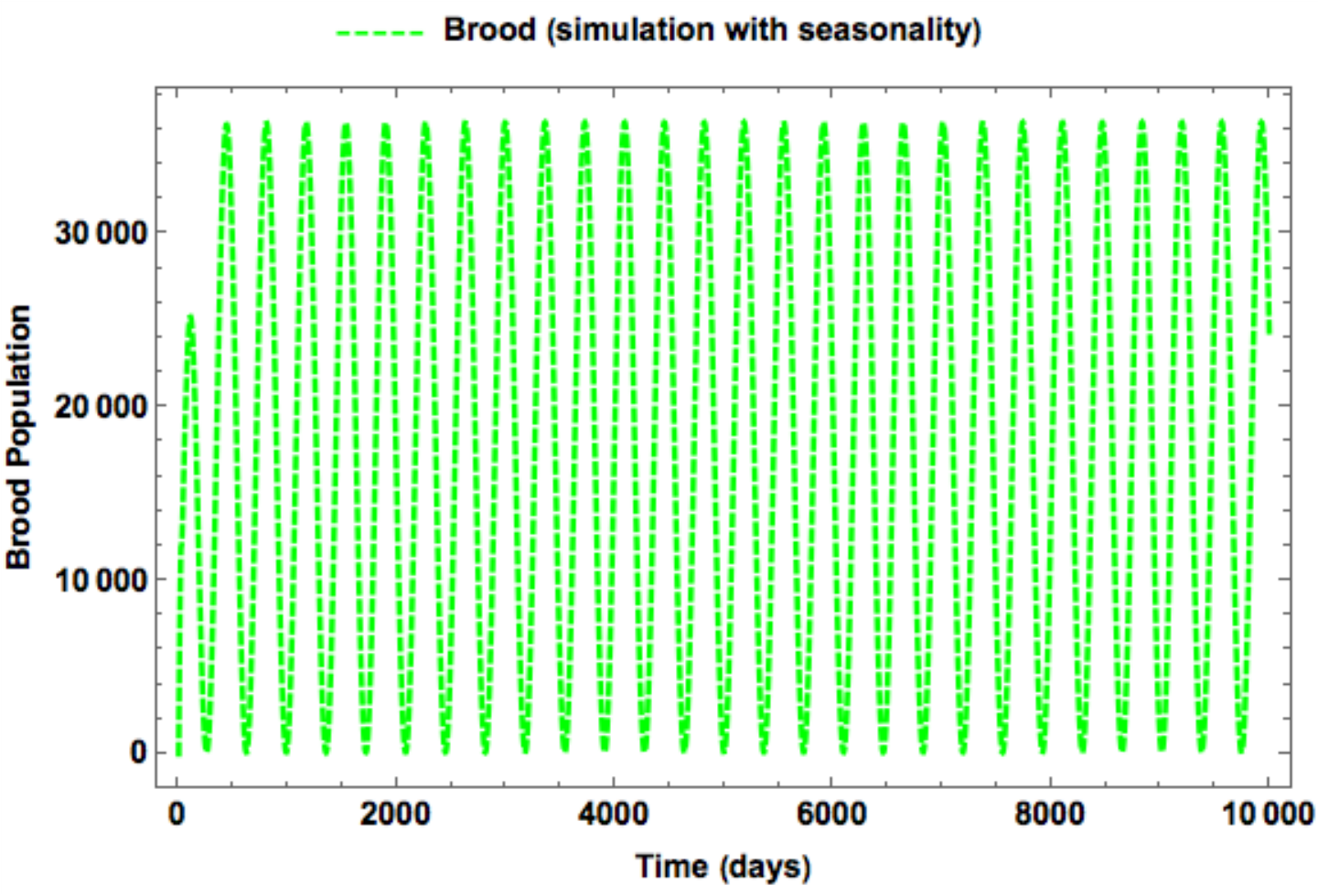}\label{fig_Brood_Oscillation_S}}\\
\subfigure
{\includegraphics[height = 55mm, width = 60mm]{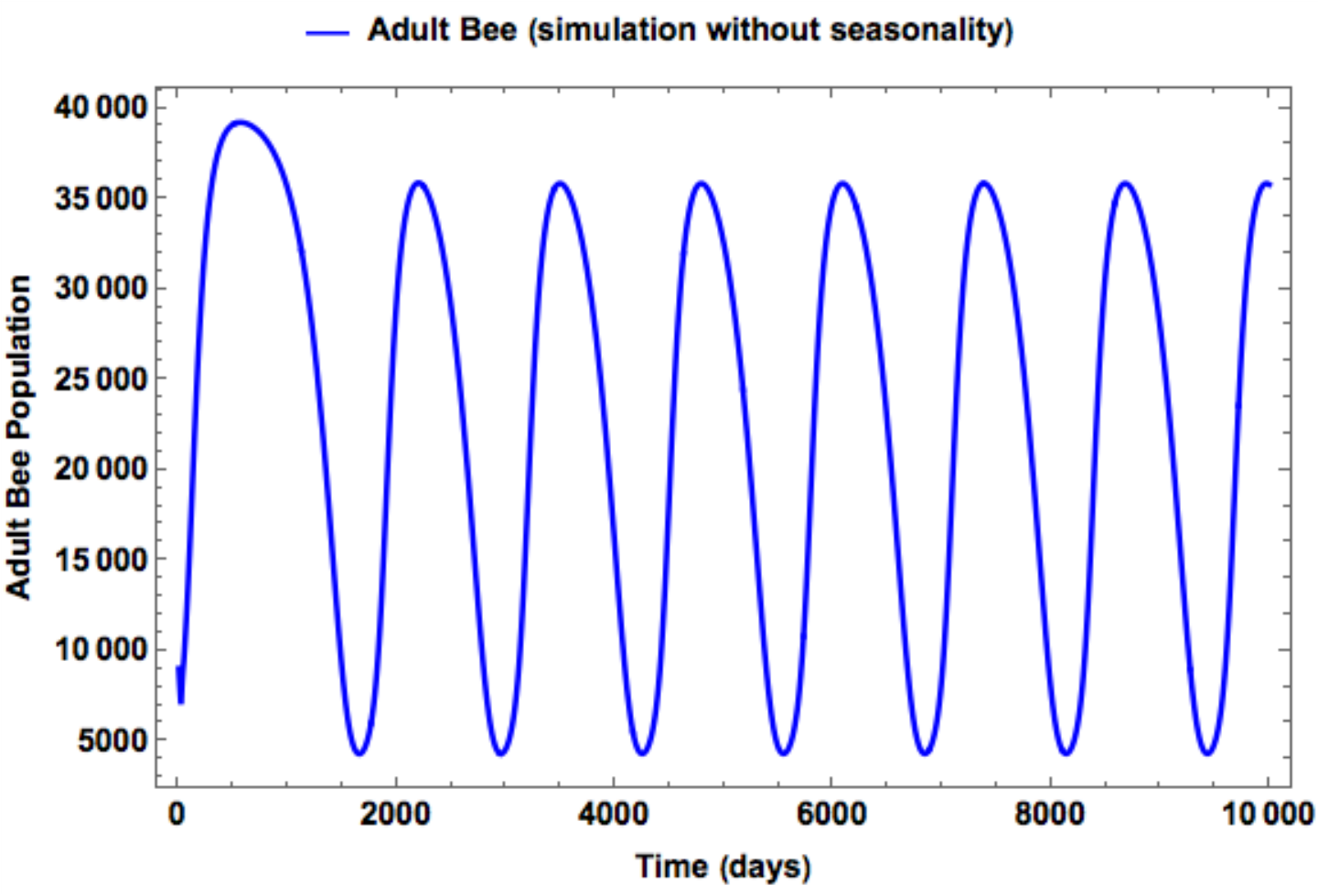}\label{fig_Bee_Oscillation_NoS}}\hspace{5mm}
\subfigure
{\includegraphics[height = 55mm, width = 60mm]{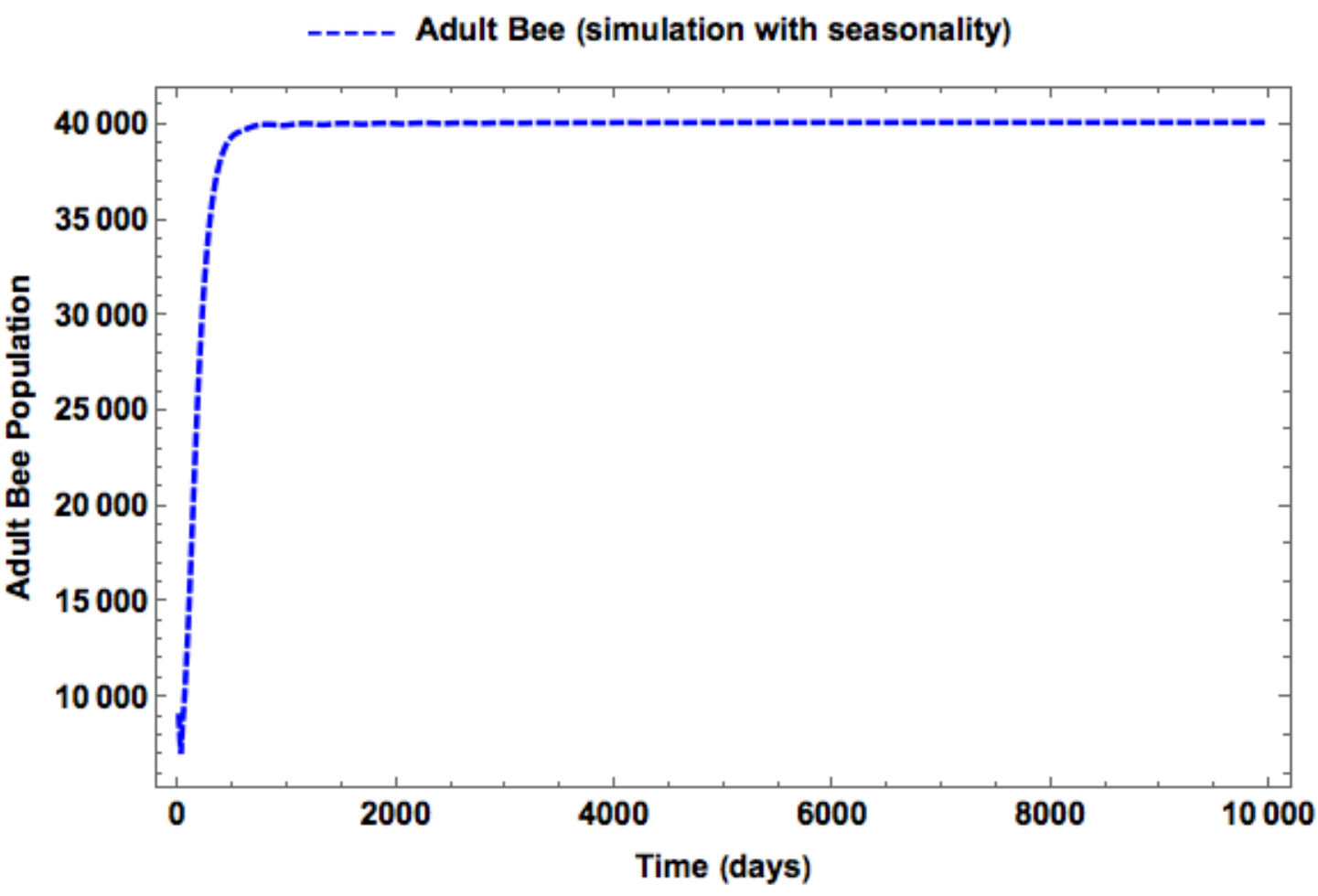}\label{fig_Bee_Oscillation_S}}\\
\subfigure
{\includegraphics[height = 55mm, width = 60mm]{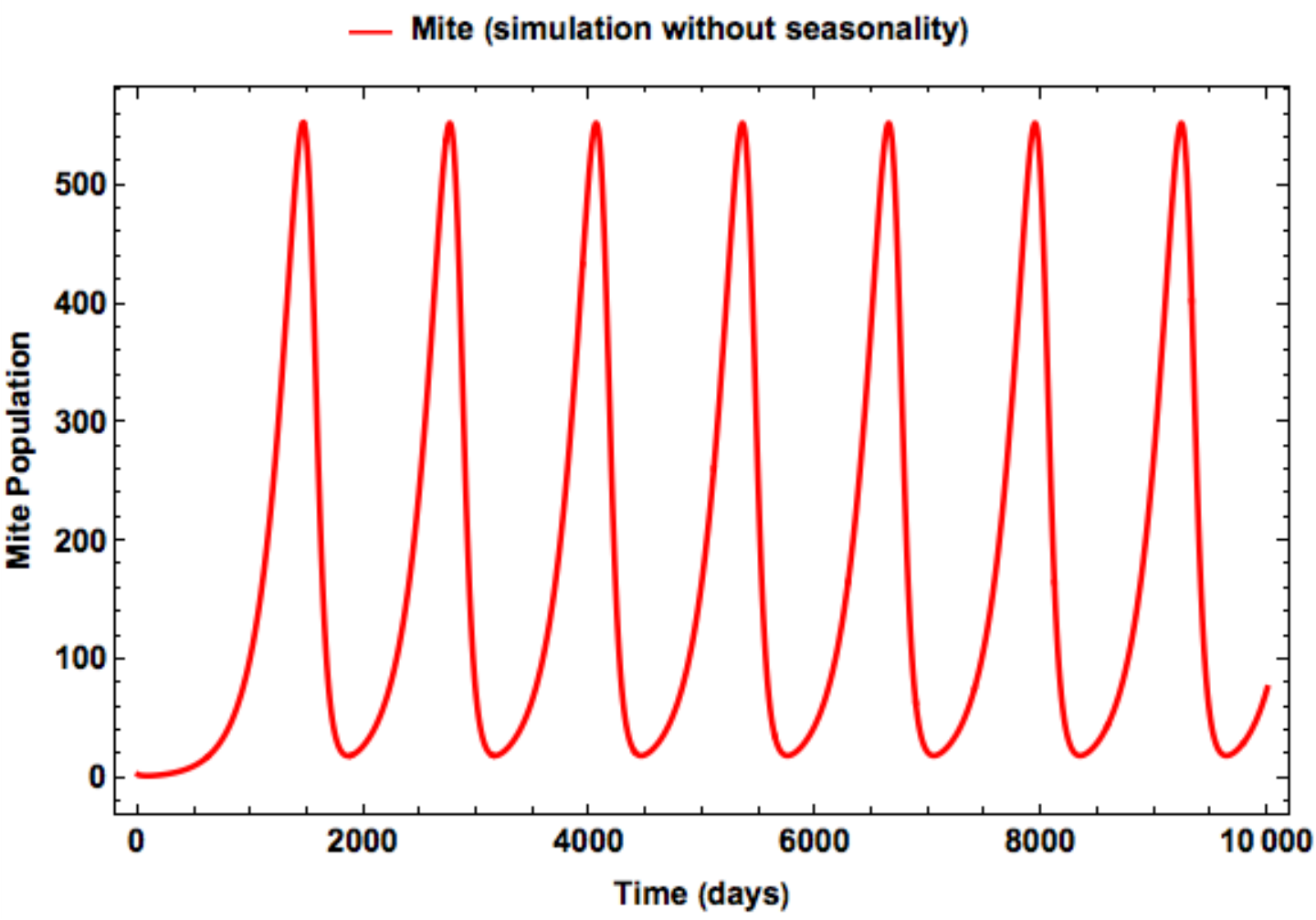}\label{fig_Mite_Oscillation_NoS}}\hspace{5mm}
\subfigure
{\includegraphics[height = 55mm, width = 60mm]{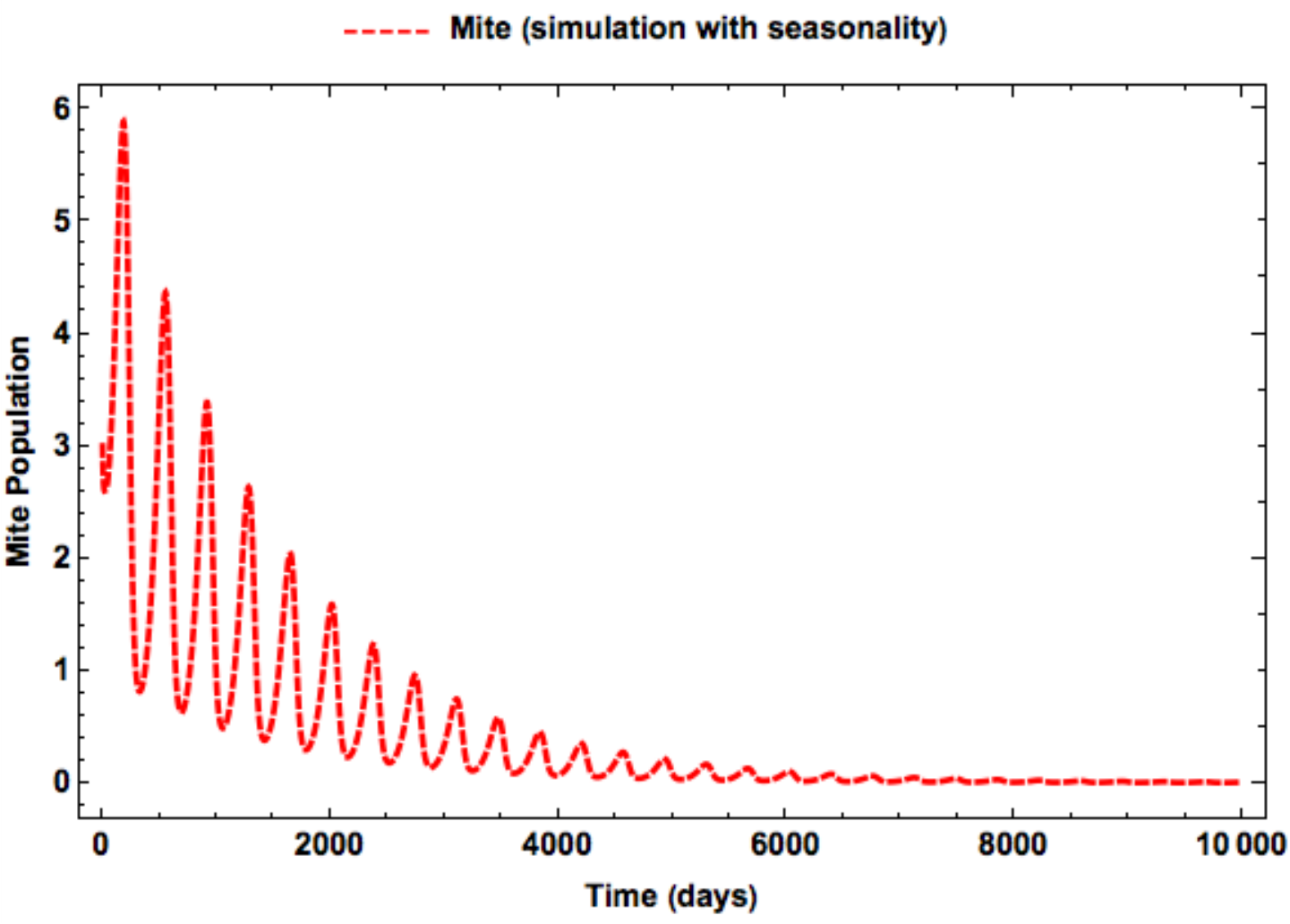}\label{fig_Mite_Oscillation_S}}\\
$\alpha_b=0.024$
\end{center}
\vspace{-10pt}
\caption[Time Series: Brood, Adult Bee, and Mite with $\alpha_b=0.024$.]{{\small  Time series of the brood, adult bee, and mites simulation using $r=1500$, $K=95000000$, $d_b =0.051$, $d_h= 0.0121$, $d_m=0.027$, $\alpha_h=0.8$, $c=1.9$, $a=8050$, $\tau=21$, $\Phi=65$, $B_0(t)=B(0)=0$, $H(0)= 9000$, and $M(0) = 3$ when the queen's eggs laying rate is constant in figures on the left column (i.e. no seasonality) and when the queen's eggs laying rate has seasonality in figures on the right column with $\alpha_b =0.024$.}} 
\label{fig:TimeSerieOscillation}
\end{figure}



\begin{figure}[H]
\begin{center}
\subfigure
{\includegraphics[height = 55mm, width = 60mm]{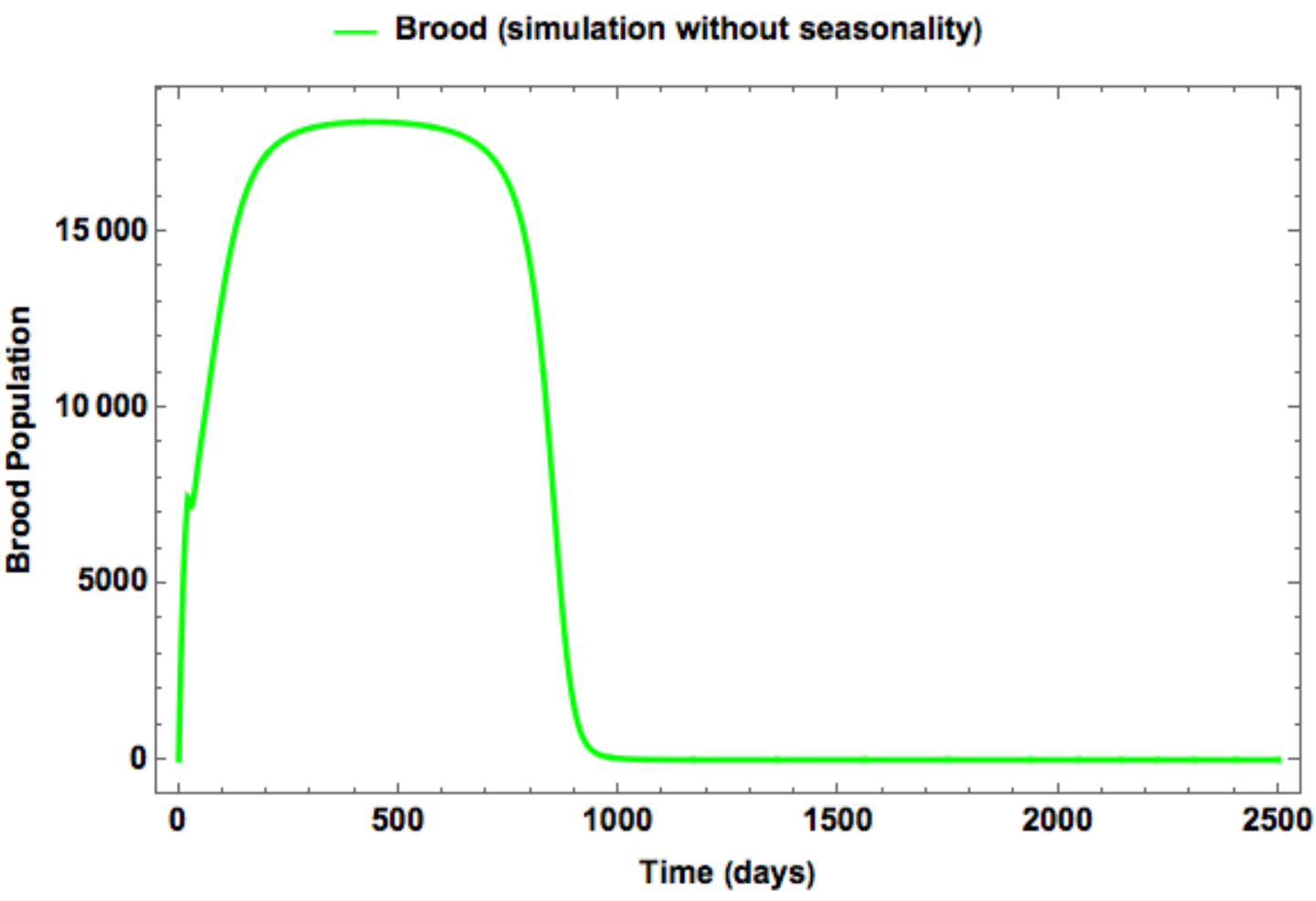}\label{fig_Brood_Extinction_NoS2}}\hspace{5mm}
\subfigure
{\includegraphics[height = 55mm, width = 60mm]{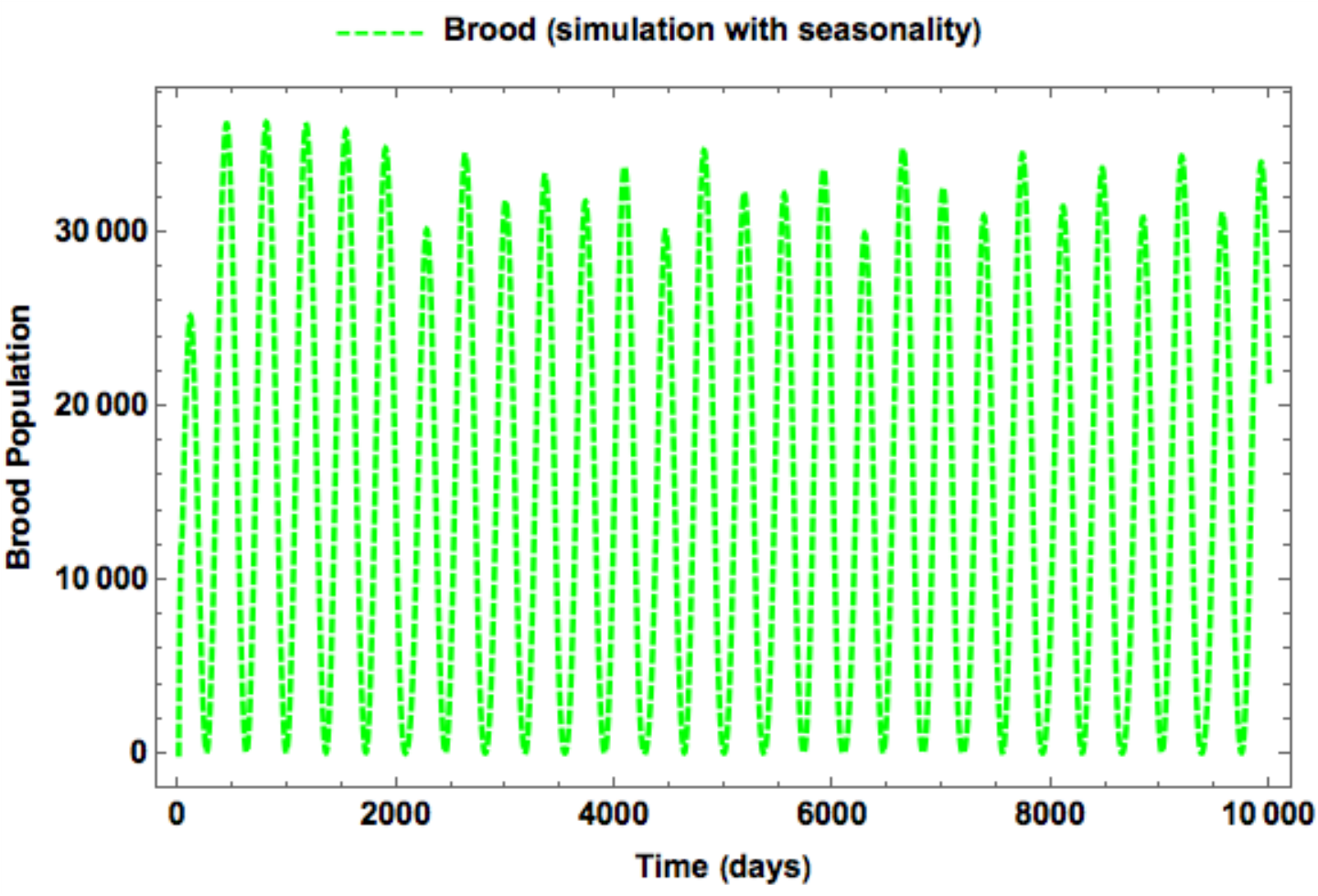}\label{fig_Brood_Extinction_S2}}\\
\subfigure
{\includegraphics[height = 55mm, width = 60mm]{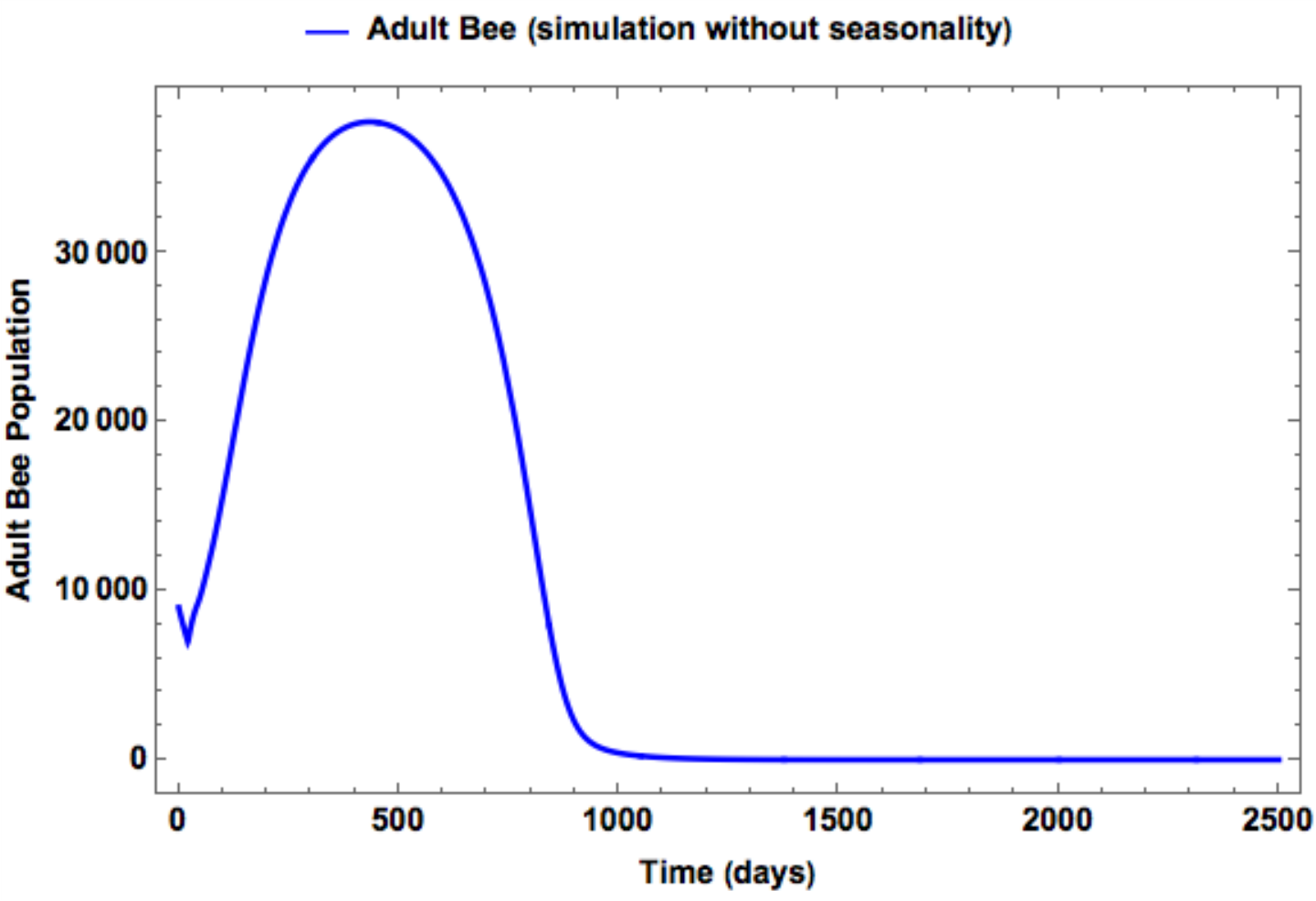}\label{fig_Bee_Extinction_NoS2}}\hspace{5mm}
\subfigure
{\includegraphics[height = 55mm, width = 60mm]{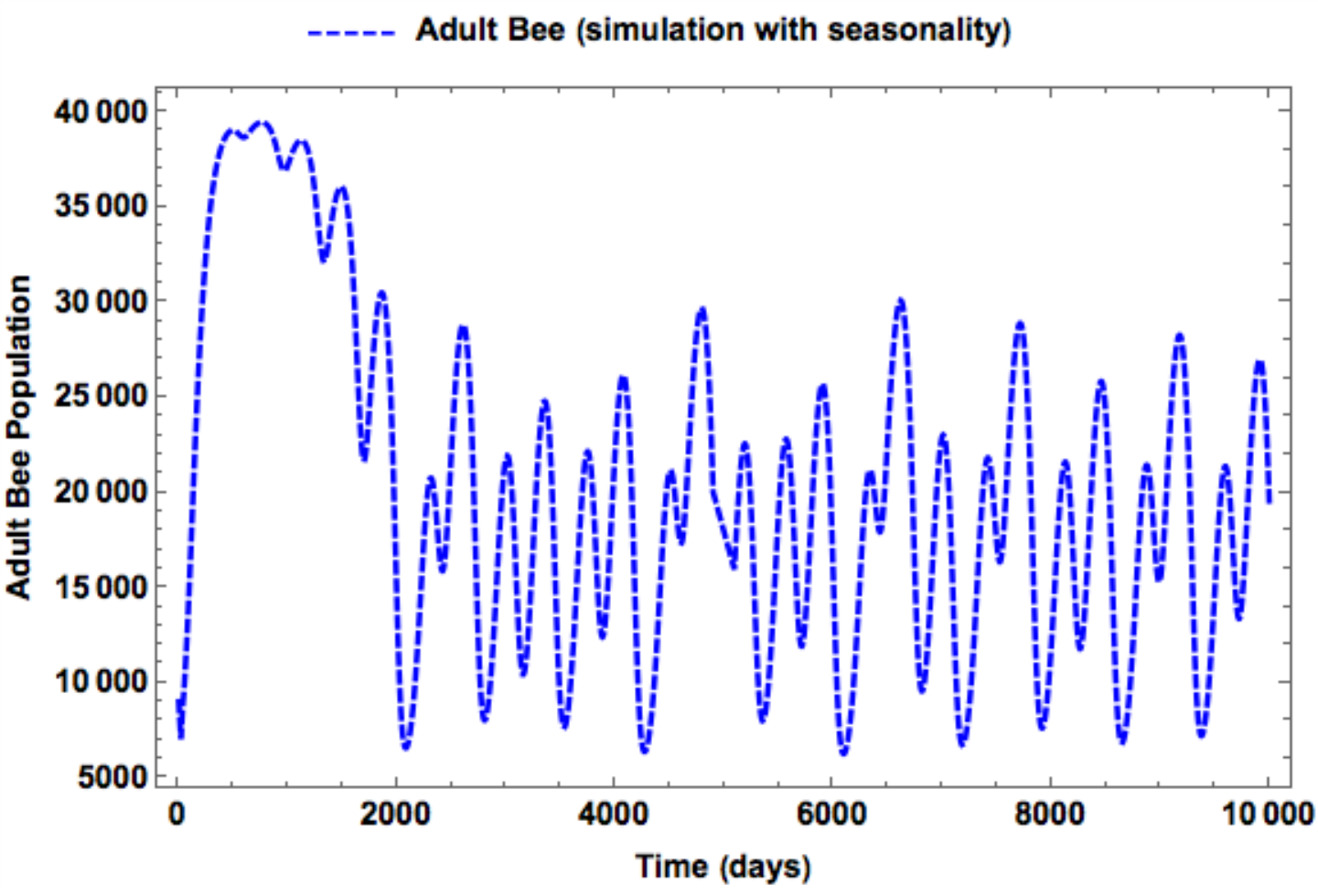}\label{fig_Bee_Extinction_S2}}\\
\subfigure
{\includegraphics[height = 55mm, width = 60mm]{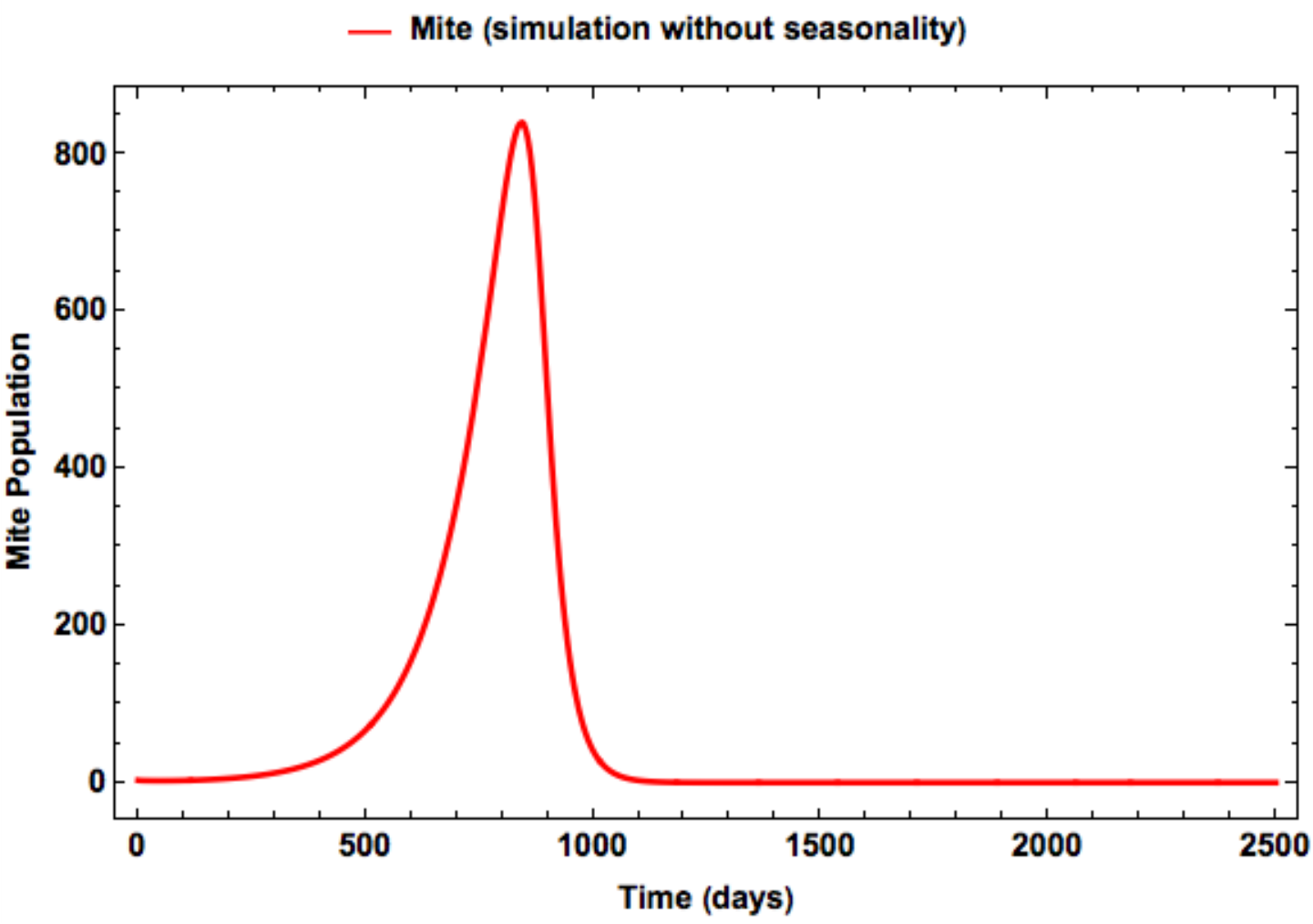}\label{fig_Mite_Extinction_NoS2}}\hspace{5mm}
\subfigure
{\includegraphics[height = 55mm, width = 60mm]{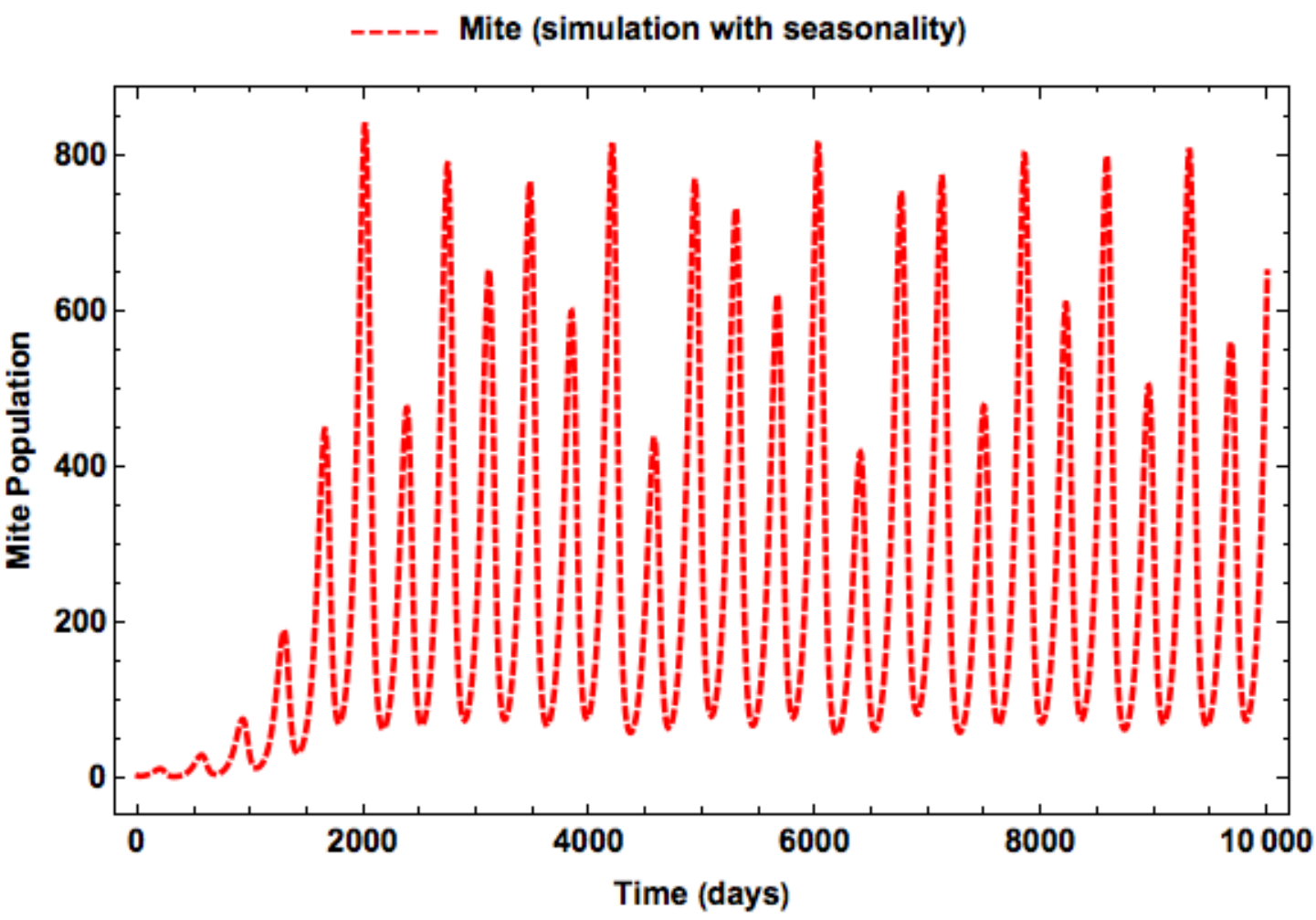}\label{fig_Mite_Extinction_S2}}\\
$\alpha_b=0.027$
\end{center}
\vspace{-10pt}
\caption[Time Series: Brood, Adult Bee, and Mite Simulation with $\alpha_b=0.027$.]{{\small  Time series of the brood, adult bee, and mites simulation using $r=1500$, $K=95000000$, $d_b =0.051$, $d_h= 0.0121$, $d_m=0.027$, $\alpha_h=0.8$, $c=1.9$, $a=8050$, $\tau=21$, $\Phi=65$, $B_0(t)=B(0)=0$, $H(0)= 9000$, and $M(0) = 3$ when the queen's eggs laying rate is constant in figures on the left column  (i.e. no seasonality) and when the queen's eggs laying rate has seasonality in figures on the right column with with $\alpha_b =0.027$.}}
\label{fig:TimeSerieNonperiodic}
\end{figure}



\begin{figure}[H]
\begin{center}
\subfigure
{\includegraphics[height = 55mm, width = 60mm]{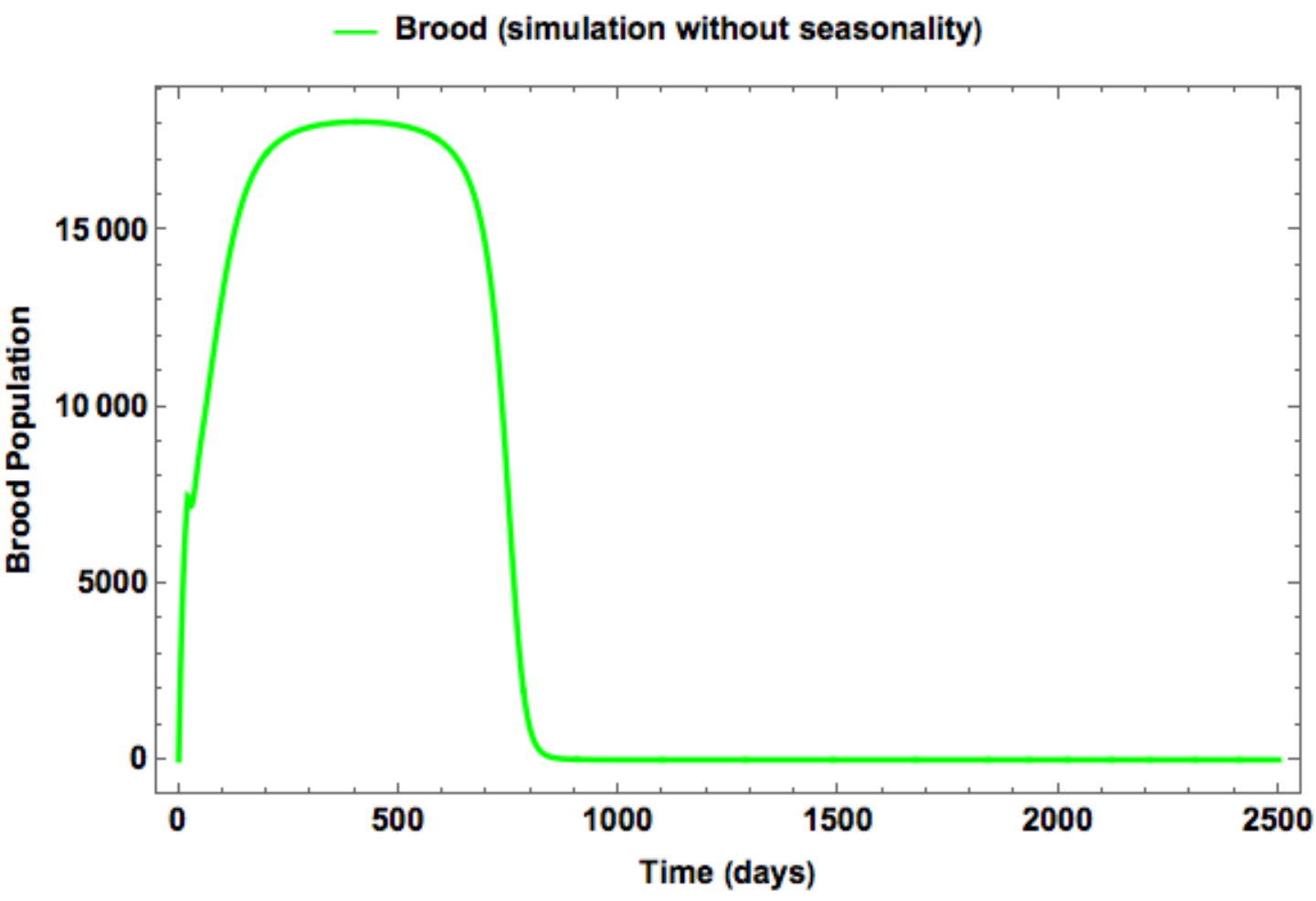}\label{fig_Brood_Extinction_NoS}}\hspace{5mm}
\subfigure
{\includegraphics[height = 55mm, width = 60mm]{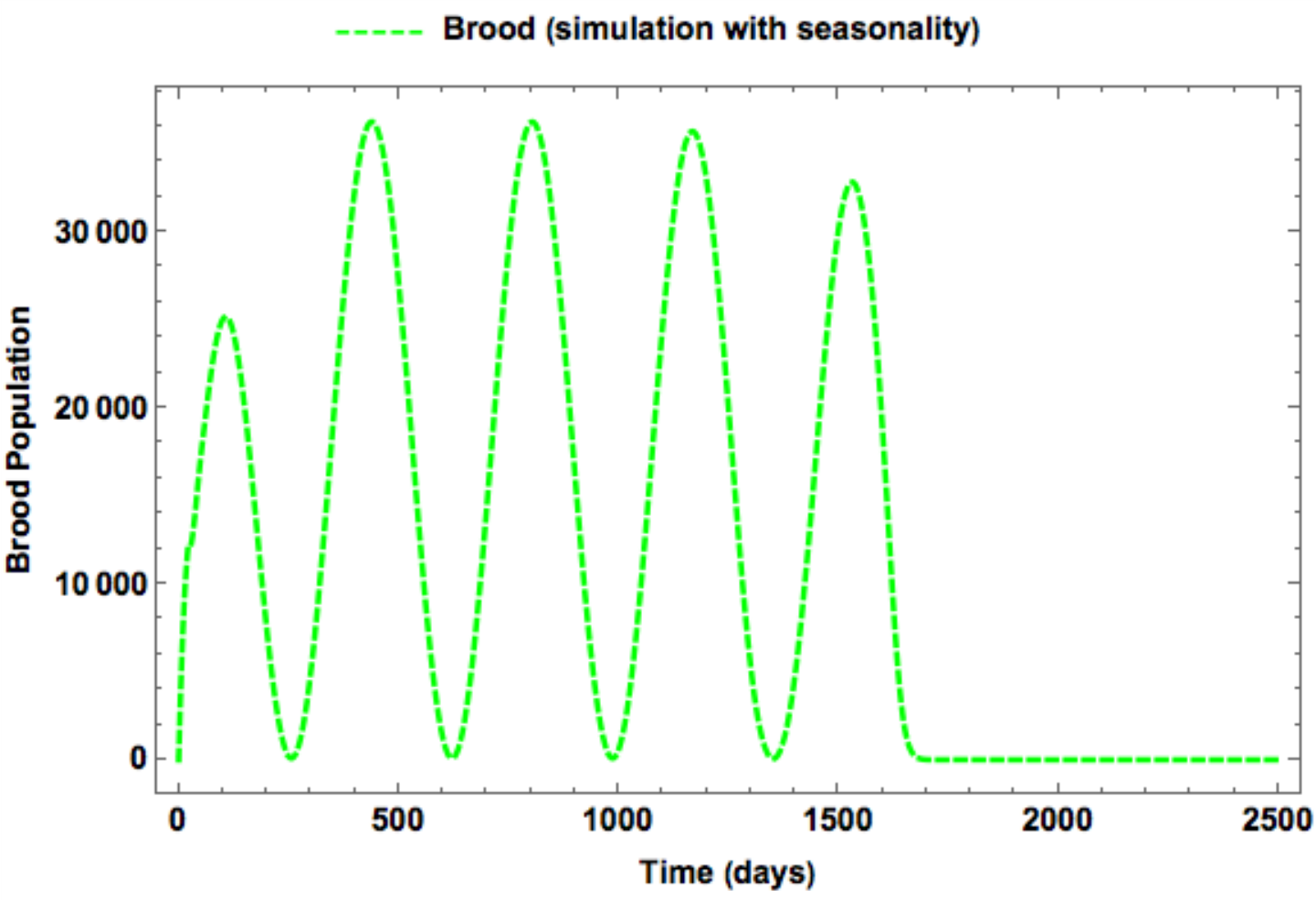}\label{fig_Brood_Extinction_S}}\\
\subfigure
{\includegraphics[height = 55mm, width = 60mm]{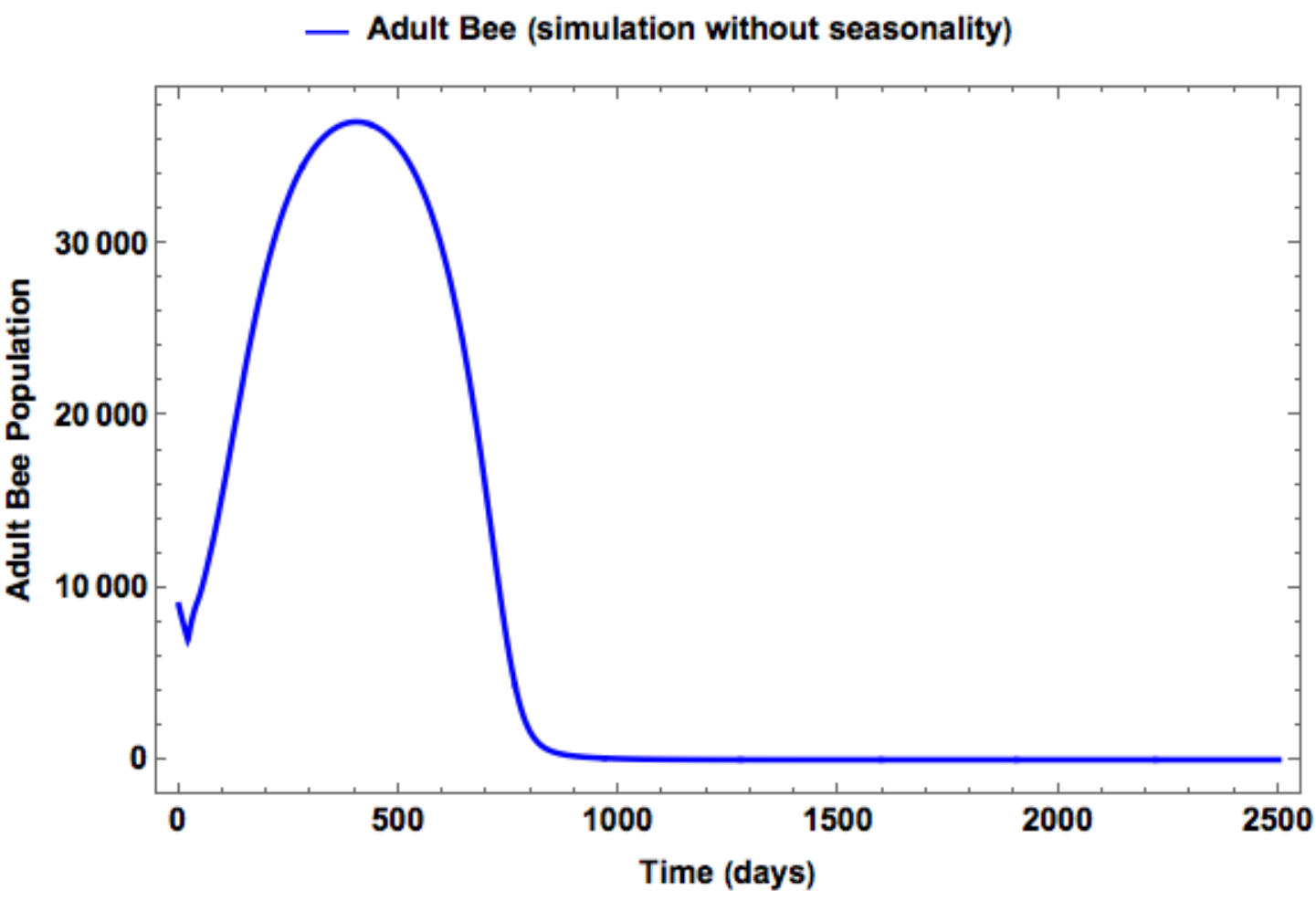}\label{fig_Bee_Extinction_NoS}}\hspace{5mm}
\subfigure
{\includegraphics[height = 55mm, width = 60mm]{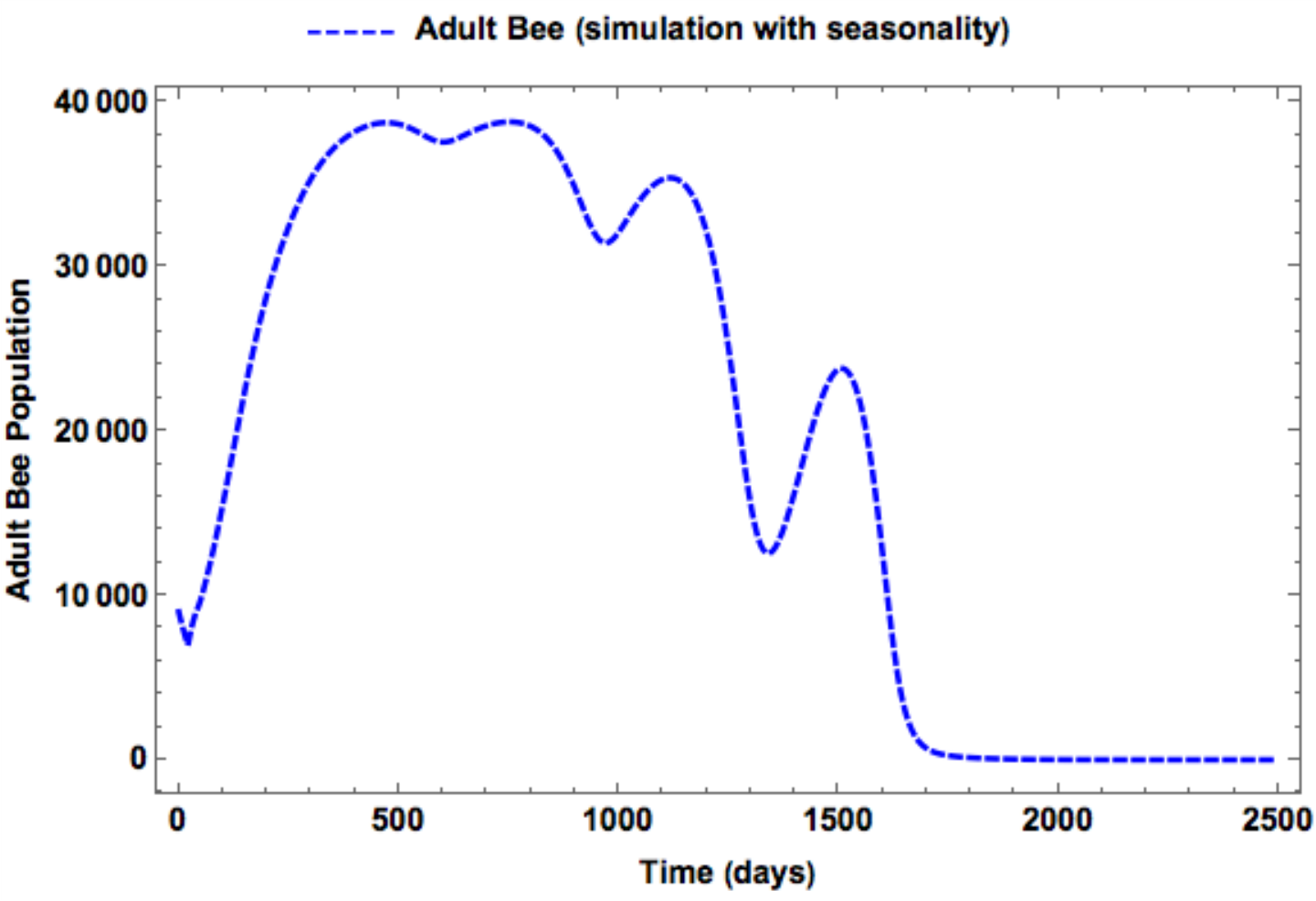}\label{fig_Bee_Extinction_S}}\\
\subfigure
{\includegraphics[height = 55mm, width = 60mm]{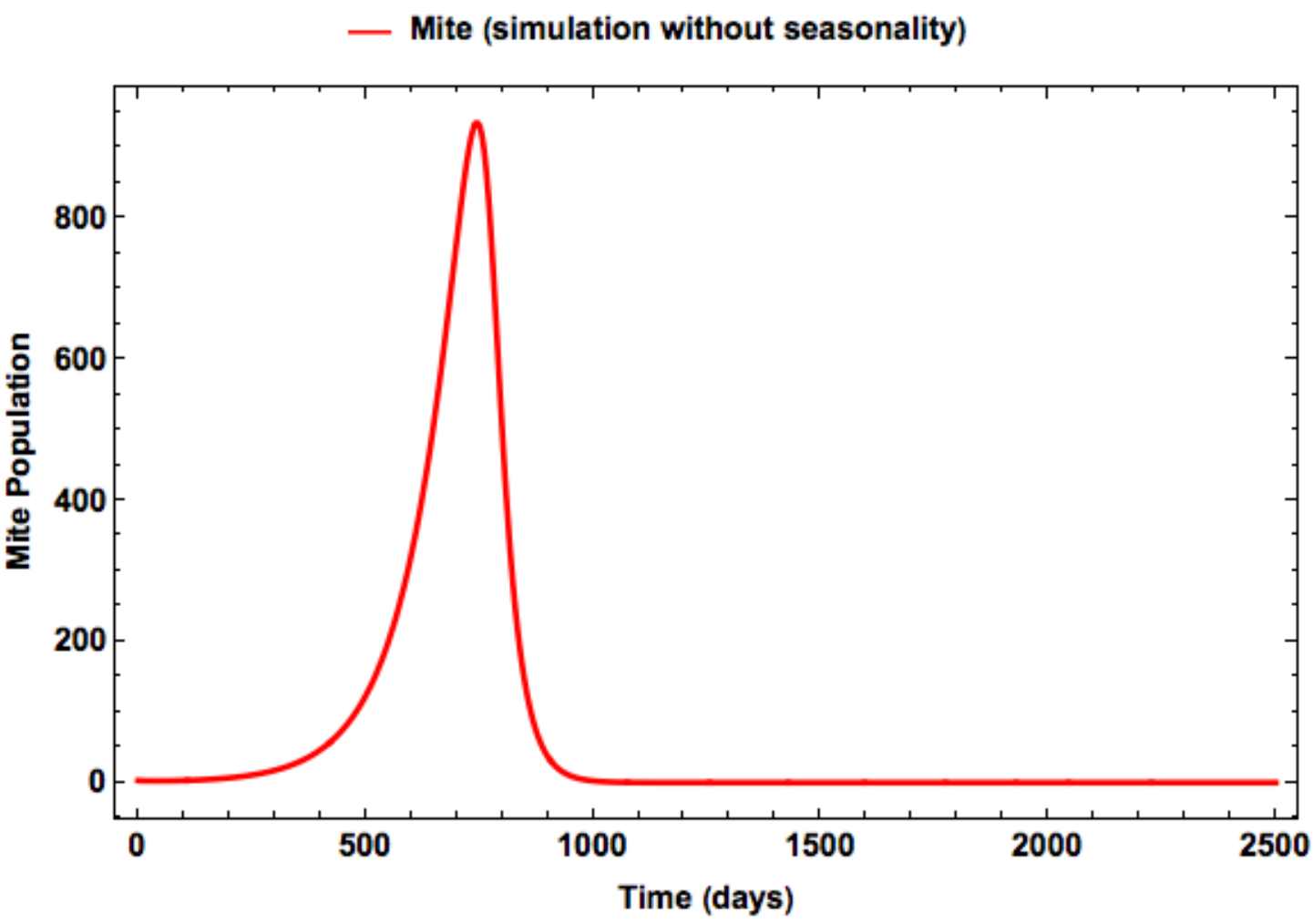}\label{fig_Mite_Extinction_NoS}}\hspace{5mm}
\subfigure
{\includegraphics[height = 55mm, width = 60mm]{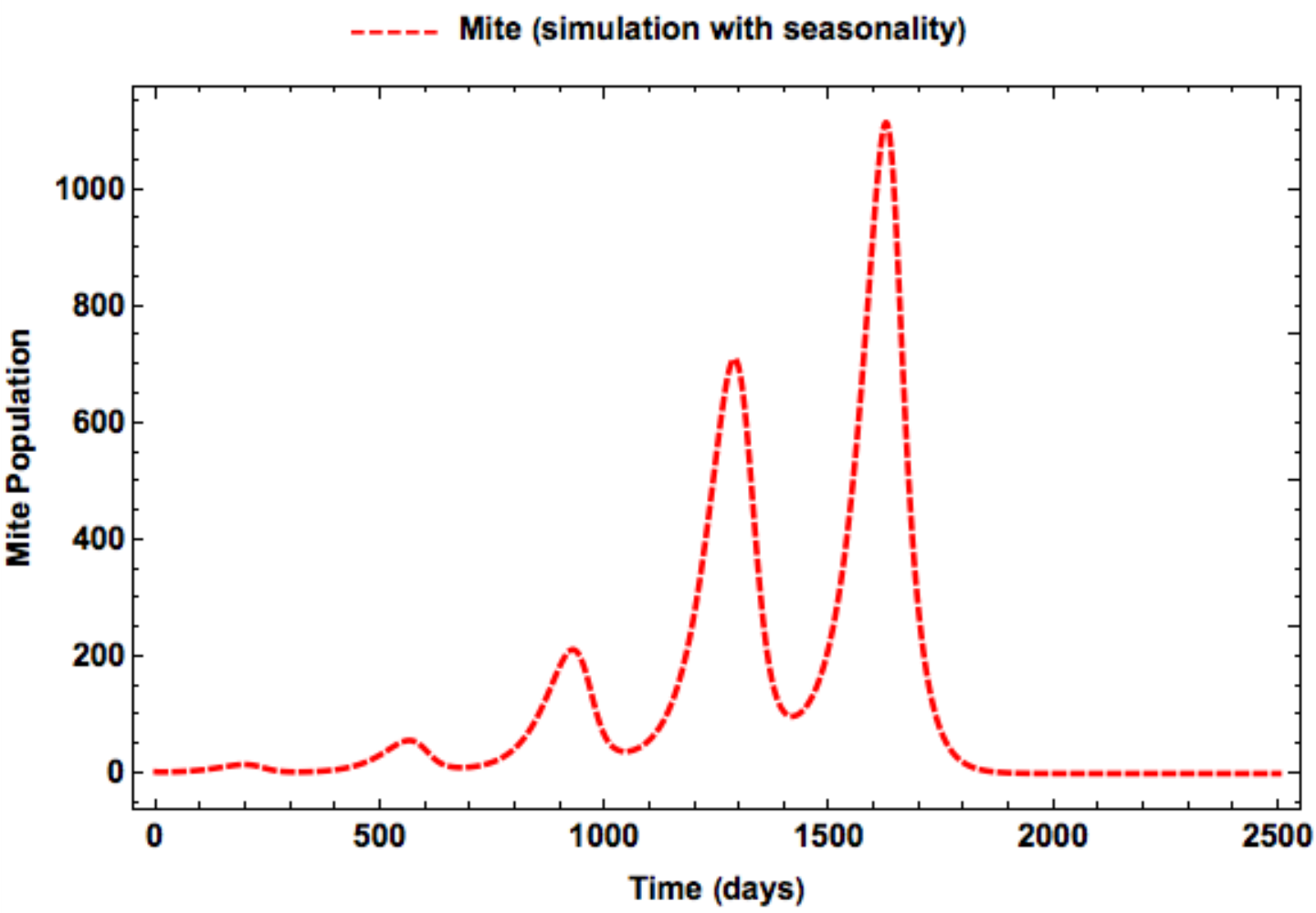}\label{fig_Mite_Extinction_S}}\\
$\alpha_b=0.028$
\end{center}
\vspace{-10pt}
\caption[Time Series: Brood, Adult Bee, and Mite Simulation with $\alpha_b=0.028$.]{{\small  Time series of the brood, adult bee, and mites simulation using $r=1500$, $K=95000000$, $d_b =0.051$, $d_h= 0.0121$, $d_m=0.027$, $\alpha_h=0.8$, $c=1.9$, $a=8050$, $\tau=21$, $\Phi=65$, $B_0(t)=B(0)=0$, $H(0)= 9000$, and $M(0) = 3$ when the queen's eggs laying rate is constant in figures on the left column (i.e. no seasonality) and when the queen's eggs laying rate has seasonality in figures on the right column with $\alpha_b =0.028$.}}
\label{fig:TimeSerieExtinction}
\end{figure}


{\scriptsize
  \begin{table}[ht]
  \centering
 {\small \begin{tabular}{|>{\bfseries}c|*{4}{c|}}\hline
    \multirow{2}{*}{\bfseries Parameters} & \multicolumn{2}{c|}{\bfseries PRCC} &
                                               \multicolumn{2}{c|}{\bfseries eFAST} \\\cline{2-5}
                       & \textbf{sensitivity index} & \textbf{p-value} & \textbf{first-order $S_i$}       & \textbf{total-order $S_{Ti}$} \\ \hline
    $r$         & 0.98441$^{***}$        & 0             &   0.67936            & 0.68784             \\ \hline
    $\Phi$ & 0.48556 $^{***}$     & 2.7807e-60        &   0.0068948             & 0.0082977             \\ \hline
    $d_b$    & -0.96544 $^{***}$     & 0           & 0.3079             & 0.31461             \\ \hline
    $d_h$        & -0.51226 $^{***}$    &  5.3765e-68            &    0.0080172           & 0.0095901             \\ \hline
    $d_m$           & 0.012098      & 0.70239            & 3.7727e-05 & 0.00089457            \\ \hline
    $\alpha_b$           & -0.055847     & 0.077527            &  0.00012458 & 0.0011038            \\ \hline
    $\alpha_h$           & 0.012944      & 0.68266            & 1.2504e-05 &  0.00081607            \\ \hline
    $a$           & 0.0469      &  0.13832            &  3.3803e-05 & 0.001148            \\ \hline
    $K$           & -0.70529$^{***}$     & 2.8295e-151           & 0.023235 & 0.024764            \\ \hline
    $c$           & -0.074858 $^{*}$     & 0.017905           & 0.00014943 & 0.0010502            \\ \hline
    $\tau$           & -0.36885$^{***}$   &  1.3841e-33            & 0.0056511 & 0.006688            \\ \hline
  \end{tabular}}
  \vspace{0.1in}
    \caption[Comparison of PRCC and eFAST Values at Time 96]{\small Comparison of PRCC and eFAST Values at Time 96 and $^{*}$ implies significance at 0.05 (i.e. $p<0.05$), $^{**}$ is the significance at 0.01 (i.e. $p<0.01$), and $^{***}$ implies significance at 0.001 (i.e. $p<0.001$).}
    \label{fig:PRCCeFAST96}
  \end{table}
}


{\scriptsize
  \begin{table}[H]
  \centering
{ \small \begin{tabular}{|>{\bfseries}c|*{4}{c|}}\hline
    \multirow{2}{*}{\bfseries Parameters} & \multicolumn{2}{c|}{\bfseries PRCC} &
                                               \multicolumn{2}{c|}{\bfseries eFAST} \\\cline{2-5}
                       & \textbf{sensitivity index} & \textbf{p-value} & \textbf{first-order $S_i$}       & \textbf{total-order $S_{Ti}$} \\ \hline
    $r$         & 0.86644$^{***}$        & 8.2575e-314             & 0.37861      & 0.4092             \\ \hline
    $\Phi$ & 0.067352$^{*}$     & 0.033204             & 0.00082837             & 0.0034278             \\ \hline
    $d_b$    & 0.17398$^{***}$     & 3.0723e-08          & 0.0046109   & 0.010879             \\ \hline
    $d_h$        & -0.70591$^{***}$    &  1.1773e-151            &  0.12011    & 0.1243           \\ \hline
    $d_m$           & 0.35619 $^{***}$      & 2.7923e-31            & 0.040672 &  0.10706            \\ \hline
    $\alpha_b$           & -0.67827$^{***}$     & 1.0389e-135            & 0.097734 & 0.1356           \\ \hline
    $\alpha_h$           & -0.098217 $^{**}$     & 0.0018739            & 0.0040036 & 0.010628            \\ \hline
    $a$           & 0.28282$^{***}$      &  7.56e-20            & 0.019395 & 0.043743     \\ \hline
    $K$           & -0.35489$^{***}$     & 4.7513e-31           &  0.020604 & 0.027903          \\ \hline
    $c$           & -0.67647$^{***}$     &  9.8499e-135           & 0.12116 & 0.21043            \\ \hline
    $\tau$           & -0.5618 $^{***}$   &  2.9213e-84           &  0.01145 &  0.014629            \\ \hline
  \end{tabular}}
    \vspace{0.1in}
    \caption[Comparison of PRCC and eFAST Values at Time 132]{\small Comparison of PRCC and eFAST Values at Time 132 and $^{*}$ implies significance at 0.05 (i.e. $p<0.05$), $^{**}$ is the significance at 0.01 (i.e. $p<0.01$), and $^{***}$ implies significance at 0.001 (i.e. $p<0.001$).}
        \label{fig:PRCCeFAST132}
  \end{table}
}

{\scriptsize
  \begin{table}[H]
  \centering
 {\small \begin{tabular}{|>{\bfseries}c|*{4}{c|}}\hline
    \multirow{2}{*}{\bfseries Parameters} & \multicolumn{2}{c|}{\bfseries PRCC} &
                                               \multicolumn{2}{c|}{\bfseries eFAST} \\\cline{2-5}
                       & \textbf{sensitivity index} & \textbf{p-value} & \textbf{first-order $S_i$}       & \textbf{total-order $S_{Ti}$} \\ \hline
    $r$         & 0.70392 $^{***}$        & 1.9182e-150             & 0.068869     & 0.11504             \\ \hline
    $\Phi$ & 0.2048 $^{***}$     & 6.248e-11             & 0.004525       & 0.013015             \\ \hline
    $d_b$    & -0.59208 $^{***}$     & 1.2156e-95          & 0.030786      & 0.057749             \\ \hline
    $d_h$        & -0.13905$^{***}$    &  1.0195e-05            & 0.0021218     & 0.0058844             \\ \hline
    $d_m$           & -0.77909 $^{***}$      & 1.3251e-204            & 0.05834 & 0.10418            \\ \hline
    $\alpha_b$           & 0.93967$^{***}$     & 0            &  0.3487 & 0.010895            \\ \hline
    $\alpha_h$           & -0.0751$^{*}$      & 0.017538            & 0.0016142 & 0.041224            \\ \hline
    $a$           & -0.5804 $^{***}$      &  4.1474e-91            & 0.016902 &  0.0093502           \\ \hline
    $K$           & -0.27377$^{***}$     & 1.1928e-18           & 0.0037531 & 0.41943           \\ \hline
    $c$           & 0.93868 $^{***}$     &  0            & 0.27371 & Yes            \\ \hline
    $\tau$           & -0.1642  $^{***}$   &  1.7756e-07           & 0.0012931 & 0.0043944           \\ \hline
  \end{tabular}}
    \vspace{0.1in}
    \caption[Comparison of PRCC and eFAST Values at Time 183]{\small Comparison of PRCC and eFAST Values at Time 183 and $^{*}$ implies significance at 0.05 (i.e. $p<0.05$), $^{**}$ is the significance at 0.01 (i.e. $p<0.01$), and $^{***}$ implies significance at 0.001 (i.e. $p<0.001$).}
        \label{fig:PRCCeFAST183}
  \end{table}
}

\end{appendices}


 \section*{Acknowledgments}
This research is partially supported by NSF-DMS (Award Number 1716802);  NSF- IOS/DMS (Award Number 1558127) and The James S. McDonnell Foundation 21st Century Science Initiative in Studying Complex Systems Scholar Award (UHC Scholar Award 220020472). The research of K.M is also partially supported by the Department of Education GAANN (P200A120192). G.DH is partially supported by USDA-Areawide Research Grant.



 %
%
%



\end{document}